\documentclass[%
pre,
twocolumn,
 amsmath,amssymb,
 aps,
]{revtex4-2}

\usepackage{color,soul}

\usepackage[normalem]{ulem}
\usepackage{soul}
\usepackage{amsmath}
\usepackage{amssymb}
\usepackage{color}
\usepackage{graphicx}
\usepackage{dcolumn}
\usepackage{bm}
\usepackage{verbatim}

\usepackage[draft]{todonotes}

\begin{document}

\title{{Particles} on Demand method: theoretical analysis, simplification techniques and model extensions}

\author{N. G. Kallikounis}
\affiliation
{Department of Mechanical and Process Engineering, ETH Zurich, 8092 Zurich, Switzerland}
\author{I. V. Karlin}\thanks{Corresponding author}
 \email{ikarlin@ethz.ch}
\affiliation
{Department of Mechanical and Process Engineering, ETH Zurich, 8092 Zurich, Switzerland}

\date{\today}

\begin{abstract}
    The Particles on Demand method [B. Dorschner, F. B\"{o}sch and I. V. Karlin, {\it Phys. Rev. Lett.} {\bf 121}, 130602 (2018)] was recently formulated with a conservative finite volume discretization and validated against challenging benchmarks. In this work, we rigorously analyze the properties of the reference frame transformation and its implications on the accuracy of the model. Based on these considerations, we propose strategies to boost the efficiency of the scheme and to reduce the computational cost. Additionally, we generalize the model such that it includes a tunable Prandlt number via quasi-equilibrium relaxation. Finally, we adapt concepts from the multi-scale semi-Lagrangian lattice Boltzmann formulation to the proposed framework, further improving the potential and the operating range of the kinetic model. Numerical simulations of high Mach compressible flows demonstrate excellent accuracy and stability of the model over a wide range of conditions.
\end{abstract}

\maketitle

\section{Introduction}

The understanding of the nature of high-speed compressible flows has been a long sought goal in the scientific and engineering community. An accurate prediction of complex hydrodynamic features is crucial in modern research, as well as in technology, with examples such as the interpretation of astrophysical jets, captured in the images of deep space telescopes \cite{1DJetPaper, JET_GARDNER_2009} and the design of air-frames and propulsion systems of high-Mach  low-altitude flying vehicles \cite{Urzay_2018}. Throughout the history of computational fluid dynamics (CFD), a number of numerical approaches has been suggested for the simulation of high-speed flows, including artificial viscosity methods \cite{vonNeumann_1950}, total variation diminishing (TVD) \cite{TVD_HARTEN}, essentially non-oscillatory (ENO) schemes \cite{ENO_Harten_Osher_1987,ENO_2_1987_HartenETAL} and weighted ENO (WENO) schemes \cite{WENO_1_1984,WENO_2_1996}. The challenging nature of these flows renders the field an active research area \cite{Pirozzoli,EKATERINARIS2005192}, with developments such as positivity preserving limiters and targeted ENO (TENO) schemes \cite{LinFu_DMR}, extending the domain of CFD towards even more exotic hydrodynamics \cite{ZhangShu2010, LinFuAllSpeed, ZhangShu2012}.

In contrast to conventional CFD, the lattice Boltzmann method (LBM) addresses the evolution of  hydrodynamic fields through the dynamics of a fully discrete kinetic system of designed particles associated with the discrete velocities $\bm{c}_i$, $i=0,\dots,Q-1$. The state is described in terms of the {populations} $f_i(\bm{x},t)$, which evolve in time and space by a simple algorithm ``stream along links $\bm{c}_i$ and collide at the nodes $\bm{x}$ in discrete time $t$". LBM has evolved into a versatile tool for the simulation of complex flows including  transitional flows \citep{Dorschner2017JFM}, flows in complex moving geometries \citep{dorschner2016entropic}, thermal and convective flows \cite{he1998,guo2007twopop,karlinConsistent}, multiphase and multicomponent flows \citep{Mazloomi2015prl,Mazloomi2017JFM,MultiPhase_Wohrwag,NileshJFM1}, reactive flows \cite{NileshReactive} and rarefied gas {\citep{shan2006kinetic}}, to mention a few recent instances; see \cite{SHARMAReview,KrugerBook,SucciBook} for a discussion of LBM and its application areas. However, despite the high efficiency and low numerical dissipation of LBM for nearly incompressible flows, the domain of high-speed compressible flows presents a number of severe challenges \cite{Qian_Orszag_1993,Guo_2007_Thermal,he1998,McNamara_1995,Shan1998}. The main directions to extend conventional LBM towards the compressible realm includes standard lattices LBM augmented with correction terms \cite{Prasianakis2007,Saadat2019,HosseinExtended1,HosseinExtended2}, multi-speed lattices \cite{Chikatamarla2006,Chikatamarla2009,Alexander1993,Frappoli_2015,Frappoli_2016} and hybrid approaches \cite{FENG_2016_hybrid,FENG_2019_hybrid,Guo_2020_Hybrid,RENARD_2021_hybrid}.

A common feature of the conventional LBM is the propagation of the populations with fixed discrete velocities, which translates as fixing the reference frame "at rest". It is well known that, when the fluid velocity significantly deviates from the frame velocity, errors and numerical instabilities corrupt the solution, impeding the applicability of LBM to high-Mach flows \cite{Qian_Orszag_1993, Prasianakis2007, Hosseini_ShiftedAnalysis}.  A remedy was the introduction of {\it uniformly} shifted lattices, which amounts to a constant shift of the reference frame, at every grid point of the numerical domain \cite{Frapolli_ShiftedLattices}. The concept demonstrated excellent performance for predominately unidirectional compressible flows, shifting the operational domain of the method in par with the chosen reference velocity shift \cite{Frapolli_ShiftedLattices,Frappoli_2016,Saadat2019}. While the concept of the uniform frame shift maintains key advantages of the scheme, such as simplicity and exact propagation, its potential diminishes for flows exhibiting large variations in flow velocity and temperature, due to the inevitable presence of strong deviations between the velocity of the actual flow and the imposed reference frame.

In contrast with the conventional LBM, 
the recently proposed Particles on Demand (PonD) method reformulates the kinetic equations in a space-time adaptive reference frame, dictated by the actual local fluid velocity and temperature \cite{Pond}. Two key elements were introduced with the PonD method: Firstly, PonD uses a consistent representation of  populations  in different reference frames, an operation termed as reference frame transformation. Secondly, a predictor-corrector iteration loop was applied, which ensured the realization of the propagation and collision step in the local co-moving reference frame, thereby optimizing accuracy and stability. Early realizations of PonD employed a semi-Lagrangian discretization, providing off-lattice flexibility to accommodate a varying reference frame, and validated the central concepts with a series of benchmarks, including multiphase and rarefied flows \cite{Ehsan2020,Ehsan2021,EhsanThesis,NileshPonD,Multiscale2021,RegPond,PondReg2}. However, the semi-Lagrangian method is prone to errors in conservation of mass, momentum and energy, deteriorating the accuracy of the solution in the presence of discontinuities (shock waves) \cite{PonD_DUGKS}. As a remedy to these shortcomings, a finite volume formulation of PonD was proposed in \cite{PonD_DUGKS}, following the discretization of the discrete unified gas kinetic scheme (DUGKS) \cite{Guo_DUGKS_IsoT,Guo_DUGKS_Compres,Guo_DUGKS_Rev}. The resulting conservative scheme, combined with a reference frame transformation based on Grad's projection of particles populations, demonstrated excellent performance in an array of challenging hypersonic compressible benchmark flows, including extreme hydrodynamic features such as the formation of near-vacuum regions.

In this paper, we aim at a further development of the finite-volume formulation of PonD, targeting strategies that simplify the scheme and enhance efficiency. A detailed analysis of the solution methodology is presented, along with the requirements to be met by the reference frame transformation. The scheme is extended to include a forcing term, as well as a variable Prandtl number. Finally, we combine the idea of the multiscale framework suggested in \cite{Multiscale2021} , with the Grad's projection frame transformation. The theoretical findings are validated in a series of numerical experiments along with extensive benchmarking of the scheme in challenging hydrodynamic flows.

The paper is organized as follows. The formulation of the kinetic equations in an adaptive reference frame is laid out in detail in Sec.\ \ref{sec::AdaptiveFormulation}. Sec.\ \ref{sec::Model} presents the kinetic model, which allows for variable adiabatic exponent and Prandlt number. Subsequently, Sec.\ \ref{sec::NumericalImplementation} describes the numerical discretization of the model. The model is extensively benchmarked in Sec.\ \ref{sec::results}, along with demonstration of important notions of the reference frame transformation. Finally, concluding remarks are provided in Sec.\ \ref{sec::Conclusions}.

\section{Adaptive reference frame formulation}
\label{sec::AdaptiveFormulation}

\subsection{Discrete velocities}

Without a loss of generality, we consider discrete speeds in two dimensions formed by tensor products of roots of Hermite polynomials $c_{i\alpha}$,
\begin{equation}
	\label{eq:ci}
	\bm{c}_i=(c_{ix}, c_{iy}).
\end{equation}
The model is characterized by the lattice temperature $T_L$ and the weights $W_i$ associated with the vectors (\ref{eq:ci}),
\begin{equation}
	\label{eq:wi}
	W_i=w_{ix}w_{iy},
\end{equation}
where $w_{i\alpha}$ are weights of the Gauss--Hermite quadrature. The discrete velocities and the associated weights are shown in Table \ref{tab:GaussHermiteVelSets}. With the discrete speeds (\ref{eq:ci}), the particles' velocities $\bm{v}_i^{\lambda_{\rm ref}} $ are defined relative to a reference frame $\lambda_{\rm ref}$,
specified by the frame velocity $\bm{u}_{{\rm ref}}$ and the frame temperature $T_{{\rm ref}}$,
\begin{align}\label{eq:veli}
        \lambda_{\rm ref} &=\{\bm{u}_{\rm ref},T_{\rm ref}\},\\
	\bm{v}_i^{\lambda_{\rm ref}} &=\sqrt{\frac{T_{\rm ref}}{T_L}}\bm{c}_i+\bm{u}_{\rm ref}.
\end{align}
The optimal reference frame is the comoving reference frame, which is specified by the \emph{local} temperature $T_{\rm ref}=T(\bm{x},t)$ and the \emph{local} flow velocity $\bm{u}_{\rm ref}=\bm{u}(\bm{x},t)$.

\begin{table}[h] \centering
	\caption{Lattice temperature $T_L$, roots of Hermite polynomials $c_{i\alpha}$ and weights $w_{i\alpha}$ of the $D=1$ Gauss--Hermite quadrature, and nomenclature.}
	\label{tab:GaussHermiteVelSets}
	\begin{tabular}{l|l|l|l|l}
		Model & $T_L$   &$c_{i\alpha}$      & $w_{i\alpha}$    & $D=2$     \\ 
		      &         &                   &                   &    \\
		$D1Q3 $ & $1$ &  $0,$     & $2/3$   & $D2Q9$          \\ 
		&    &$\pm\sqrt{3}$   & $1/6$       &       \\ 
		&        &                &          &           \\
		$D1Q4$ & $1$  &$\pm \sqrt{3-\sqrt{6}} $     & $(3+\sqrt{6})/12$   &   $D2Q16$         \\
		&   &$\pm \sqrt{3+\sqrt{6}}$     & $(3-\sqrt{6})/12$           &   \\
		&   &                            &                            &    \\
		$D1Q5$ & $1$   &$0$   & $8/15$          & $D2Q25$    \\ 
		&$ $     &$\pm \sqrt{5-\sqrt{10}}$   & $(7+2\sqrt{10})/60$    &          \\ 
		&$ $     &$\pm \sqrt{5+\sqrt{10}}$   & $(7-2\sqrt{10})/60$    &          \\ 
	\end{tabular}
\end{table}

\subsection{Reference frame transformation}
\label{subsec::ReferenceFrameTransformation}

A critical element of our construction is the transformation of the populations $f_i^{\lambda}$, defined with respect to a $\lambda$ reference frame, to a different reference frame $\lambda'$,
\begin{equation}\label{eq:refprime}
	\lambda'=\{\bm{u}',T'\}.
\end{equation}
In this work, we follow the strategy of \cite{PonD_DUGKS}. Let us denote $\bm{M}_k^{\lambda}$ a moment tensor of order $k$,
\begin{equation}
	\bm{M}_k^{\lambda} =\sum_{i=0}^{Q-1}f_i^{\lambda} \underbrace{ \bm{v}_{i}^\lambda  \bm{v}_{i}^\lambda \cdots \bm{v}_{i}^\lambda}_{k}.
\end{equation}
The reference frame transformation is then defined by the condition of invariance of the moments of orders $k=0,1, \dots, K$,
\begin{equation}\label{eq:Mcondition}
    \bm{M}_k^{\lambda'}=\bm{M}_k^{\lambda},\ k=0,1, \dots, K,
\end{equation}
where $K$ denotes the maximal moment order which is required to be frame invariant. The transformed populations are then sought as a Grad's projection,
\begin{equation}
\label{GradsProjection}
    f_i^{\lambda'}=W_i \sum_{n=0}^{K}\frac{1}{n!}\bm{\alpha}^{(n)}(\bm{m};\lambda')\bm{H}^{(n)}(\bm{c}_i),
\end{equation}
where $\bm{H}^{(n)}(\bm{c}_i)$ correspond to the Hermite polynomials of the lattice velocities and the expansion coefficients $\bm{\alpha}^{(n)}(\bm{m};\lambda')$ are calculated such that the moment invariant system \eqref{eq:Mcondition} is satisfied (detailed in Appendix\ \ref{HermiteAppendix}). The latter depend on the vector of frame invariant moments $\bm{m}=\{ \bm{M}_0,\dots, \bm{M}_{K} \}$ and the target reference frame $\lambda'$. As a shorthand notation for the reference frame transformation, we use the following formula,
\begin{equation}
     f_i^{\lambda'}=\mathcal{G}_{i,\lambda}^{\lambda'}f^{\lambda}.
\end{equation}

\subsection{Solution methodology}

We consider a simple kinetic model, which recovers compressible hydrodynamics under the restriction of fixed adiabatic exponent and Prandtl number. The kinetic evolution can be formulated in an arbitrary constant, four-parametric, reference frame $\overline{\lambda}$,
\begin{equation}
\label{eq::StartingEquationPonD}
    \partial_tf_i^{\overline{\lambda}}+\bm{v}_{i }^{\overline{\lambda}}  \cdot \nabla f_i^{\overline{\lambda}}=\Omega_{f,i}^{\overline{\lambda}},
\end{equation}
where $\Omega_{f,i}^{\overline{\lambda}}$ is a collision kernel of the populations. The populations $f_i^{\lambda(\bm{x},t)}$, are described with respect to a local reference frame $\lambda(\bm{x},t)$, which generally differs from the monitoring frame $\overline{\lambda}$. The reference frame transformation connects the populations between these two reference frames,
\begin{equation}
\label{eq::TransformedEqs}
f_i^{\overline{\lambda}}=\mathcal{G}_{i,\lambda(\bm{x},t)}^{\overline{\lambda}} f^{\lambda(\bm{x},t)}.
\end{equation}
To simplify the notation, we drop the space-time dependence of the reference frame and $\lambda$ is reserved for the local reference frame, $\lambda=\lambda(\bm{x},t)$. The overbar and subscripts shall be used to denote monitoring reference frames, uniform throughout the domain (e.g. $\overline{\lambda}$). Inserting the transformed populations \eqref{eq::TransformedEqs} into the evolution \eqref{eq::StartingEquationPonD} recovers the final equation of PonD,

\begin{equation}
\label{MainPondEquation_f}
  \partial_t(\mathcal{G}_{i,\lambda}^{\overline{\lambda}} f^{\lambda})+\bm{v}_{i }^{\overline{\lambda}} \cdot \nabla (\mathcal{G}_{i,\lambda}^{\overline{\lambda}} f^{\lambda})= \mathcal{G}_{i,\lambda}^{\overline{\lambda}}\Omega_f^{\lambda}.
\end{equation}

We emphasize that the kinetic equation \eqref{MainPondEquation_f} is formulated with respect to a uniform frame $\overline{\lambda}$. Therefore, Eq.\ \eqref{MainPondEquation_f} constitutes a typical kinetic equation, with constant characteristics, amenable to usual numerical realizations in the context of LBM, such as integration along characteristics. The necessary element is the introduction of the frame transformation operator. The effect of the varying reference frame is evident in the operation $\mathcal{G}_{i,\lambda}^{\overline{\lambda}} f^{\lambda}$, inside the non-local gradient operations. The above observation is crucial and determines the requirements that must be satisfied by the reference frame transformation.

Comments are in order:
\begin{itemize}

\item The kinetic equations can be formulated in principle with respect to any arbitrary reference frame. As such, different monitoring points can employ different reference frames. The consistency of the evolution in different reference frames is established by proper reference frame transformations. In the limit of infinite discrete velocities, the solution in every frame would be identical, i.e. no reason to do that. But in discrete systems, the accuracy of the solution depends on the proximity of the imposed frame with the actual local frame, dictated by the local flow conditions. Thus, with this procedure, we maximize the accuracy of a given model across the domain.

\item

The direct formulation of a kinetic equation with 
adaptive velocities leads to additional "forcing" terms,
containing derivatives with respect to the particles velocities.
A thorough discussion in this direction can be found in \cite{KaufThesis}. In the context of PonD, the solution methodology consists of a set of equations, each one at its own, spatially uniform, reference frame. This strategy avoids the explicit requirement for the computation of "forcing" terms.
The price to be paid instead, amounts to the operation of reference frame transformations.

\item Which domain needs to be transformed around a "frame-generating" monitoring point? Thanks to the hyperbolicity of the system, only the numerical domain of dependence needs to be transformed. Of course, this procedure for elliptic type of equations would be prohibitively computationally demanding.

\end{itemize}

\subsection{Hydrodynamic limit analysis}
\label{subsec::HydroLimit}

We analyze the governing kinetic equation \eqref{MainPondEquation_f} with the Chapman--Enskog method and investigate the consistency requirements for the moment invariant system. We rewrite Eq.\ \eqref{MainPondEquation_f} in terms of a Bhatnagar--Gross--Krook (BGK) collision operator and a small parameter $\epsilon$ for the relaxation time $\tau_1$,
\begin{equation}
    \partial_t(\mathcal{G}_{i,\lambda}^{\overline{\lambda}} f^\lambda) +\bm{v}^{\overline{\lambda}}_i  \cdot \nabla (\mathcal{G}_{i,\lambda}^{\overline{\lambda}} f^\lambda) =\frac{1}{\epsilon \tau_1} \mathcal{G}_{i,\lambda}^{\overline{\lambda}}(f^{\lambda,{\rm eq}}-f^\lambda).
\end{equation}
Following the conventional notation, we introduce the following multiscale expansion,
\begin{align}
\partial_t &= \partial_t^{(1)}+\epsilon\partial_t^{(2)}+\mathcal{O}(\epsilon^2), \\
f_i^\lambda &=f_i^{\lambda,(0)}+\epsilon f_i^{\lambda,(1)}+\epsilon^2 f_i^{\lambda,(2)}+\mathcal{O}(\epsilon^3).
\end{align}
We inject the expansions into the governing equations and separate the dynamics according to different orders of $\epsilon$, 
\begin{align}
\label{eq::CE_KineticEqs1}
\begin{split}
&\mathcal{O}(\epsilon^0):\quad \mathcal{G}_{i,\lambda}^{\overline{\lambda}}f^{\lambda,(0)}=\mathcal{G}_{i,\lambda}^{\overline{\lambda}}f^{\lambda,{\rm eq}},
\end{split}\\
\begin{split}
\label{eq::CE_KineticEqs2}
\mathcal{O}(\epsilon^1):\quad    \partial_t^{(1)}(\mathcal{G}_{i,\lambda}^{\overline{\lambda}} f^{\lambda,(0)})+ \bm{v}^{\overline{\lambda}}_i  \cdot \nabla  (\mathcal{G}_{i,\lambda}^{\overline{\lambda}} f^{\lambda,(0)}) \\= -\frac{1}{\tau_1}\mathcal{G}_{i,\lambda}^{\overline{\lambda}}f^{\lambda,(1)},
\end{split}\\
\begin{split}
\label{eq::CE_KineticEqs3}
\mathcal{O}(\epsilon^2):\quad         \partial_t^{(2)}(\mathcal{G}_{i,\lambda}^{\overline{\lambda}} f^{\lambda,(0)})+ \partial_t^{(1)}(\mathcal{G}_{i,\lambda}^{\overline{\lambda}} f^{\lambda,(1)}) \\+ \bm{v}^{\overline{\lambda}}_i  \cdot \nabla  (\mathcal{G}_{i,\lambda}^{\overline{\lambda}} f^{\lambda,(1)}) = -\frac{1}{\tau_1}\mathcal{G}_{i,\lambda}^{\overline{\lambda}}f^{\lambda,(2)}.
\end{split}
\end{align}
At the $\mathcal{O}(\epsilon^0)$ order we obtain the equilibrium populations,
\begin{equation}
    \mathcal{G}_{i,\lambda}^{\overline{\lambda}}f^{\lambda,(0)}=\mathcal{G}_{i,\lambda}^{\overline{\lambda}}f^{\lambda,{\rm eq}} \Leftrightarrow  f^{\lambda,(0)}_i=f^{\lambda,{\rm eq}}_i,
\end{equation}
which implies the following solvability constraints,
\begin{equation}
    \sum_{i=0}^{Q-1}  \{1,\bm{v}^{\overline{\lambda}}_i,({v}^{\overline{\lambda}}_i)^2 \} \mathcal{G}_{i,\lambda}^{\overline{\lambda}}f^{\lambda,(k)}  = \{0,\bm{0},0 \},\ k\ge 1.
\end{equation}

\subsubsection{Equilibrium moments}
\label{subsubsec::EqmMoments}
The functional form of the equilibrium moments is the basic element of the analysis and determines the recovered hydrodynamic equations. We underline that all the lattices discussed in this work and listed in Table \ref{tab:GaussHermiteVelSets}, reproduce the pertinent equilibrium moments as their Maxwell--Boltzmann (MB) continuous counterparts  in the comoving reference frame. For example, even for the standard $D2Q9$ lattice, the evaluation with the comoving reference frame frame 
\begin{align}
    \lambda &=\{\bm{u},T\}, \\
   \bm{v}_i^{\lambda} &=\sqrt{T/T_L}\bm{c}_i+\bm{u}, \\
    f_i^{\mathrm{eq},\lambda} &=\rho W_i,
\end{align}
retrieves the following moments,
\begin{align}
    \bm{J}^{\mathrm{eq},\lambda} &= \sum_{i=0}^{Q-1} \bm{v}_i^{\lambda}f_i^{\mathrm{eq},\lambda}&&=\bm{J}^{\mathrm{MB}},\\
    \bm{P}^{\mathrm{eq},\lambda} &= \sum_{i=0}^{Q-1} \bm{v}_i^{\lambda}\bm{v}_i^{\lambda}f_i^{\mathrm{eq},\lambda}&&=\bm{P}^{\mathrm{MB}},\\
\bm{Q}^{\mathrm{eq},\lambda} &=\sum_{i=0}^{Q-1} \bm{v}_i^{\lambda}\bm{v}_i^{\lambda}\bm{v}_i^{\lambda}f_i^{\mathrm{eq},\lambda}&&=\bm{Q}^{\mathrm{MB}},    \\
   \bm{q}^{\mathrm{eq},\lambda} &= \sum_{i=0}^{Q-1} \bm{v}_i^{\lambda}({v}_i^{\lambda})^2f_i^{\mathrm{eq},\lambda}&&=\bm{q}^{\mathrm{MB}},\\
\bm{R}^{\mathrm{eq},\lambda} &=\sum_{i=0}^{Q-1} (v_i^{\lambda})^2\bm{v}_i^{\lambda}\bm{v}_i^{\lambda}f_i^{\mathrm{eq},\lambda}&&=\bm{R}^{\mathrm{MB}}.    
\end{align}
where the MB moments are,
\begin{align}
\label{MB_Moments_FIRST}
\bm{J}^{\mathrm{MB}} &=\rho \bm{u},\\
\bm{P}^{\mathrm{MB}} &=\rho T \bm{I}+\rho \bm{u}\bm{u},\\
\bm{Q}^{\mathrm{MB}} &=\rho T\overline{\bm{u}\bm{I}}+\rho\bm{u}\bm{u}\bm{u}, \\
\bm{q}^{\mathrm{MB}} &=\rho\bm{u}(u^2+T(D+2)),\\
\label{MB_Moments_LAST}
\bm{R}^{\mathrm{MB}} &=\rho T((D+2)T+u^2)\bm{I}+\rho((D+4)T+u^2)\bm{u}\bm{u}.
\end{align}
Here, overline denotes symmetrization. While all pertinent equilibrium moments in the comoving reference frame are accurate, the same conclusion does not necessarily hold when a different, non comoving, arbitrary reference frame $\overline{\lambda}=\{\overline{\bm{u}},\overline{T}\}$ is used for the evaluation. Indeed, the crucial difference between various lattices rests with the frame invariance of the equilibrium moments. 

In the non comoving reference frame $\overline{\lambda}$, the equilibrium populations $f^{{\rm eq}, \overline{\lambda}}$ are no longer given by the simple expression $\rho W_i$, but must be computed. This can easily be accomplished by the reference frame transformation,
\begin{equation}
\label{NonComovingEq}
   f^{{\rm eq}, \overline{\lambda}}=\mathcal{G}_{\lambda}^{\overline{\lambda}}   f^{{\rm eq}, {\lambda}}=\rho \mathcal{G}_{\lambda}^{\overline{\lambda}}W,
\end{equation}
operating on the vector of the lattice weights $W$, from the comoving frame $\lambda$ to the $\overline{\lambda}$ frame. By construction, the equilibrium moments in the $\overline{\lambda}$ frame match the MB moments, if they are frame invariant. For example, the following relation holds for a third-order Grad's projection, sustained by the $D2Q16$ lattice,
\begin{equation}
    \bm{Q}^{\mathrm{eq},\overline{\lambda}}=\sum_{i=0}^{Q-1} \bm{v}_i^{\overline{\lambda}}\bm{v}_i^{\overline{\lambda}}\bm{v}_i^{\overline{\lambda}}f_i^{\mathrm{eq},\overline{\lambda}}=\bm{Q}^{\mathrm{eq},{\lambda}}=\bm{Q}^{\mathrm{MB}}.
\end{equation} 
However, deviations occur for higher order moments, not included in the set of frame invariant moments. Continuing with the same example, the fourth-order equilibrium moment in the $\overline{\lambda}$ frame becomes,
\begin{equation}
\bm{R}^{\mathrm{eq},\overline{\lambda}}=\bm{R}^{\rm MB}+\bm{R}'.
\end{equation}
The explicit form of the deviation can be computed via algebraic manipulations of Eqs.\ \eqref{MB_Moments_LAST}-\eqref{NonComovingEq},
\begin{equation}
\label{DeviationR}
\begin{split}
    \bm{R}'=-\rho (\bm{\xi} \bm{\xi} \xi^2+ \theta((D+2)\theta \bm{I}+  \xi^2 \bm{I}+ (D+4)\bm{\xi} \bm{\xi})),
\end{split}    
\end{equation}
where,
\begin{equation}
    \bm{\xi} = \bm{u}-\overline{\bm{u}}, \
    \theta = T-\overline{T}.
\end{equation}
As indicated by the previous expressions, the deviations vanish when the monitoring reference frame $\overline{\lambda}$ approach the comoving reference frame $\lambda$.

\subsubsection{Full invariant moment system}
\label{subsubsec::FullMomSystem}
Let us consider the case where the isotropy of the lattice supports the frame invariance of moments, up to fourth-order.
This case corresponds, in particular, to the $D2Q25$ lattice mentioned in Table\ \ref{tab:GaussHermiteVelSets}. The zeroth order moment evaluation of Eq.\ \eqref{eq::CE_KineticEqs2} leads to the following equation, 
\begin{equation}
\label{euler_continuity}
     \partial_t^{(1)}\rho^{\mathrm{eq},\overline{\lambda}} + \nabla \cdot \bm{J}^{\mathrm{eq},\overline{\lambda}}=0,
\end{equation}
where,
\begin{align}
    \rho^{\mathrm{eq},\overline{\lambda}} &=\sum_{i=0}^{Q-1} \mathcal{G}_{i,\lambda}^{\overline{\lambda}}f^{\mathrm{eq},\lambda}, \\
    \bm{J}^{\mathrm{eq},\overline{\lambda}} &=\sum_{i=0}^{Q-1} \bm{v}_i^{\overline{\lambda}}\mathcal{G}_{i,\lambda}^{\overline{\lambda}}f^{\mathrm{eq},\lambda}.
\end{align}
Both moments belong to the frame invariant system of the reference frame transformation. As such, they can be evaluated equally well at the comoving reference frame,
\begin{align}
\label{mom_0_invariant}
    \rho^{\mathrm{eq},\overline{\lambda}} &=\rho^{\mathrm{eq},\lambda}=\rho, \\
    \label{mom_1_invariant}
\bm{J}^{\mathrm{eq},\overline{\lambda}} &=\bm{J}^{\mathrm{eq},\lambda}=\rho \bm{u}.    
\end{align}
We substitute Eqs.\ \eqref{mom_0_invariant}, \eqref{mom_1_invariant} into Eq.\ \eqref{euler_continuity} and recover the continuity equation,
\begin{equation}
     \partial_t^{(1)}\rho + \nabla \cdot (\rho \bm{u}) = 0.
\end{equation}
The same reasoning applies to the rest of the conserved moments of Eq.\ \eqref{eq::CE_KineticEqs2}, since all pertinent moments are frame invariant. The momentum and energy conservation laws at the Euler level are as follows,
\begin{align}
\label{FullMoment_1order_First1}
      \partial_t^{(1)}(\rho \bm{u}) + \nabla \cdot \bm{P}^{\mathrm{MB}} &= 0,\\
\label{FullMoment_1order_Last1}      
    \partial_t^{(1)}(2\rho E) + \nabla \cdot \bm{q}^{\mathrm{MB}} &= 0.
\end{align}
Analogously, the moments of the second-order equation \eqref{eq::CE_KineticEqs3}
recover the Navier--Stokes--Fourier (NSF) contributions, 
\begin{align}
      \label{FullMoment_2order_Density}
     \partial_t^{(2)}\rho &= 0,\\
           \label{FullMoment_2order_momentum}
      \partial_t^{(2)}(\rho \bm{u}) + \nabla \cdot \bm{P}^{(1)} &= 0,\\
      \label{FullMoment_2order_Energy}
    \partial_t^{(2)}(2\rho E) + \nabla \cdot \bm{q}^{(1)} &= 0, 
\end{align}
where
\begin{align}
    \bm{P}^{(1)} &=-\tau_1\left( \partial_t^{(1)}\bm{P}^{\mathrm{MB}}+\nabla \cdot \bm{Q}^{\mathrm{MB}} \right),\\
    \bm{q}^{(1)} &=-\tau_1 \left( \partial_t^{(1)}\bm{q}^{\mathrm{MB}}+\nabla \cdot \bm{R}^{\mathrm{MB}} \right).
\end{align}
The compressible NSF equations are recovered from the summation of the $\mathcal{O}(\epsilon^1), \mathcal{O}(\epsilon^2)$ contributions,
\begin{align}
   \partial_t \rho &= -\nabla \cdot (\rho \bm{u}), \\
   \partial_t (\rho \bm{u}) &=-\nabla \cdot (\rho \bm{u}  \bm{u})-\nabla \cdot \bm{\pi}, \\
   \partial_t (\rho E) &= -\nabla \cdot (\rho E \bm{u})-\nabla \cdot \bm{q}-\nabla \cdot (\bm{\pi} \cdot \bm{u}),
\end{align}
where $\bm{\pi}$ is the pressure tensor,
\begin{equation}
  \bm{\pi}=\rho T \bm{I}-\mu \left(\bm{S}-\frac{2}{D}(\nabla \cdot \bm{u}) \bm{I} \right),
\end{equation}
$\bm{S}$ the strain rate tensor,
\begin{equation}
   \bm{S}=\nabla\bm{u}+\nabla\bm{u}^{\top}, 
\end{equation}
$\bm{q}$ is the heat flux,
\begin{equation}
    \bm{q} = - \kappa \nabla T,
\end{equation}
and 
\begin{align}
    \mu &= \tau_1 \rho T, \\
    \kappa &= \tau_1\rho C_p T.
\end{align}

\subsubsection{Third-order invariant moment system}
\label{subsubsec::ThirdOrderSys}
We continue with the analysis of a third order moment invariant system, which corresponds to the $D2Q16$ lattice. The Euler level dynamics and the NSF density and momentum contributions include moments which are frame invariant. Thus, Eqs.\ \eqref{FullMoment_1order_First1}-\eqref{FullMoment_1order_Last1} and the NSF density and momentum contributions \eqref{FullMoment_2order_Density}-\eqref{FullMoment_2order_momentum} are obtained accurately. However, the NSF energy contribution includes the flux of energy flux tensor, which is not frame invariant. The evaluation of this moment at the monitoring frame $\overline{\lambda}$, instead of the comoving frame $\lambda$, induces an error, as seen from Eq.\ \eqref{DeviationR}. This deviation gives rise to a diffusive error term at the NSF energy equation \eqref{FullMoment_2order_Energy},
\begin{equation}
     \label{3order_Energy_equation_OE2}
     \begin{split}
    \partial_t^{(2)}(2\rho E) =  \nabla  \cdot \tau_1 \left( \partial_t^{(1)}\bm{q}^{\text{MB}}+\nabla \cdot \bm{R}^{\text{MB}}+  \right. \\ \left. \frac{\partial \bm{R}'}{\partial \bm{u}}\cdot \nabla  \bm{u}+\frac{\partial \bm{R}'}{\partial T}\cdot \nabla  T\right).
      \end{split}
\end{equation}
As expected from this analysis, the $D2Q16$ lattice demonstrates excellent performance in inviscid Euler gas dynamic systems, even at the presence of very strong discontinuities. In the presence of important viscous effects, the error term (last term in Eq.\eqref{3order_Energy_equation_OE2}) will affect the accuracy of the solution. The magnitude of the error scales with the spatial variation of the reference frame, or in other words the gradients of the velocity and temperature field. As shown in subsequent numerical simulations, benchmark viscous hydrodynamic flows can be accurately captured with $D2Q16$ lattice, suggesting that the magnitude of the error term is rather weak. However, as the velocity and temperature gradients grow, the error terms manifests in the solution.
Sec. \ref{sec::viscousflows} provides further discussion on this topic, with the aid of numerical simulations.

\subsubsection{Second-order invariant moment system}

Finally, we examine the hydrodynamic properties of a second order frame invariant moment system, with a representative example being the $D2Q9$ lattice. Such a model cannot support the full energy flux tensor and higher order tensors ($\bm{Q},\bm{R})$ as frame invariant moments. Following the same reasoning with the previous cases, we observe that error terms are introduced in the energy equation of the Euler-level dynamics \eqref{FullMoment_1order_Last1} and the momentum and energy equations of the NSF-level dynamics \eqref{FullMoment_2order_momentum}, \eqref{FullMoment_2order_Energy}. For relatively smooth flows without shocks, numerical evidence suggest that the effect of the error terms of the $D2Q9$ model is rather small. A prominent example is the case of an advected vortex, which has been shown to be captured accurately even for vortex and advection Mach numbers at the range of $\text{Ma}_{v}=0.8$ and $\text{Ma}_{a}=100$ \cite{Pond}. However, for hydrodynamic flows with shocks, the errors due to the frame variation are non-negligible. Importantly, the shock dynamics at the Euler level are not described accurately, which translate into errors in the shock propagation speed.

\subsubsection{Summary of observations}

We summarize the domain of validity of the kinetic model, according to the order of the frame invariant moment system:
\begin{itemize}
    \item Second-order ($D2Q9$): Appropriate for smooth regions of compressible flows.
    \item Third-order ($D2Q16$): Appropriate for shocked compressible flows, with small dissipation effects (Euler flows).
    \item Fourth-order ($D2Q25$): Generally valid for NSF flows.
\end{itemize}
We take advantage of the above hierarchy to achieve the best efficiency with our framework. In particular, the low-order model can be applied to the smooth regions of the flow and the high-order model in the regions of steep hydrodynamic gradients. In the following section, we discuss the coupling of the different models, in the spirit of \cite{Multiscale2021}. 

\subsection{Multiscale frame transformation}
\label{subsec::Multiscale}

The Grad's projection approach for the reference frame transformation is advantageous, in terms of stability and efficiency. In the following, we demonstrate an additional benefit, which is the deployment of different lattices throughout the domain, with minimal change in the framework and limited computational overhead. In essence, we combine the core idea of the multiscale concept \cite{Multiscale2021} along with the Grad projection frame transformation. 

Let us consider two velocity sets of different order,
		\begin{align*}
			\mathcal{V}_q&=\{\bm{c}_{i}^{q}, i=0, \dotsc, q-1\},\\
			\mathcal{V}_Q&=\{\bm{c}_{i}^{Q}, i=0, \dotsc, Q-1\},
		\end{align*} 
where $q<Q$. We distinguish two different operations coupled with the frame transformation from $\lambda$ to $\lambda'$:

\begin{itemize}
		\item Lifting: The lifting operation switches from the lower-order $q$-model to the higher-order $Q$-model, requiring thus a map,
		\begin{align}
		f_q^{\lambda} \to f_Q^{\lambda'}.	\label{eq:lifting1}
		\end{align}
		\item Projection:  The projection operation switches from the higher-order $Q$-model to the lower-order $q$-model, requiring thus a map,
		\begin{align}
			f_Q^{\lambda} \to f_q^{\lambda'}.	\label{eq:projection1}
		\end{align}
	\end{itemize}

The construction of both operations amounts to identifying the proper expansion coefficients of the Grad's expansion. We recall that the expansion coefficients are function of the moments and the reference frame $\lambda'$.

\subsubsection{Lifting}

For the lifting operation, we can identify the list of moments $\bm{m}_{q \to Q}$, required for the reference frame transformation , as a composition,
\begin{equation}
    \bm{m}_{q \to Q}=\{ \bm{m}_q, \bm{m}_{Q-q}\},
\end{equation}
where $\bm{m}_q$ is operationally available from $f_q^{\lambda}$ and $\bm{m}_{Q-q}$ constitutes the remaining unknown higher order moments. The lifting operation consists in specifying the respective contributions as,
\begin{align}
	&\bm{m}_q=\mathcal{M}_{q,\lambda}f_q^{\lambda},\\
	&\bm{m}_{Q-q}=\bm{m}_{Q-q}^{\rm eq}.
\end{align}
where $\mathcal{M}_{q,\lambda}$ is the $q \times q$ matrix of the populations to moments map. With the required moments identified, the construction of the lifted populations proceeds similarly with Sec.\ \ref{subsec::ReferenceFrameTransformation}. The expansion coefficients are computed from the moments $ \bm{m}_{q \to Q}$ and the reference frame $\lambda'$,
\begin{equation}
   \bm{\alpha}_{n}=\bm{\alpha}_{n}( \bm{m}_{q \to Q};\lambda'). 
\end{equation}
The lifted populations $f_Q^{\lambda'}$ can then be found from Grad's expansion,
\begin{equation}
     f_{Q,i}^{\lambda'} =W_{i}^Q \sum_{n=0}^{K}\frac{1}{n!}\bm{\alpha}^{(n)}( \bm{m}_{q \to Q};\lambda')\bm{H}^{(n)}(\bm{c}_{i}^{Q}),   
\end{equation}
where $W_{i}^Q$, $\bm{H}^{(n)}(\bm{c}_{i}^{Q})$ and $K$ are the weights, the Hermite polynomials and the order of expansion of the high-order model respectively.

\subsubsection{Projection}

In the projection step, the high-order population $f_Q^{\lambda}$ contains the subset of the $q$ linearly independent moments, which is required for the construction of the low-order population $f_q^{\lambda'}$. Hence, in contrast to the lifting procedure, there is no missing information and the low-order moment vector $\bm{m}_{Q \rightarrow q}$ is operationally available from $f_Q^{\lambda}$,
\begin{equation}
    \bm{m}_{Q \to q}=\mathcal{M}_{Q,\lambda}f_Q^{\lambda}.
\end{equation}
Similarly to the lifting operation, the projected populations $f_q^{\lambda'}$ are given by the Grad's expansion,

\begin{equation}
     f_{q,i}^{\lambda'} =W_i^q \sum_{n=0}^{k}\frac{1}{n!}\bm{\alpha}^{(n)}(\bm{m}_{Q \to q};\lambda')\bm{H}^{(n)}(\bm{c}_{i}^{q}),   
\end{equation}
where $W_i^q$, $\bm{H}^{(n)}(\bm{c}_{i}^{q})$ and $k$ are the weights, the Hermite polynomials and the order of expansion of the low-order model respectively.

\section{Variable adiabatic exponent and Prandtl number}
\label{sec::Model}
\subsection{Kinetic model}

The kinetic model can be extended towards a variable adiabatic exponent via the two-population approach \cite{Frappoli_2016}. The second set of populations ($g$-populations) is designed to carry the internal energy associated with non-translational degrees of freedom, and thus enable an adjustable adiabatic exponent $\gamma=C_p/C_v$, where $C_p=C_v+1$ is the specific heat of ideal gas at constant pressure and $C_v$ is the specific heat at constant volume \cite{Rykov,Nie_gamma}.  The governing kinetic equations can be written as follows,
\begin{align}
\label{KineticModel_f}
{\partial_t f_i}+\bm{v}_i \cdot \nabla f_i &=\Omega_{f,i}= \frac{1}{\tau_1}(f_i^{\rm eq}-f_i), \\
\label{KineticModel_g}
{\partial_t g_i}+\bm{v}_i \cdot \nabla g_i &=\Omega_{g,i}= \frac{1}{\tau_1}(g_i^{\rm eq}-g_i),
\end{align}
Additionally, the collision operators can accommodate an intermediate relaxation to quasi-equilibrium states, thus enabling a variable Prandtl number \cite{Ansumali_QuasiEq, Frappoli_2016},
\begin{align}
\Omega_{f,i} &= \frac{1}{\tau_1}(f_i^{\rm eq}-f_i)+\left(\frac{1}{\tau_1}-\frac{1}{\tau_2} \right)(f_i^{\ast}-f_i^{\rm eq}), \\
\Omega_{g,i} &= \frac{1}{\tau_1}(g_i^{\rm eq}-g_i)+\left(\frac{1}{\tau_1}-\frac{1}{\tau_2}\right)(g_i^{\ast}-g_i^{\rm eq}),
\end{align}
where $f_i^{\ast}, g_i^{\ast}$ are the quasi-equilibria of the $f-$ and $g-$ populations and the relaxation time $\tau_2$ determines the Prandtl number. The local conservation laws for the density $\rho$, momentum $\rho\bm{u}$ and the total energy $\rho E$ are,
\begin{align}
\label{eq::ConservationLaws}
    \rho &= \sum_{i=0}^{Q-1}f_i = \sum_{i=0}^{Q-1}f_i^{\rm eq}, \\
    \rho\bm{u} &=  \sum_{i=0}^{Q-1}\bm{v}_if_i = \sum_{i=0}^{Q-1}\bm{v}_if_i^{\rm eq}, \\
     \rho E &= \sum_{i=0}^{Q-1} \frac{{v}_i^2}{2}f_i
     	+  \sum_{i=0}^{Q-1} g_i  =  \sum_{i=0}^{Q-1} \frac{{v}_i^2}{2}f_i^{\rm eq}+  \sum_{i=0}^{Q-1} g_i^{\rm eq},
\end{align}
where the total energy of ideal gas is,
\begin{equation}
	\label{eq:E}
	\rho E=C_v\rho T + \frac{\rho{u}^2}{2}.
\end{equation}
The equilibrium populations in the comoving reference frame are as follows,
\begin{align}
	\label{feqPond}
	&  f_i^{\rm eq} = \rho W_i, \\
	\label{geqPond}
	&  g_i^{\rm eq} = \left(C_v-\frac{D}{2}\right)T\rho W_i.
\end{align}
The expressions for the dynamic viscosity, bulk viscosity and thermal conductivity are \cite{Frappoli_2016},
\begin{align}
    \mu &= \tau_1 \rho T, \\
    \varsigma &= \left(\frac{1}{C_v}-\frac{2}{D}\right)\mu, \\
    \kappa &= \tau_2\rho C_p T.
\end{align}
The Prandtl number is therefore,
\begin{equation}
    \rm Pr =\frac{C_p \mu}{\kappa}=\frac{\tau_1}{\tau_2}.
\end{equation}
For $\rm Pr < 1$, the quasi-equilibria are designed to conserve the centered heat flux, resulting in the following expressions,
\begin{align}
\label{eq::quasi_f}
   f_i^{\ast} &= f_i^{\rm eq}+ W_i \bm{Q} \cdot (\bm{e}_i\bm{e}_i\bm{e}_i-3T\bm{e}_i \bm{I})/6T^3, \\
   g_i^{\ast} &= g_i^{\rm eq}+ W_i \bm{\varsigma} \cdot \bm{e}_i/T, 
\end{align}
where $\bm{e}_i=\bm{v}_i-\bm{u}$, $\bm{Q}$ is the non-equilibrium third-order flux tensor and $ \bm{\varsigma}$ is the energy flux associated with the internal degrees of freedom,
\begin{align}
    \bm{Q} &= \sum_{i=0}^{Q-1} \bm{e}_i \bm{e}_i \bm{e}_i (f_i-f_i^{\rm eq}), \\
    \bm{\varsigma} &= \sum_{i=0}^{Q-1} \bm{e}_i (g_i-g_i^{\rm eq}). 
\end{align}

\subsection{Comments on $g-$ populations}

We note that the concepts presented so far apply equally well for the $g-$ populations, with the sole difference being the required frame invariant moments, which have to be supported by the corresponding $g-$ lattice. A Chapman--Enskog analysis \cite{Frappoli_2016} shows that $g-$ equilibrium moments up to second order are enough to recover the NSF equations. Therefore, the $D2Q9$ is safely employed in this work for the $g-$ populations. The reference frame transformation (Sec.\ \ref{subsec::ReferenceFrameTransformation}) and its multiscale realization (Sec.\ \ref{subsec::Multiscale}) apply equally well for the $g-$ populations, taking into account that the maximal frame invariant moment is second order.

\section{Numerical implementation}
\label{sec::NumericalImplementation}

\subsection{Finite volume discretization}

We proceed with the finite-volume discretization, in the spirit of PonD-DUGKS framework \cite{Guo_DUGKS_IsoT,PonD_DUGKS}. In accord with the notions above, the kinetic equation can be formulated in an arbitrary reference frame $\lambda$. We first present the discretization for ${\rm Pr}=1$. The extension for variable Prandtl number is explained in the following section.

\subsubsection{Updating rule}

The evolution of the kinetic model \eqref{KineticModel_f}-\eqref{KineticModel_g} can be discretized as follows,
\begin{align}
    \label{eq:FinalUpdateDUGKS}
    \begin{split}
    \Tilde{f}_{i}^\lambda(\bm{x}_j,t_{n+1})&=\left( \frac{2\tau-\delta t}{2\tau+\delta t} \right) \Tilde{f}_{i}^\lambda(\bm{x}_j,t_{n})+ \\&
    \left(\frac{2\delta t}{2\tau+\delta t} \right){f}_{i}^{{\rm eq,\lambda}}(\bm{x}_j,t_{n})
-\frac{\delta t}{V_j} F_{f,i}^\lambda(\bm{x}_j,t_{n+1/2}),
    \end{split} \\
    \label{eq:FinalUpdateDUGKS_g}
     \begin{split}
    \Tilde{g}_{i}^\lambda(\bm{x}_j,t_{n+1})&=\left( \frac{2\tau-\delta t}{2\tau+\delta t} \right) \Tilde{g}_{i}^\lambda(\bm{x}_j,t_{n})+\\&
    \left(\frac{2\delta t}{2\tau+\delta t} \right){g}_{i}^{{\rm eq,\lambda}}(\bm{x}_j,t_{n})-\frac{\delta t}{V_j} F_{g,i}^\lambda(\bm{x}_j,t_{n+1/2}). 
     \end{split}
\end{align}
The update equations are derived from the integration of the continuous equations \eqref{KineticModel_f}-\eqref{KineticModel_g}, formulated in the reference frame $\lambda$, in a control volume centered at $\bm{x}_j$, with volume $V_j$, from time $t_n$ to $t_{n+1}=t_n+\delta t$, using the midpoint rule for the convection term and the trapezoidal rule for the collision term \cite{Guo_DUGKS_IsoT}. To remove the implicitness, DUGKS scheme adopts the variable transformation from the standard LBM practice, \cite{LBM_VarTransform1,LBM_VarTransform2}
\begin{equation}
    \label{eq:variableTransform}
    \Tilde{\phi}_{i}^\lambda={\phi}_{i}^\lambda-\frac{\delta t}{2}\Omega_{\phi,i}^\lambda={\phi}_{i}^\lambda-\frac{\delta t}{2\tau}(\phi_i^{\rm eq,\lambda}-\phi_i^\lambda),  
\end{equation}
where $\phi$ stands for the $f$- and $g$- populations and $\Omega_{\phi,i}$ are the collision BGK kernels defined in Eqs. \eqref{KineticModel_f}-\eqref{KineticModel_g}. The fluxes of the populations $F_{\phi,i}^\lambda(\bm{x}_j,t_{n+1/2})$ across the surface of the control volume are defined as,
\begin{equation}\label{eq:fluxesDUGKS}
    F_{\phi,i}^\lambda(\bm{x}_j,t_{n+1/2})=\int_{\partial V_j}(\bm{v}_i^\lambda \cdot \bm{n})\phi_{i}^\lambda(\bm{x},t_{n+1/2})d\bm{S},
\end{equation}
where $\bm{n}$ is the outward unit vector normal to the surface. Finally, we remark that within the finite volume context, the populations and the collision terms are cell-averaged quantities,
\begin{equation}
     \phi_{i}^\lambda(\bm{x}_j,t_{n})=\frac{1}{V_j}\int_{V_j}{\phi_{i}^\lambda(\bm{x},t_n)d\bm{x}}.
\end{equation}
The reference frame  which is used for the evolution of the populations at $(\bm{x}_j,t_{n})$, is set to the comoving frame, from the known flow velocity and temperature,
\begin{equation}
    \lambda(\bm{x}_j,t_{n}) =\{\bm{u}(\bm{x}_j,t_{n}),T(\bm{x}_j,t_{n})\}.
\end{equation}

\subsubsection{Flux evaluation}
\label{Comoving Flux}
The key element of the update equations \eqref{eq:FinalUpdateDUGKS},\eqref{eq:FinalUpdateDUGKS_g} is the evaluation of the flux term, $F_{\phi,i}^\lambda(\bm{x}_j,t_{n+1/2})$, which contains the unknown populations ${\phi}_{i}^\lambda(\bm{x}_b,t_{n+1/2})$ at the cell interface $\bm{x}_b$ and time $t_{n+1/2}$. The frame which shall be used for the flux evaluation is $\lambda_F= \{ \bm{u}_F,T_F \}$, with the frame velocity $\bm{u}_F$ and temperature $T_F$  constructed by the average frame of the adjacent cell centers to the interface $\bm{x}_{1}, \bm{x}_{2}$,
\begin{align}
    \label{eq::AvgFrame}
    \bm{u}_F &=\frac{1}{2}\left(\bm{u}(\bm{x}_1,t_{n})+\bm{u}(\bm{x}_2,t_{n}) \right), \\ 
   T_F &=\frac{1}{2}\left(T(\bm{x}_1,t_{n})+T(\bm{x}_2,t_{n}) \right).
\end{align}
The integration of Eqs.\ \eqref{KineticModel_f}-\eqref{KineticModel_g} along the characteristics for half-time step shows that the required populations ${\phi}_{i}^{\lambda_F}(\bm{x}_b,t_{n+1/2})$, are connected with the known populations at time $t_{n}$ through the following equation \cite{Guo_DUGKS_IsoT},
\begin{equation}
    \label{eq:InterfacePop}
    \bar{\phi}^{ \lambda_F}_{i}(\bm{x}_b,t_{n+1/2})=\bar{\phi}_{i}^{+, \lambda_F}(\bm{x}_b-\bm{v}^{\lambda_F}_i \delta t/2,t_{n}),
\end{equation}
where,
\begin{align}\label{eq:FluxTransf}
    \bar{\phi}_{i}^{ \lambda_F} &={\phi}_{i}^{ \lambda_F}-\frac{\delta t/2}{2}\Omega_{\phi,i}^{ \lambda_F},\\ \label{eq:Fbarplus}
    \bar{\phi}_{i}^{+, \lambda_F} &={\phi}_{i}^{ \lambda_F}+\frac{\delta t/2}{2}\Omega_{\phi,i}^{ \lambda_F}.
\end{align}
Eq.\ \eqref{eq:InterfacePop} is essentially a half-time semi-Lagrangian step, with the final point located at the interface $\bm{x}_b$, at $t_{n+1/2}$. The populations $\bar{\phi}_{i}^{+, \lambda_F}$ and the spatial gradients $\bm{\sigma}_{i}^{\lambda_F} =\nabla \bar{\phi}_{i}^{+,\lambda_F} $ are subsequently evaluated in the neighbouring cells of the interface, at time $t_n$. In this work, Van Leer and minmod slope limiters were used for the computation of the spatial derivatives \cite{VanLeerLimiter,RoeLimiter}. We also note that the reference frame transformation is applied, to express the required populations from their original reference frame to the target reference frame $\lambda_F$. The populations are reconstructed at the departure point $\bm{x}'=\bm{x}_b-\bm{v}_i^{\lambda_F}\delta t/2$, with the MUSCL scheme \cite{VanLeerMuscl},
\begin{equation}
    \bar{\phi}_{i}^{+,\lambda_F}(\bm{x}',t_n)=\bar{\phi}_{i}^{+,\lambda_F}(\bm{x}_j,t_n)+(\bm{x}'-\bm{x}_j)\cdot \bm{\sigma}_{i}^{\lambda_F}(\bm{x}_j,t_n).
\end{equation}
According to Eq. \eqref{eq:InterfacePop}, we obtain the $\bar{\phi}_{i}^{\lambda_F}$ populations at the interface $\bm{x}_b$ and time $t_{n+1/2}$,
\begin{equation}
    \bar{\phi}_{i}^{\lambda_F}(\bm{x}_b,t_{n+1/2})=\bar{\phi}_{i}^{+,\lambda_F}(\bm{x}',t_n).
\end{equation}
The density, momentum and temperature at $(\bm{x}_b, t_{n+1/2})$ are finally computed by 
\begin{align}
\begin{split}
    \rho & =\sum_{i=0}^{Q-1} \bar{f}_{i}^{\lambda_F}(\bm{x}_b,t_{n+1/2}) ,
    \end{split}
     \\
     \begin{split}
    \rho \bm{u} & =\sum_{i=0}^{Q-1} \bm{v}_i^{\lambda_F} \bar{f}_{i}^{\lambda_F}(\bm{x}_b,t_{n+1/2}), 
      \end{split}
      \\
    \begin{split}  
    C_v\rho T & =\sum_{i=0}^{Q-1} \frac{{({v}_i^{\lambda_F})}^2}{2} \bar{f}_{i}^{\lambda_F}(\bm{x}_b,t_{n+1/2}) \\ & +  \sum_{i=0}^{Q-1}\bar{g}_{i}^{\lambda_F}(\bm{x}_b,t_{n+1/2})- \frac{\rho{u}^2}{2}.
    \end{split}     
\end{align}
With the calculated macroscopic fields $(\rho, \bm{u}, T)$ at $(\bm{x}_b,t_{n+1/2})$, the equilibrium  populations $\phi_i^{\rm eq,\lambda_F}(\rho, \bm{u}, T)$ can be computed and subsequently also the populations $\phi_i^{\lambda_F}(\bm{x}_b,t_{n+1/2})$, after inversion of Eq.\ \eqref{eq:FluxTransf}. We remind that the equilibrium populations can be obtained through the reference frame transformation \eqref{NonComovingEq},
\begin{equation}
    \phi^{\rm eq,\lambda_F}(\rho, \bm{u}, T)=\rho \mathcal{G}_{\{\bm{u},T\}}^{\{\bm{u}_F,T_F\}}W.
\end{equation}
Finally, the fluxes which are required to update the cell centers populations can be found from summation over the faces of the cell and proper reference frame transformation,
\begin{equation} \label{fluxEq}
    F_{\phi,i}^{\lambda}(\bm{x}_j,t_{n+1/2})=\sum_{c}(\bm{v}_i^{\lambda} \cdot \bm{n}_c)\mathcal{G}_{i,\lambda_{F,c}}^\lambda \phi^{\lambda_{F,c}}(\bm{x}_{b,c},t_{n+1/2}),
\end{equation}
where $\bm{x}_{b,c}$ designates the center of the $c$-th face of the cell, $\bm{n}_c$ is the outwards normal vector and $\lambda$ is the reference frame of the evolution of the cell.

\subsubsection{Summary of the algorithm}
\label{subsubsec::Summary}

Based on the previous steps, we summarize the evolution procedure from time $t_n$ to $t_{n+1}$:
\begin{enumerate}
    \item Initial data (cell centers $\bm{x}_j$)
    \begin{itemize}
        \item Given $(\rho, \bm{u}, T)$ , comoving reference frame $\lambda=\{\bm{u}(\bm{x}_j,t_{n}),T(\bm{x}_j,t_{n})\}$ and populations. ${\phi}_{i}^{\lambda}(\bm{x}_j,t_{n})$
        \item Calculation of the $\bar{\phi}_{i}^{+,{\lambda}}(\bm{x}_j,t_{n})$ populations, according to eq. \eqref{eq:Fbarplus}.
    \end{itemize}
    
    \item Calculation of the fluxes (Loop over cell faces $\bm{x}_b$)
    
    \begin{itemize}
        \item Set reference frame $\lambda_F$ at interface and time $(\bm{x}_b,t_{n+1/2})$
        \item Calculation of the populations ${\phi}_{i}^{\lambda_F}(\bm{x}_b,t_{n+1/2})$ according to procedure in Sec. \ref{Comoving Flux}.
    \end{itemize}
    \item Population update (Loop over cell centers $\bm{x}_j$)
    \begin{itemize}
        \item  Computation of the fluxes to the local reference frame of the cell, Eq.\ \eqref{fluxEq} and update the populations through Eqs. \eqref{eq:FinalUpdateDUGKS}, \eqref{eq:FinalUpdateDUGKS_g}. 
    \end{itemize}
\end{enumerate}
We stress the crucial difference between the proposed realization and the scheme suggested in \cite{PonD_DUGKS}, which is the absence of iterations within the flux evaluation step. We remind that a semi-Lagrangian step is executed to retrieve the populations at the cell faces, according to Eq.\ \eqref{eq:InterfacePop}. In this work, the reference frame for the above step is set from the average reference frames of the neighbouring cell centers (eq.\ \eqref{eq::AvgFrame}) and the flux evaluation is performed explicitly. With this approach, it is necessary to obtain non-comoving equilibrium populations, $\phi^{\rm eq,\lambda_F}(\rho, \bm{u}, T)$, to finalize the flux evaluation. The scheme in \cite{PonD_DUGKS} suggested an iterative predictor-corrector procedure, such that the flux calculation is realized in the comoving reference frame. While computationally demanding, the iteration procedure operates only with the simple comoving equilibrium populations Eqs.\ \eqref{feqPond}, \eqref{geqPond}. A further analysis of this aspect via numerical simulations is provided in Sec.\ \ref{subsec::ResultsDiscussion}.

\subsection{ Imlementation of variable Prandtl number}
\label{subsec::Prandtl}
The quasi-equilibrium relaxation can be implemented as a forcing term in the kinetic equations. We follow a typical approach in the context of DUGKS \cite{Guo_DUGKS_Rev} and realize the quasi-equilibrium relaxation via the Strang-splitting method \cite{StrangSplitting}:

\begin{enumerate}
    \item Quasi-equilibrium relaxation of the populations in the cell centers (half-time step),
\begin{equation}
\label{QE_relaxation}
    f_i' = f_i+ \frac{\delta t}{2} \left(\frac{1}{\tau_1}-\frac{1}{\tau_2}\right) (f_i^{\ast}-f_i^{\rm eq}).
\end{equation}
    \item Update step without quasi-equilibrium relaxation, Eqs.\ \eqref{eq:FinalUpdateDUGKS}, \eqref{eq:FinalUpdateDUGKS_g}.  
    \item Quasi-equilibrium relaxation of the populations in the cell centers (half-time step), as in step 1.  
\end{enumerate}
We note that half-time relaxations steps occur in each cell center and are local operations. By construction, the quasi-equilibrium relaxation conserves the flow velocity and temperature, and the populations remain in their comoving reference frame.

\subsection{Multiscale implementation}

The presented framework can be implemented in a multiscale setting with minimal changes in the algorithm. The different lattices are deployed adaptively in the simulation domain following a switching criterion. According to Sec.\ \ref{subsec::HydroLimit}, the switching criterion is a function of the hydrodynamic gradients, with the high-order lattice being activated in the regions of steep gradients. In this work, the switching function consists of threshold criteria on the numerically computed flow velocity and temperature gradients. The different lattices are updated normally as presented in the previous section, with the difference being that the reference frame transformations in the vicinity of the interface regions are replaced by the multiscale frame transformations (presented in Sec.\ \ref{subsec::Multiscale}).

\subsection{Boundary conditions}

\begin{figure}
    \centering
   \includegraphics[width=0.45\textwidth]{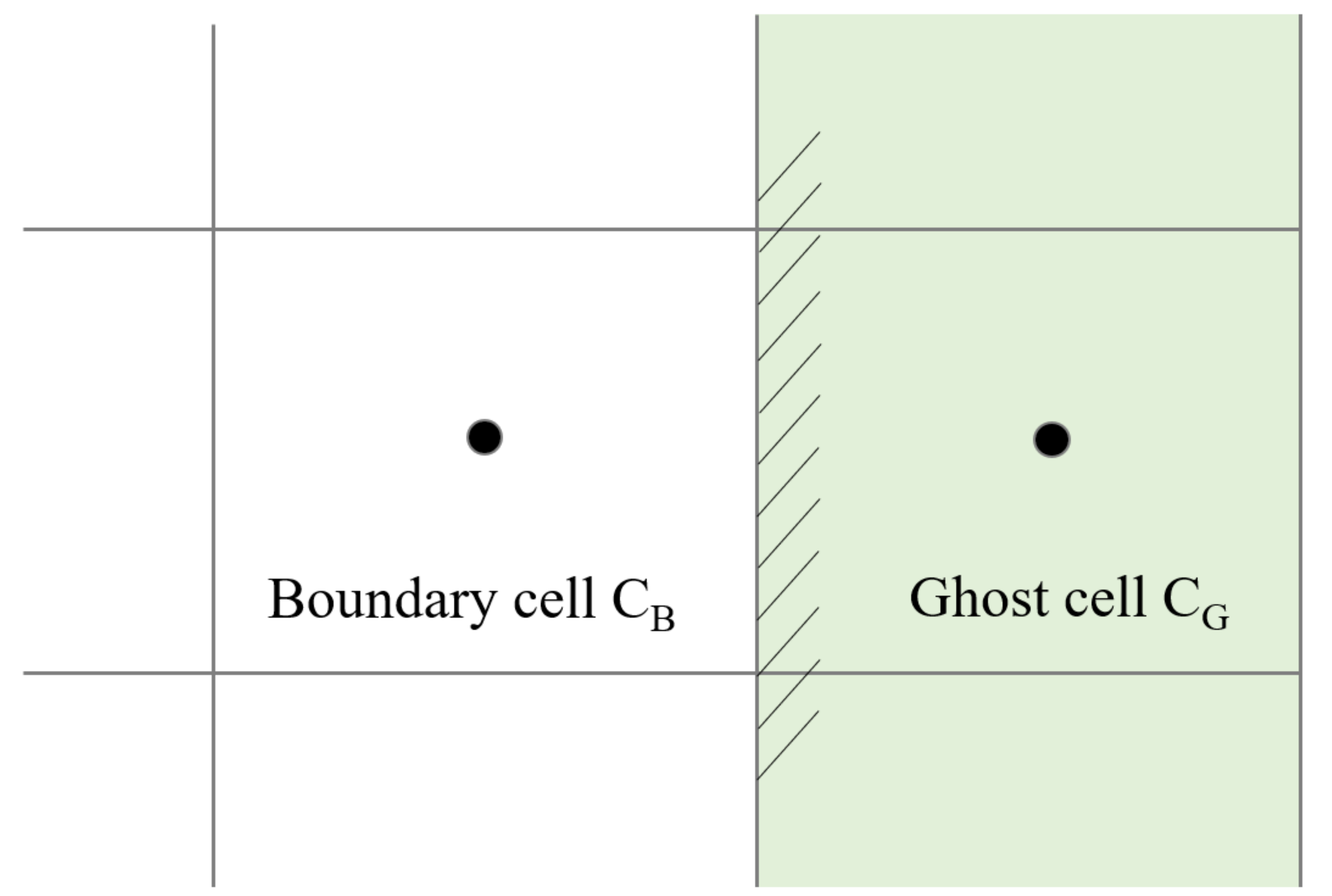}
    \caption{Schematic for the implementation of boundary conditions.}
    \label{fig::BCs}
\end{figure}

The boundary conditions (BCs) are enforced in the current work via the ghost node approach \cite{GhostNodeLBM}. First, the density, flow velocity and temperature are determined at the ghost cell $C_G$ (see Fig.\ \ref{fig::BCs}). For fixed values at the wall, e.g. no slip velocity $\bm{u}_w$, the value $\bm{u}_G$ at the ghost cell is,
\begin{equation}
    \bm{u}_G=2\bm{u}_w-\bm{u}_B,
\end{equation}
where $\bm{u}_B$ is the corresponding value at the boundary cell. To impose zero normal gradient condition, e.g. for density computation, we enforce
\begin{equation}
    \rho_G=\rho_B.
\end{equation}
With the macroscopic values ($\rho_G,\bm{u}_G,T_G$) defined, the reference frame of the ghost cell $\lambda_G$ is set to the comoving reference frame, $\lambda_G=\{{\bm{u}_G,T_G}\}$. The equilibrium populations are then,
\begin{align}
    f_i^{G,\rm eq} &= \rho_G W_i, \\
    g_i^{G,\rm eq} &=(C_v-\frac{D}{2})T_G\rho_G W_i.
\end{align}
The approximation of non-equilibrium contributions follows the implementation of \cite{Frappoli_2016}. In particular, the first-order non-equilibrium moments are estimated from the Chapman--Enskog solution, and they depend on the local hydrodynamic gradients. The pertinent non-equilibrium moments of the $f-$ populations are \cite{Frappoli_2016},
\begin{align}
    \bm{P}^{(1)} &=-\tau_1\rho_G T_G \left( \bm{S}-\frac{1}{C_v}(\nabla \cdot \bm{u}) \bm{I} \right), \\
    \bm{Q}^{(1)} &=-\tau_2\rho_G T_G(\overline{\nabla T \bm{I}})+\overline{\bm{u}\bm{P}^{(1)}}.
\end{align}
The zeroth up to second order non-equilibrium moments of the $g-$ populations, $M_{g,0}^{(1)}, \bm{M}_{g,1}^{(1)}, \bm{M}_{g,2}^{(1)}$ are estimated as \cite{Frappoli_2016},
\begin{align}
   M_{g,0}^{(1)} &= -\tau_1\rho_G T_G(2C_v-D)\left(\frac{1}{C_v}\nabla \cdot \bm{u}\right), \\
 \bm{M}_{g,1}^{(1)} &=-\tau_2\rho_G T_G(2C_v-D)\nabla T+ M_{g,0}^{(1)}\bm{u}, \\
 \bm{M}_{g,2}^{(1)} &=-\tau_1\rho_G T_G(2C_v-D)(T\bm{S}+\overline{\bm{u}\nabla T}).
\end{align}
The hydrodynamic gradients are evaluated with a second-order centered scheme, based on previous time step quantities. The non-equilibrium populations are computed from their non-equilibrium moments, according to the Grad's projection procedure,
\begin{align}
    f_i^{\lambda'} &= W_i \sum_{n=0}^{3}\frac{1}{n!}\bm{\alpha}^{(n)}(\bm{P}^{(1)},\bm{Q}^{(1)};\lambda_G)\bm{H}^{(n)}(\bm{c}_i),\\
g_i^{\lambda'} &= W_i \sum_{n=0}^{2}\frac{1}{n!}\bm{\alpha}^{(n)}_g(M_{g,0}^{(1)}, \bm{M}_{g,1}^{(1)}, \bm{M}_{g,2}^{(1)};\lambda_G)\bm{H}^{(n)}(\bm{c}_i). 
\end{align}

\section{Results and discussion}
\label{sec::results}

In this section, we validate the model with 1D/2D Euler gas dynamics benchmarks, and viscous flows to assess the Prandtl number as well as the accuracy of the wall BCs. Subsequently, we focus on the shock structure problem and demonstrate numerically the implications of the moment analysis of Sec.\ \ref{subsec::HydroLimit}. The framework is then implemented with the multiscale setting (Sec.\ \ref{subsec::Multiscale}), via the deployment of different lattices across the simulation domain. We conclude this section with a summary of our observations and discussion of the model capabilities. We remind that the $g-$ populations evolve with the $D2Q9$ lattice. Unless stated otherwise, the numerical parameters of the simulations are the following. The time step $\delta t$ is such that the Courant–Friedrichs–Lewy (CFL) number is $\text{CFL}= \max|v_{i\alpha}|(\delta t /\delta x) =0.2$, where $\delta x$ is the grid resolution. The adiabatic exponent is $\gamma=1.4$. Additionally, the viscosity for the Euler flows is low enough such that the results remain invariant (typically $\mu \sim \mathcal{O}(10^{-3}-10^{-2})$). Finally, we note that the formulation of the initial and boundary conditions are based on non-dimensional variables, scaled with appropriate reference density, velocity and pressure.

\subsection{Euler gas dynamics}

We validate the model using the $D2Q16$ lattice and a third-order Grad's projection for the moment transformation. According to the moment analysis in Sec.\ \ref{subsec::HydroLimit}, the hydrodynamics at the Euler level should be captured accurately. Indeed, the model performs very well against a series of 1D Riemann problems, involving low density-near vacuum regions and very strong discontinuities. While all benchmarks of the previous work \cite{PonD_DUGKS} were tested, we present here two representative 1D examples. The 2D cases include a high Mach Riemann problem, a Mach 3 flow over a step obstacle and a shock diffraction over a corner.

\subsubsection{Strong shock tube}
\begin{figure}
    \centering
   \includegraphics[width=0.4\textwidth]{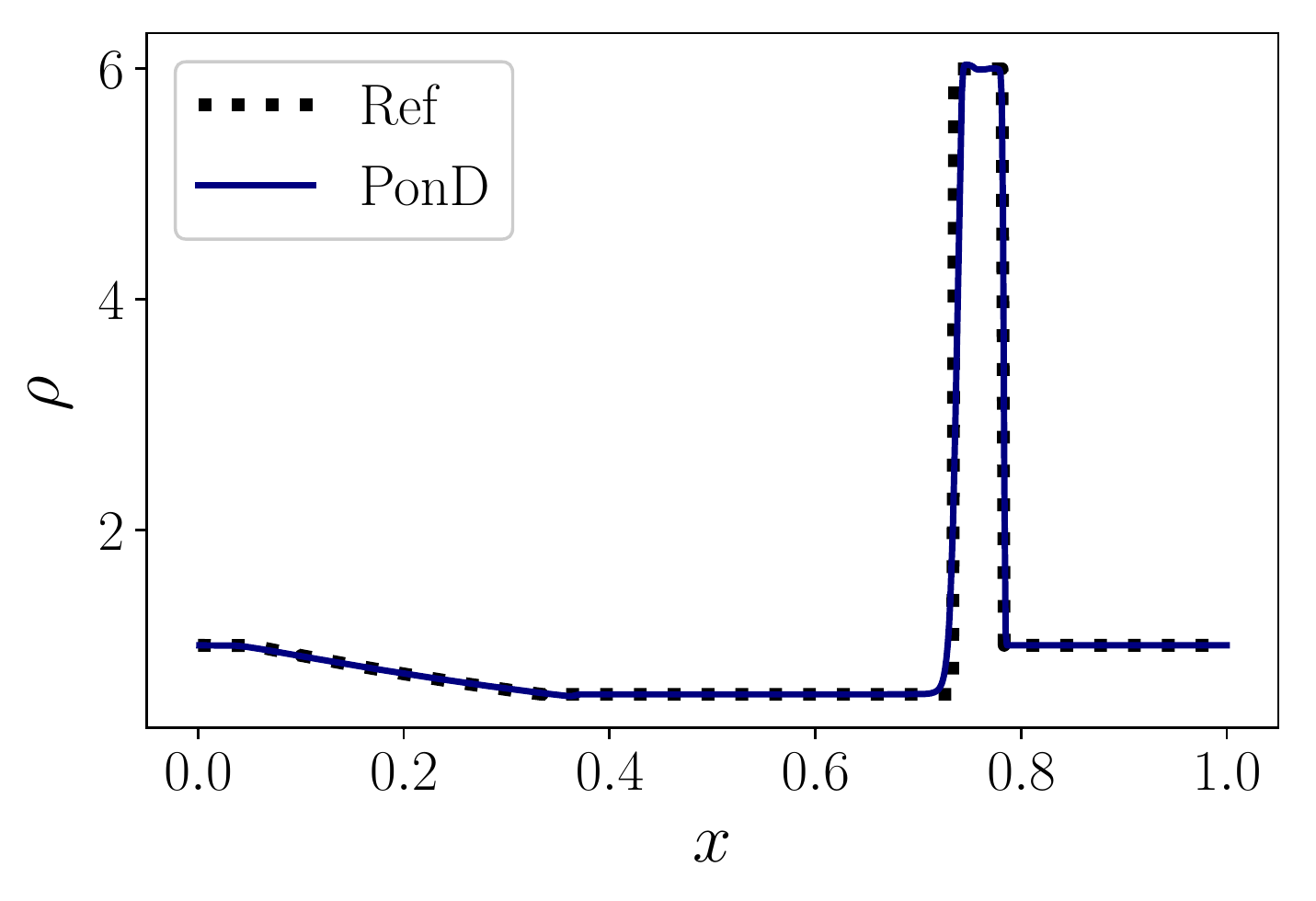}
    \includegraphics[width=0.4\textwidth]{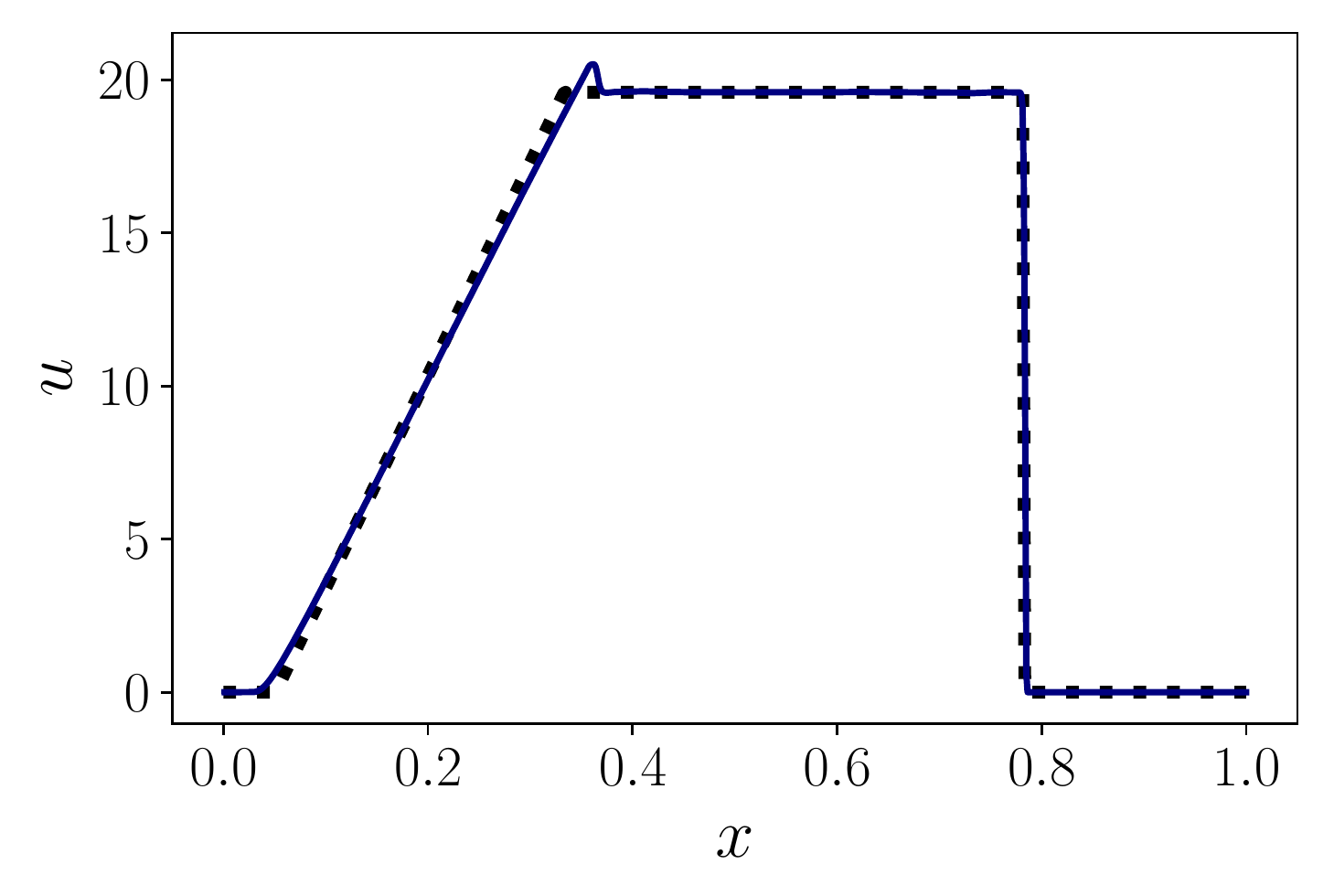}
     \includegraphics[width=0.4\textwidth]{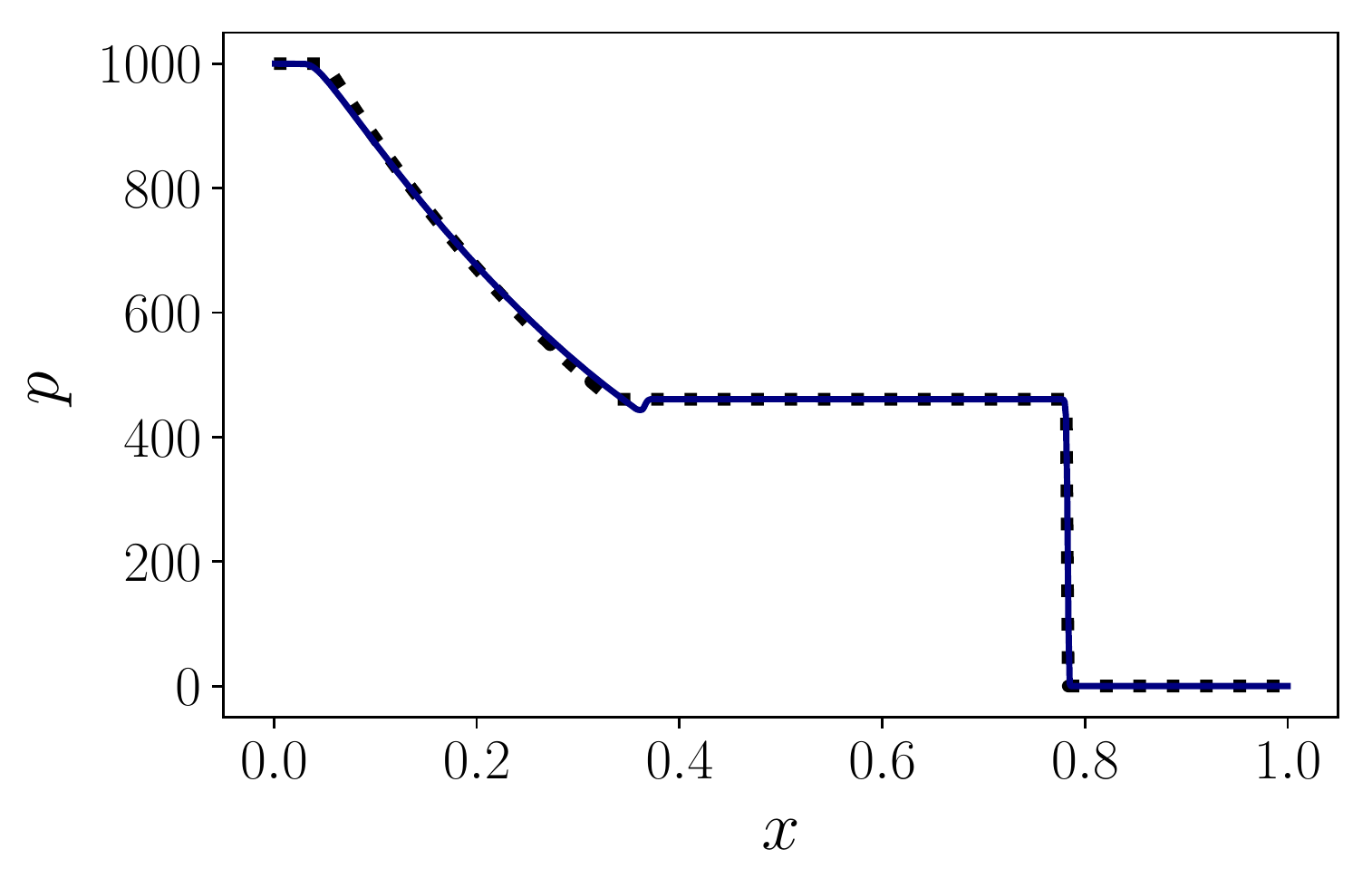}
    \caption{Density (top), velocity (middle) and pressure (bottom) profiles for the strong shock tube problem, at $t=0.012$. Solid line: PonD model. Dashed line: Reference from an exact Riemann solver. }
    \label{fig:Strong_ShockTube_Problem}
\end{figure}

 We consider the case of a strong shock tube \cite{StrongShock}, where the ratio between the temperature of the left and right side is $10^5$. The initial conditions for this problem are,
\begin{equation}
    (\rho,u_x,p)=\begin{cases}
    (1, 0, 1000),& 0 \leq x<0.5, \\
    (1, 0, 0.01),& 0.5 \leq x \leq 1. \\
    \end{cases}
\end{equation}
This problem, characterized by the strong temperature discontinuity, probes the robustness and accuracy of the numerical methods. The results of the simulation, at $t=0.012$ and $L=800$, are shown in Fig. \ref{fig:Strong_ShockTube_Problem}. Overall, a very good agreement with the exact solution is noted.

\subsubsection{Le Blanc problem}

\begin{figure}
    \centering
   \includegraphics[width=0.4\textwidth]{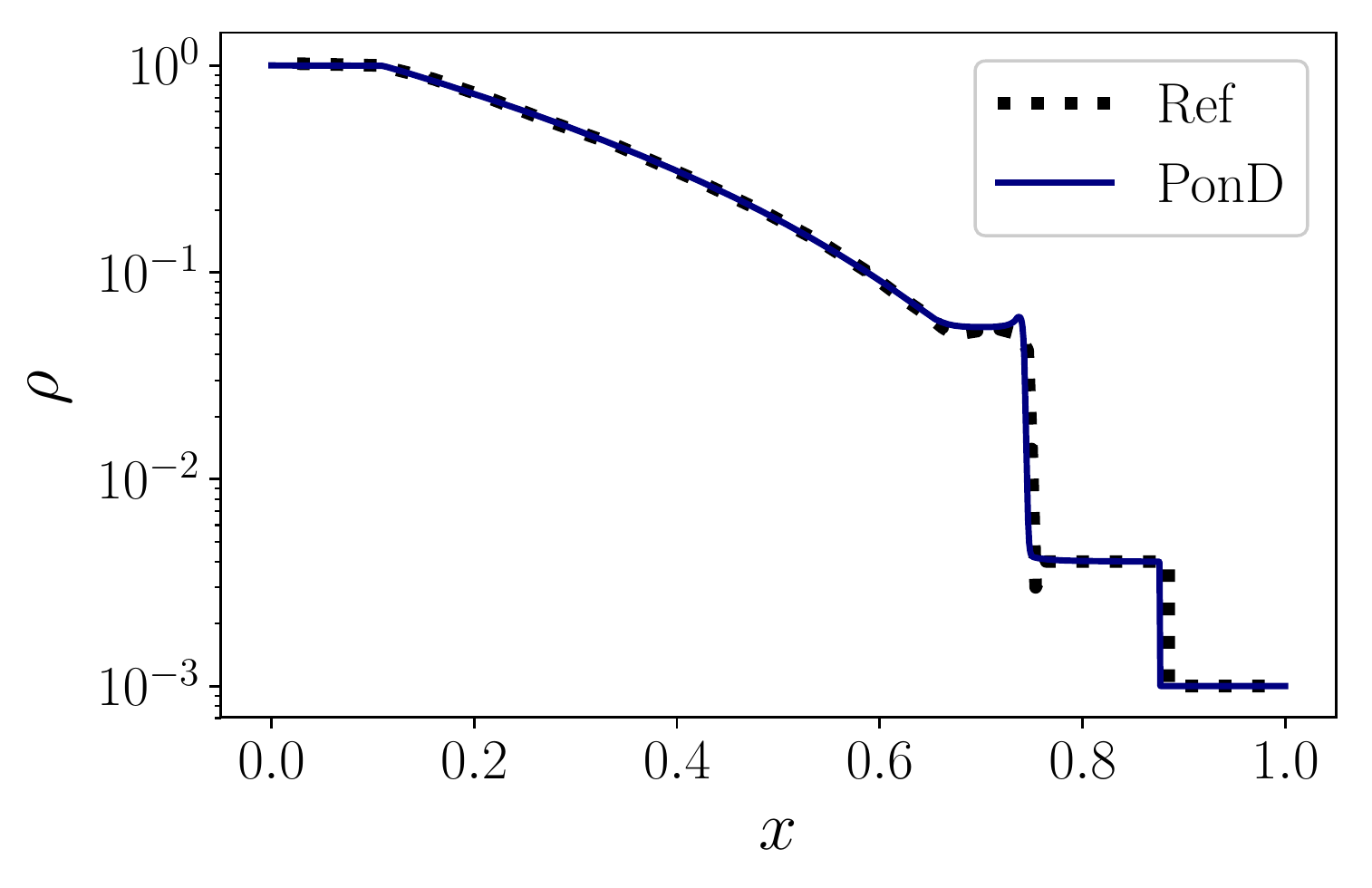}
    \includegraphics[width=0.4\textwidth]{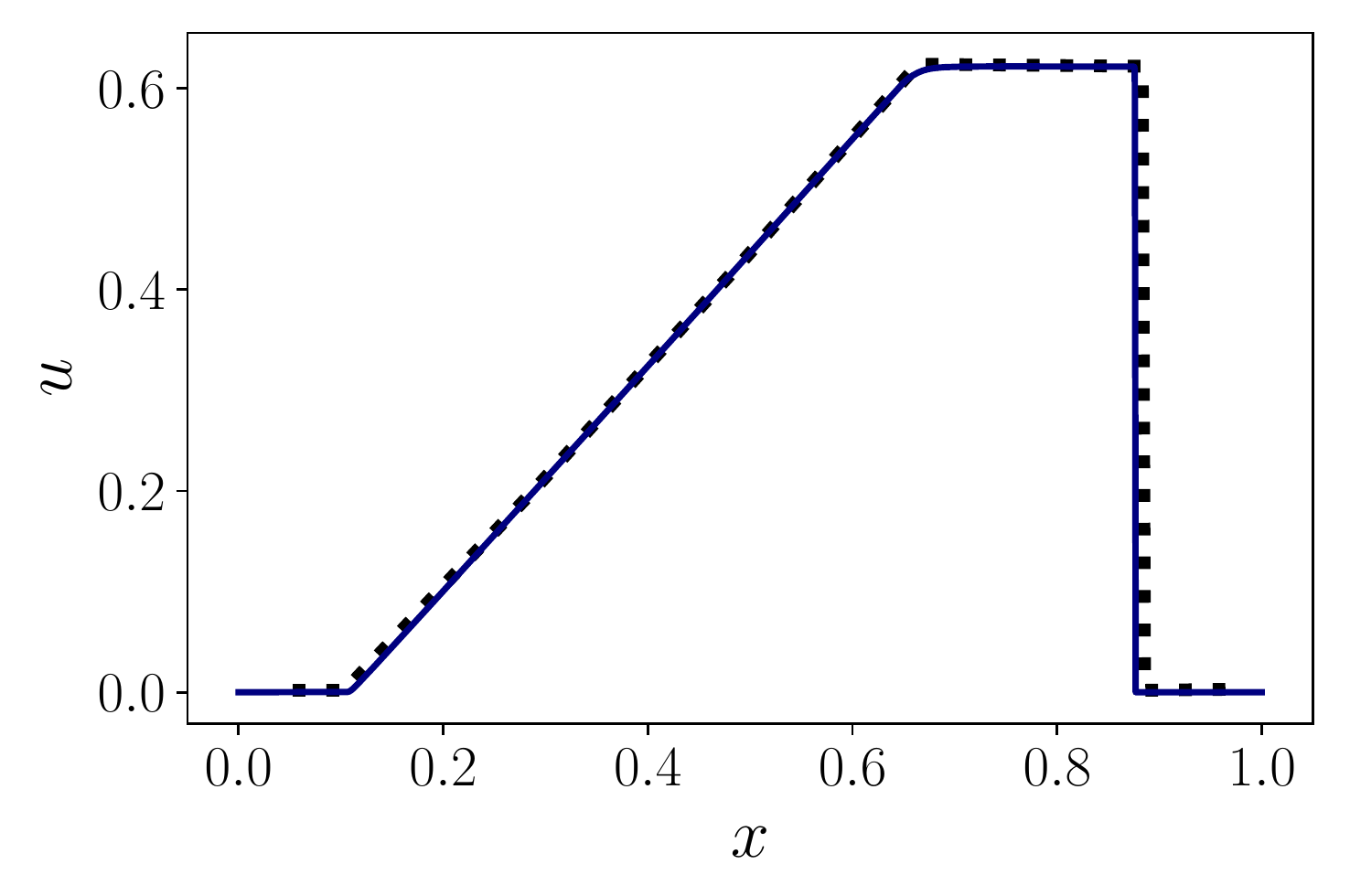}
     \includegraphics[width=0.4\textwidth]{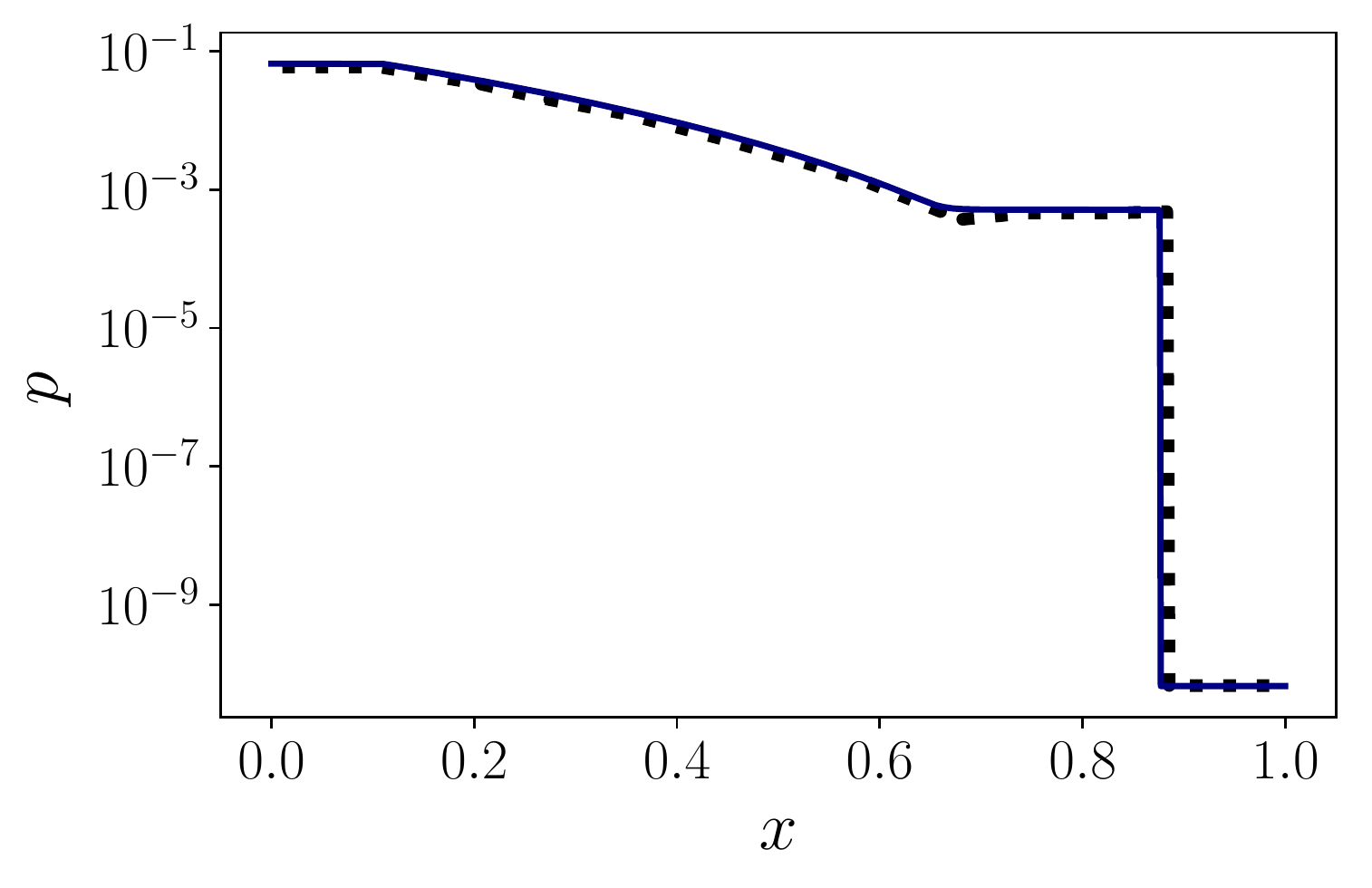}
    \caption{Density (top), velocity (middle) and pressure (bottom) profiles for the Le Blanc problem, at $t=6$. Solid line: PonD model. Dashed line: Reference solution \cite{LinFuAllSpeed}.}
    \label{fig:Le Blanc_Problem}
\end{figure}

The Le Blanc problem is considered next \cite{LeBlanc}, which involves very strong discontinuities and is initialized with the following conditions,
\begin{equation}
    (\rho,u_x,p)=\begin{cases}
    (1, 0, 2/3 \times 10^{-1}),& 0\leq x <3, \\
    (10^{-3}, 0, 2/3 \times 10^{-10}),& 3\leq x \leq 9. \\
    \end{cases}
\end{equation}
In this problem, the adiabatic exponent is fixed to $\gamma=5/3$. Fig.\ \ref{fig:Le Blanc_Problem} shows the results at $t=6$ and $L=4000$. With the exception of minor oscillations, a very good agreement of the present scheme with the reference solution \cite{LinFuAllSpeed} is observed.

\subsubsection{2D Riemann, configuration 3}

As a first validation in two dimensions we simulate a 2D Riemann problem, which is a classical benchmark for compressible flow solvers \cite{Lax_RiemannReference}. A square domain $(x,y) \in [0,1]\times [0,1]$ is divided into four quadrants, each of which is initialized with constant values of density, velocity and pressure as follows:
\begin{equation}
\begin{split}
     &(\rho,u_x,u_y,p) \\&= \begin{cases}
    (1.5, 0, 0, 1.5),&x>0.5, y>0.5, \\
    (0.5323, 1.206, 0, 0.3),&x\leq 0.5, y>0.5, \\
    (0.138, 1.206, 1.206, 0.029),&x\leq 0.5, y\leq 0.5, \\
    (0.5323, 0, 1.206, 0.3),&x>0.5, y\leq 0.5. \\
    \end{cases}
\end{split}    
\end{equation}
At the boundaries, zero-gradient BCs were imposed $\partial_{\bm{n}} f=0$, where $\bm{n}$ is the outwards unit normal vector. The simulation was performed with resolution $[500,500]$. The results of the density field, as well as  density contours near the center of the domain, are depicted in Fig.\ \ref{2D Riemann, configuration 3.}. The initial conditions of the Riemann problem lead to shock wave interaction and the formation of complex patterns. The results show a very good agreement with the reference solutions in \cite{Lax_RiemannReference,2DRiem_ref}.

\begin{figure}
    \centering
   \includegraphics[width=0.45\textwidth]{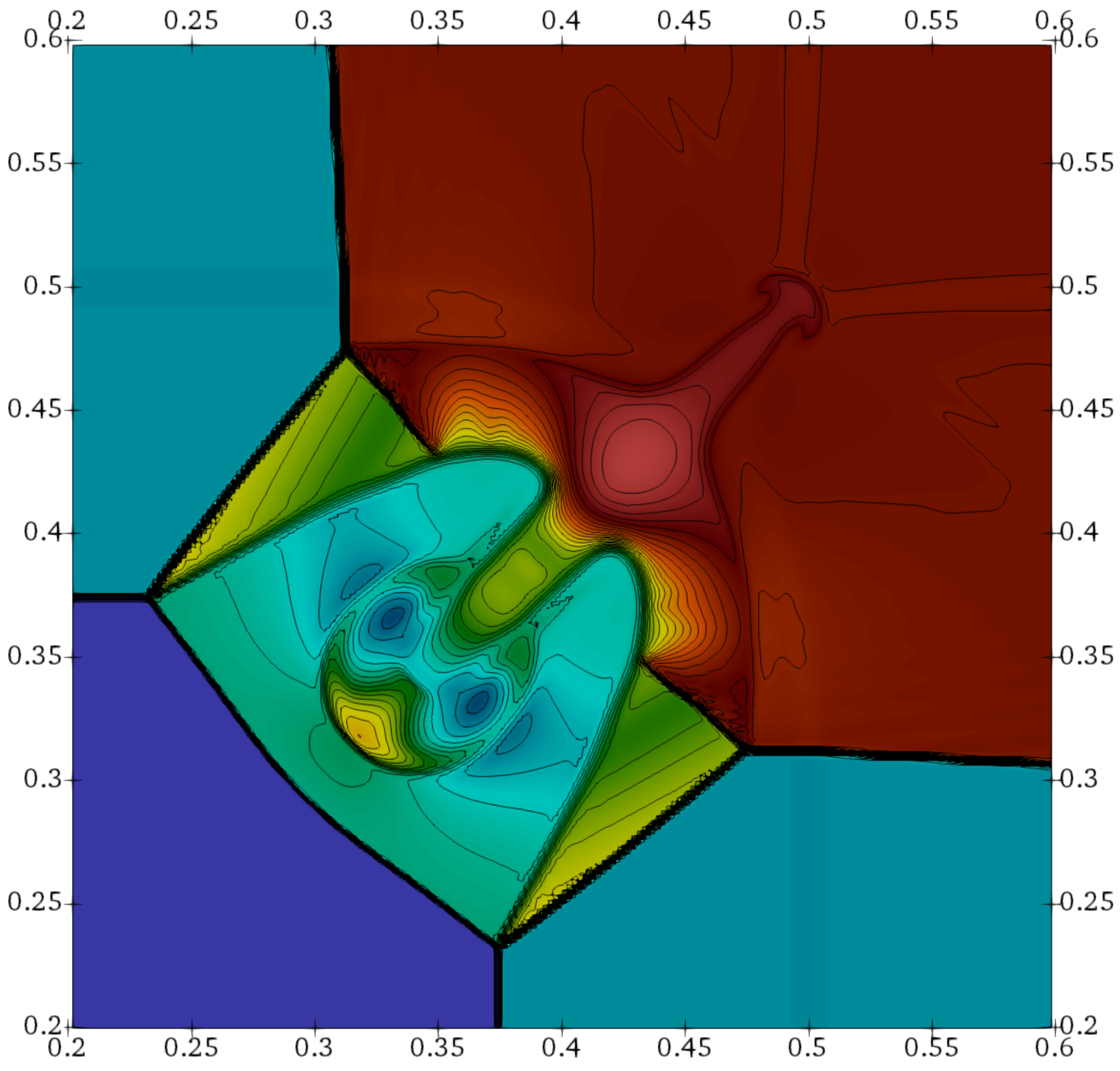}
    \caption{2D Riemann problem, with resolution of [500,500] grid points, at $t=1$. 25 equidistant density contours are superimposed on the results.}
    \label{2D Riemann, configuration 3.}
\end{figure}

\subsubsection{Mach 3 flow over step}

\begin{figure}
    \centering
   \includegraphics[width=0.45\textwidth]{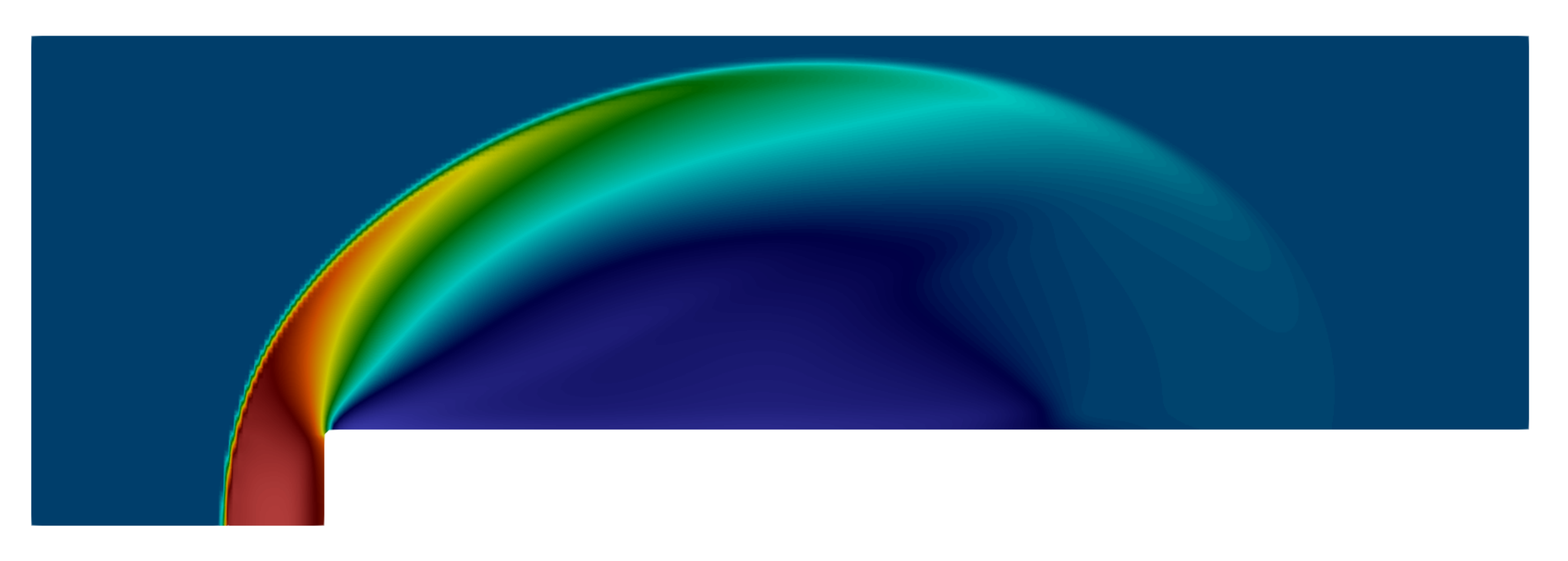}
    \includegraphics[width=0.45\textwidth]{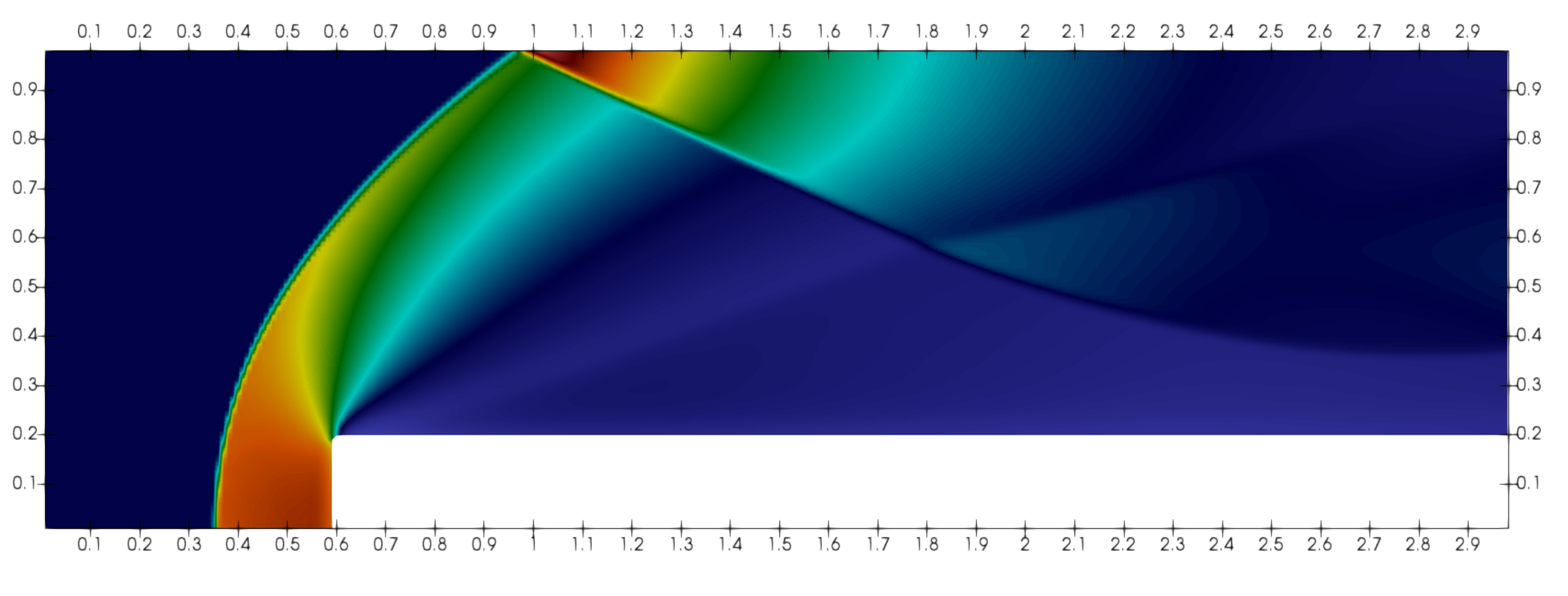}
     \includegraphics[width=0.45\textwidth]{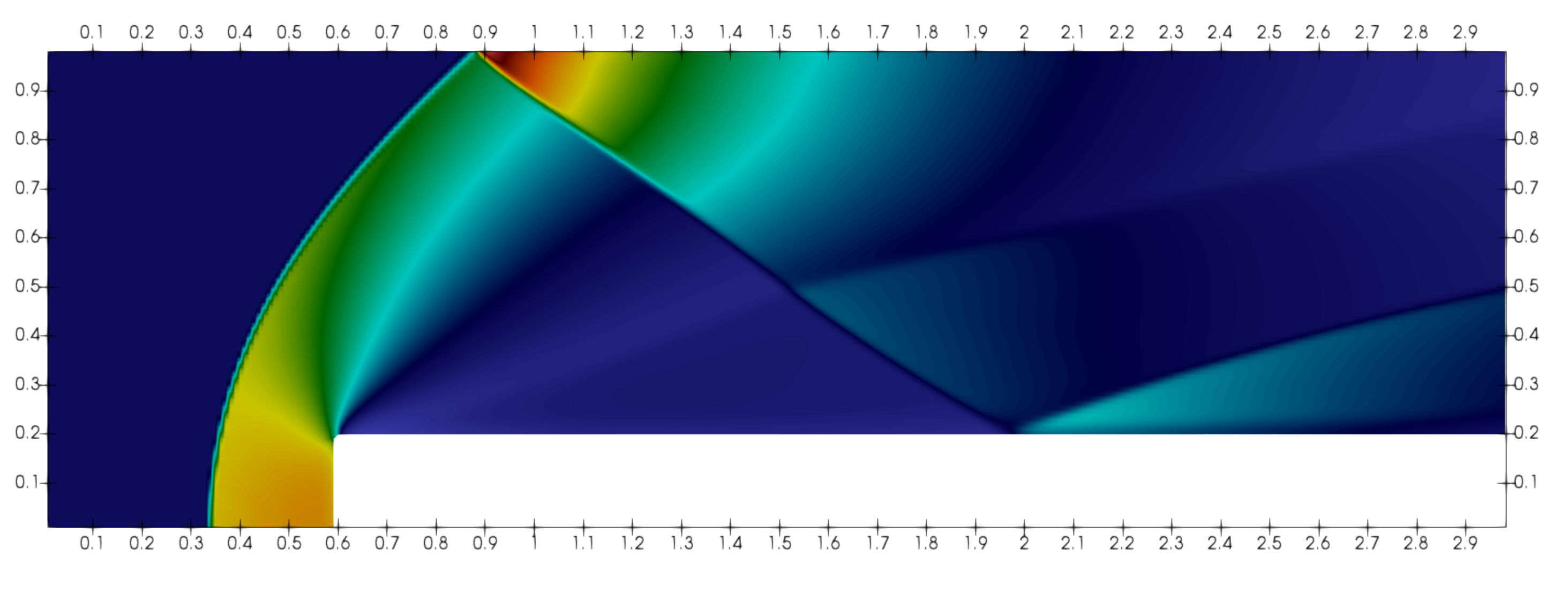}
       \includegraphics[width=0.45\textwidth]{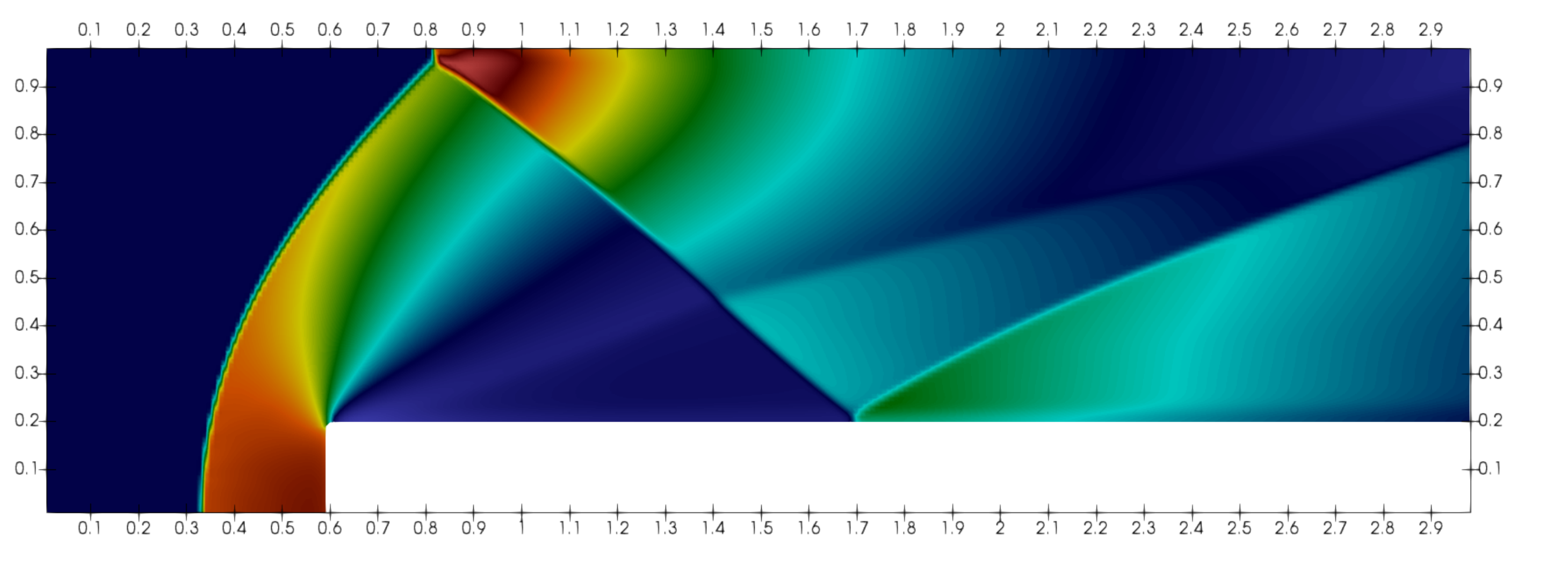}
         \includegraphics[width=0.45\textwidth]{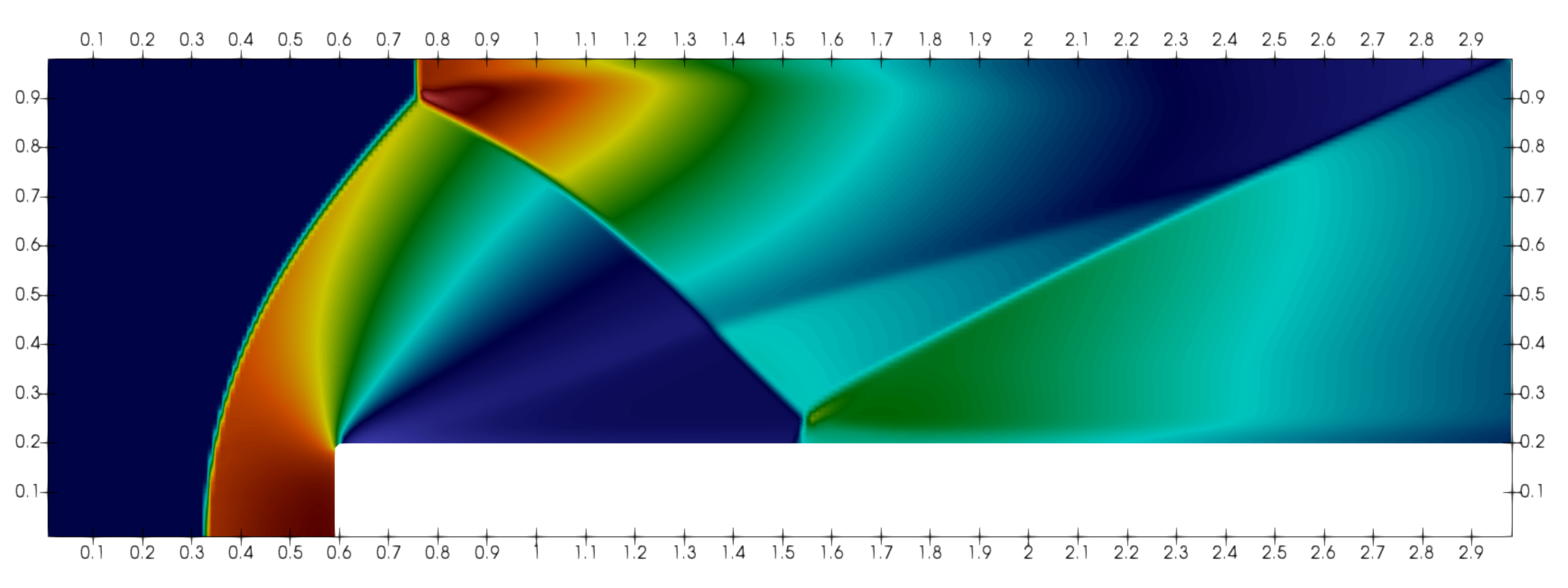}
           \includegraphics[width=0.45\textwidth]{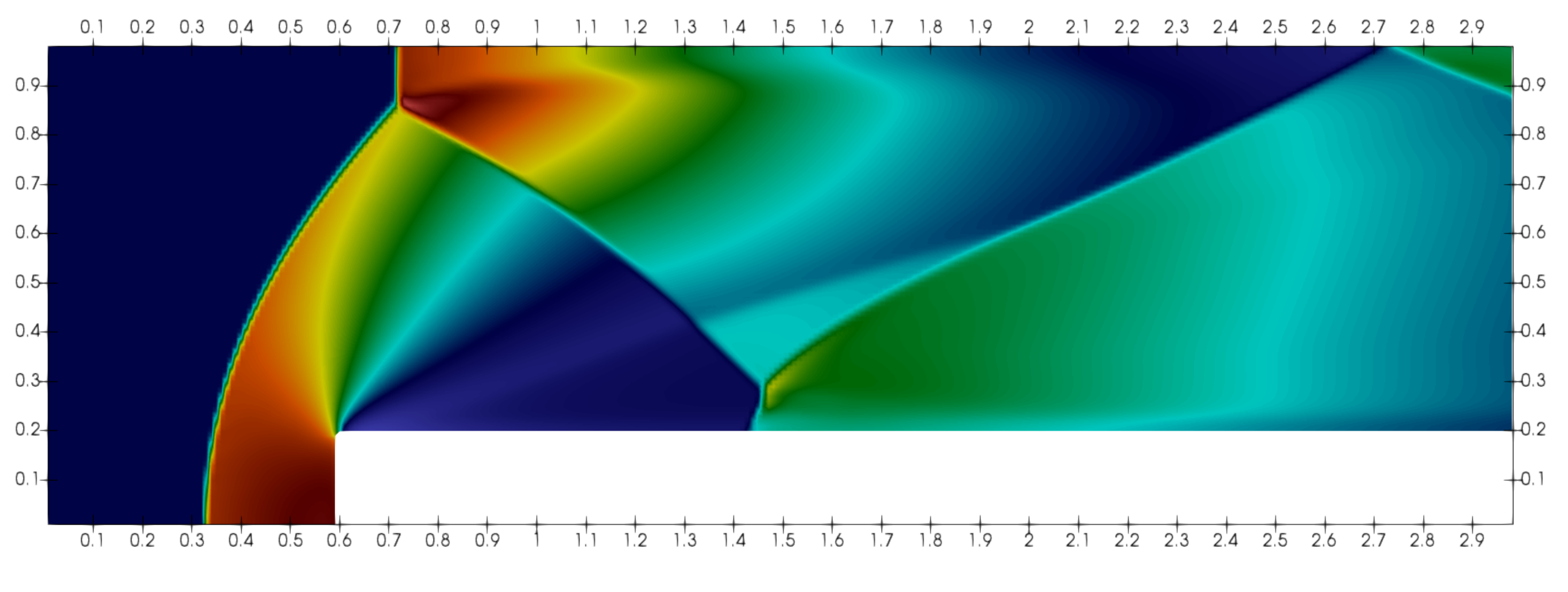}
    \caption{Density profiles for the Mach 3 flow over forward step. Six snapshots are shown at equal time intervals from $t=0$ to $t=3$.}
    \label{Mach3Step}
\end{figure}

In this problem, a uniform Mach 3 flow is imposed on a wind tunnel containing a step \cite{WoodwardCollela_DoubleMachReflection}. A transient shock wave develops from the step, reflects at the walls and forms a complicated flow pattern. The computational domain is bounded by a $[0,3]\times [0,1]$ rectangle, while the step is located at $(0.6,0)$ and has a height of $\Delta y=0.2$. Initially a gas with $\gamma=1.4$ is spatially uniform, with the following hydrodynamic conditions,
\begin{equation}
    (\rho,u_x,u_y,p)=(1.4, 3, 0, 1).
\end{equation}
The same conditions are imposed as inflow BCs at the left boundary $x=0$ and outflow BCs at the right boundary $x=3$. Reflecting BCs are applied at the walls of the domain. The results of a simulation resolved with $[300,100]$ grid points are presented in Fig.\ \ref{Mach3Step}, at six equal time intervals ($t=0.5$ to $t=3$). The flow features and dynamics are in very good agreement with the corresponding results from the literature \cite{WoodwardCollela_DoubleMachReflection}.

\subsubsection{Shock diffraction over corner}

\begin{figure}
    \centering
   \includegraphics[width=0.45\textwidth]{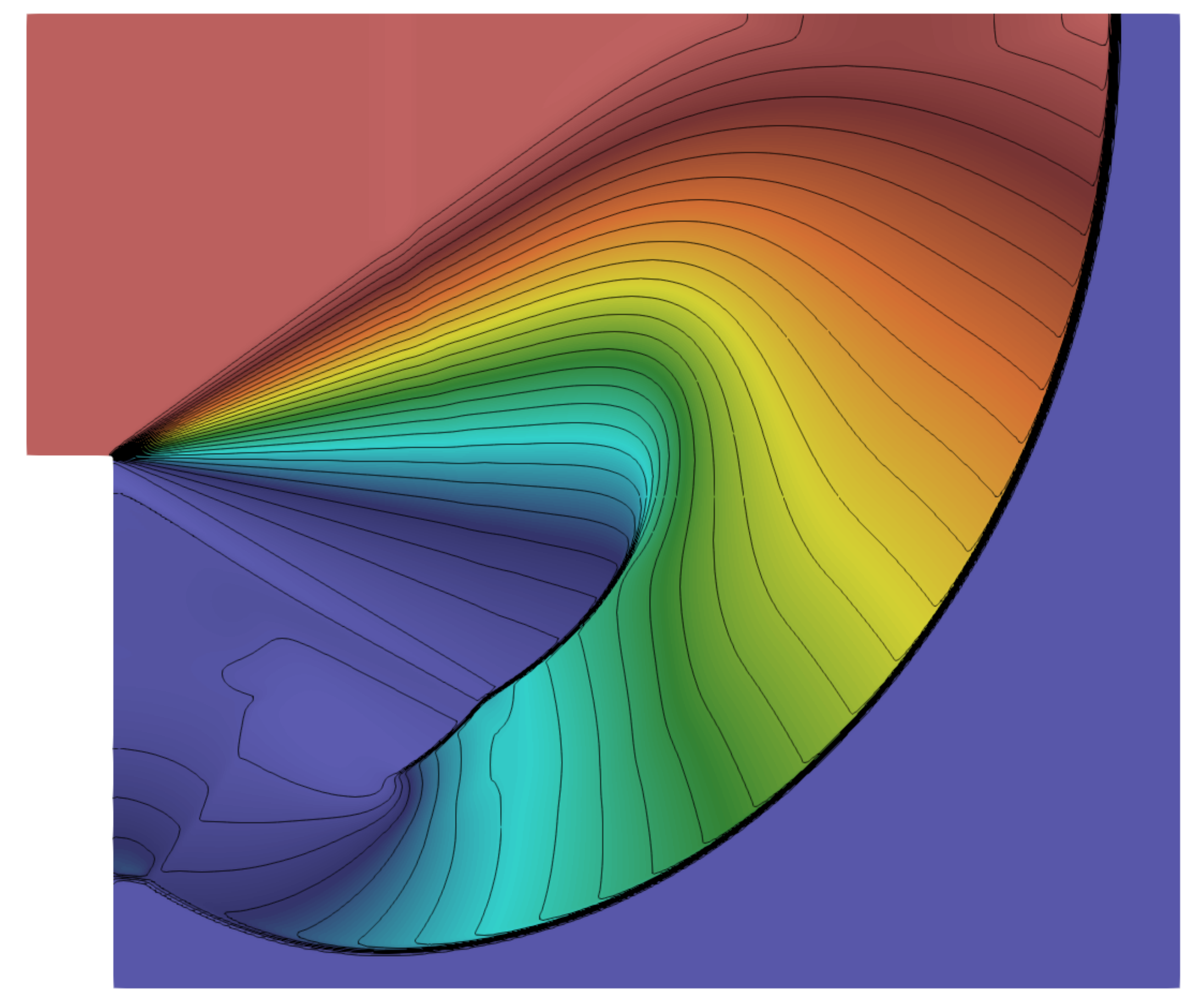}
    \includegraphics[width=0.45\textwidth]{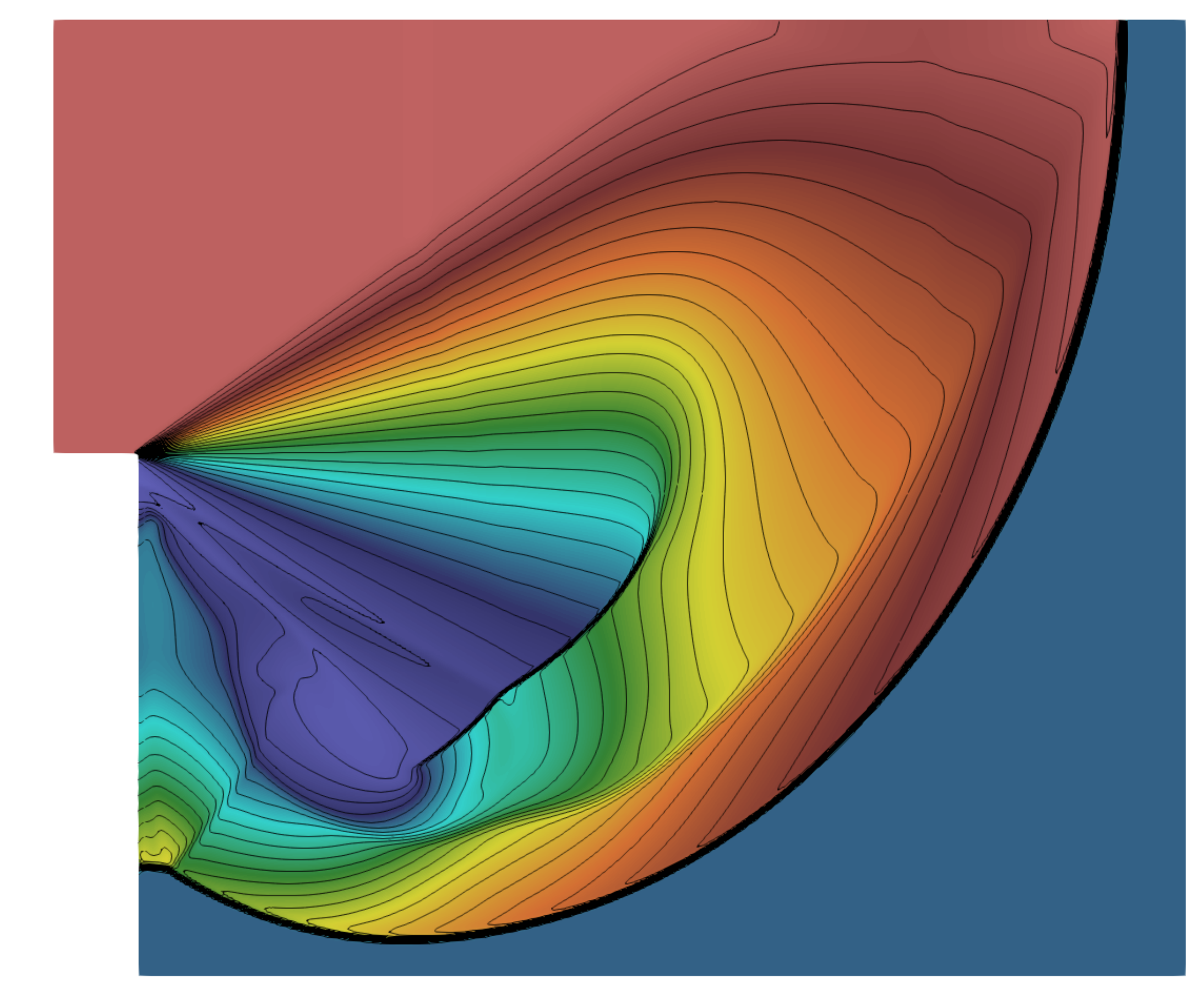}

    \caption{Pressure (top) and density (bottom) profiles for the Mach 5.09 shock diffraction problem over a corner, at $t=2.3$. 30 equidistant contours are superimposed on the fields.}
    \label{ShockDiffraction}
\end{figure}

Here we investigate the shock diffraction problem, in which a shock wave flows over a backward facing corner \cite{ZhangShu2012}. The hydrodynamic patterns of this problem have been studied theoretically, experimentally and via simulations. From the numerical standpoint however, this problem has been challenging due to the development of negative pressure and/or density around the corner. We follow the conventional setup of the problem: the computational domain consists of the union of $[0,1]\times [6,11]$ and $[1,13]\times [0,11]$ rectangles. Initially, a $\rm Ma=5.09$ right-moving shock wave is located at $x=0.5$ and $6\leq y \leq 11$ and propagates into undisturbed air, with density 1.4 and pressure 1. For the BCs, we use inflow with the initial conditions at $x=0,0\leq y\leq 11$, outflow at $x=13, 0\leq y \leq 11$, $1\leq x \leq 13, y=0$ and $0\leq x \leq 13, y=11$. Reflective BCs are applied at the walls of the domain $0\leq x \leq 1, y=6$ and $x=1, 0\leq y \leq 6$. The results, for resolution [390, 330] and $t=2.3$, are shown in Fig.\ \ref{ShockDiffraction} and compare very well with the reference results from \cite{ZhangShu2012}.

\subsection{Viscous flows}
\label{sec::viscousflows}
In this section, we focus on hydrodynamic flows with important viscous effects. The discussion pivots around the accuracy and the limitations of the third-order moment invariant system, sustained by the $D2Q16$ lattice.

\subsubsection{Channel flow}

\begin{figure}
    \centering
   \includegraphics[width=0.45\textwidth]{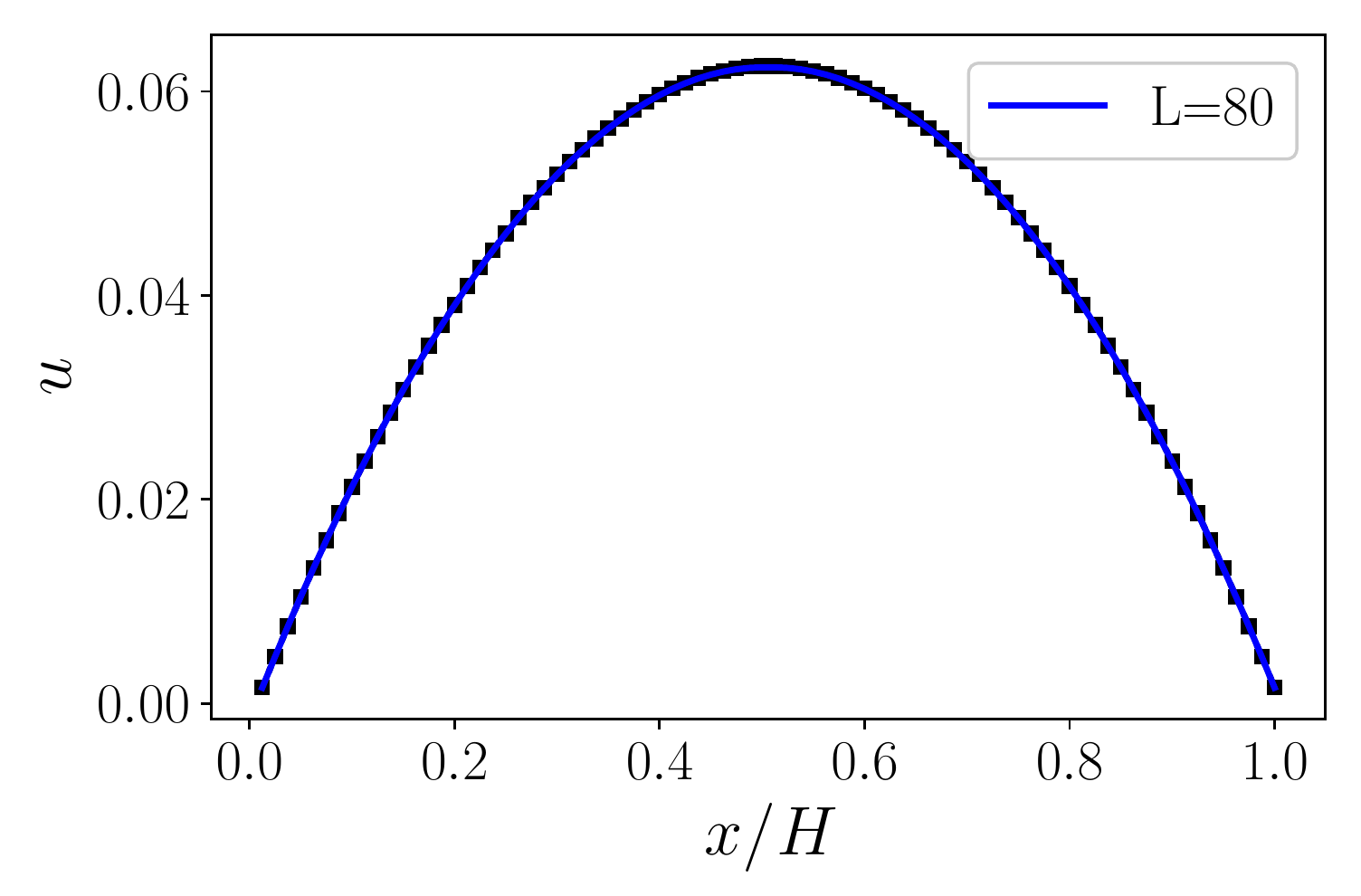}
    \includegraphics[width=0.45\textwidth]{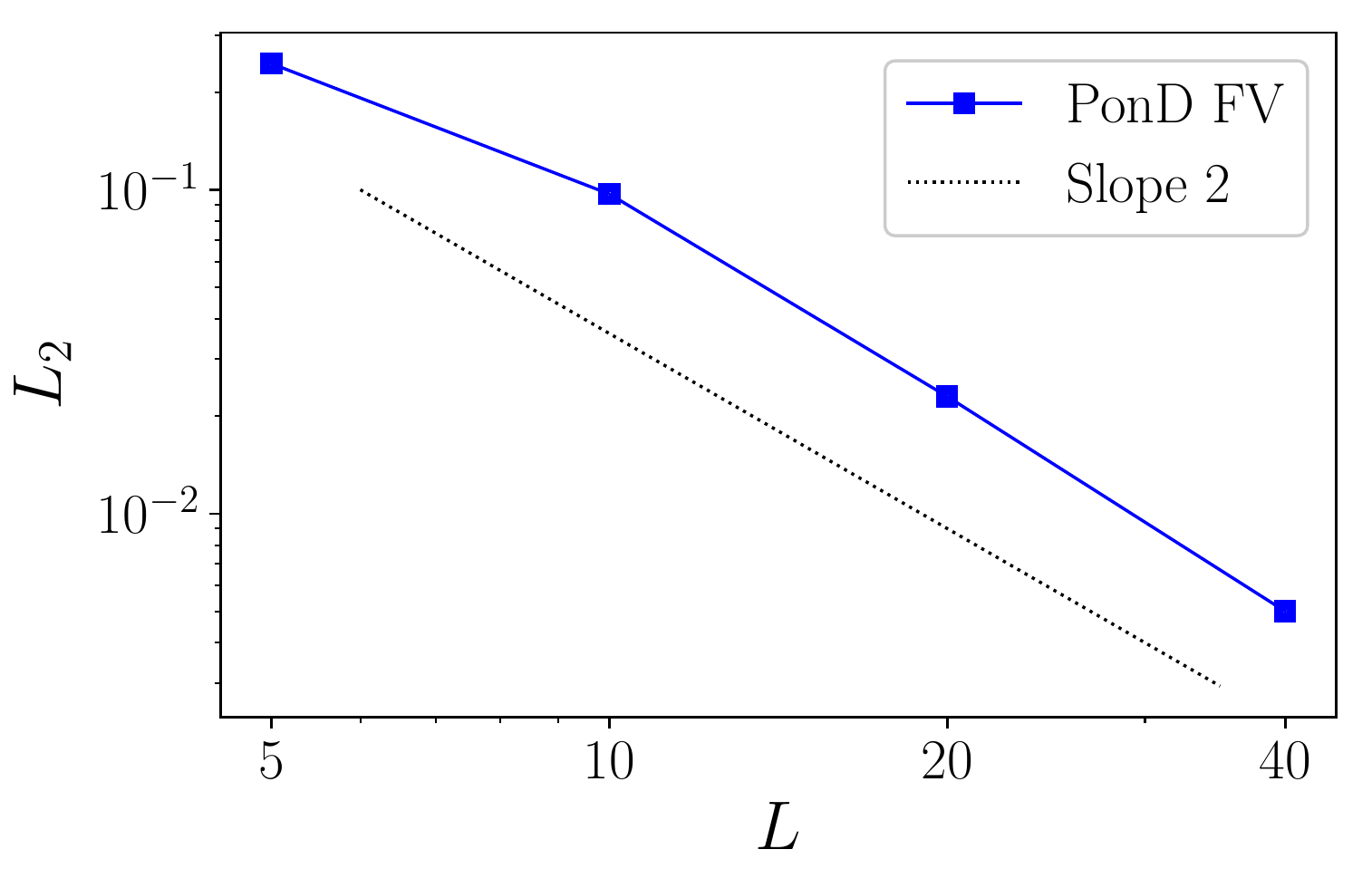}
    \caption{Top: Force driven flow over channel at $\mathrm{Re}=100$. Symbols correspond to simulation results and solid line to analytical solution. Bottom: convergence order analysis. Solid line corresponds to $L_2$ error norm. Dashed line indicates a second order $L_2$ error norm slope.}
    \label{fig::ChannelFlow}
\end{figure}

We begin with an isothermal channel flow with Reynolds number of $Re=100$, to assess the wall BCs. The results for a simulation with 80 grid points, shown in Fig.\ \ref{fig::ChannelFlow}, demonstrate an excellent agreement with the analytical solution. Additionally, a convergence order study with respect to the $L_2$ error, verifies a second order spatial convergence of the scheme.

\subsubsection{Thermal Couette flow}

\begin{figure}
    \centering
   \includegraphics[width=0.45\textwidth]{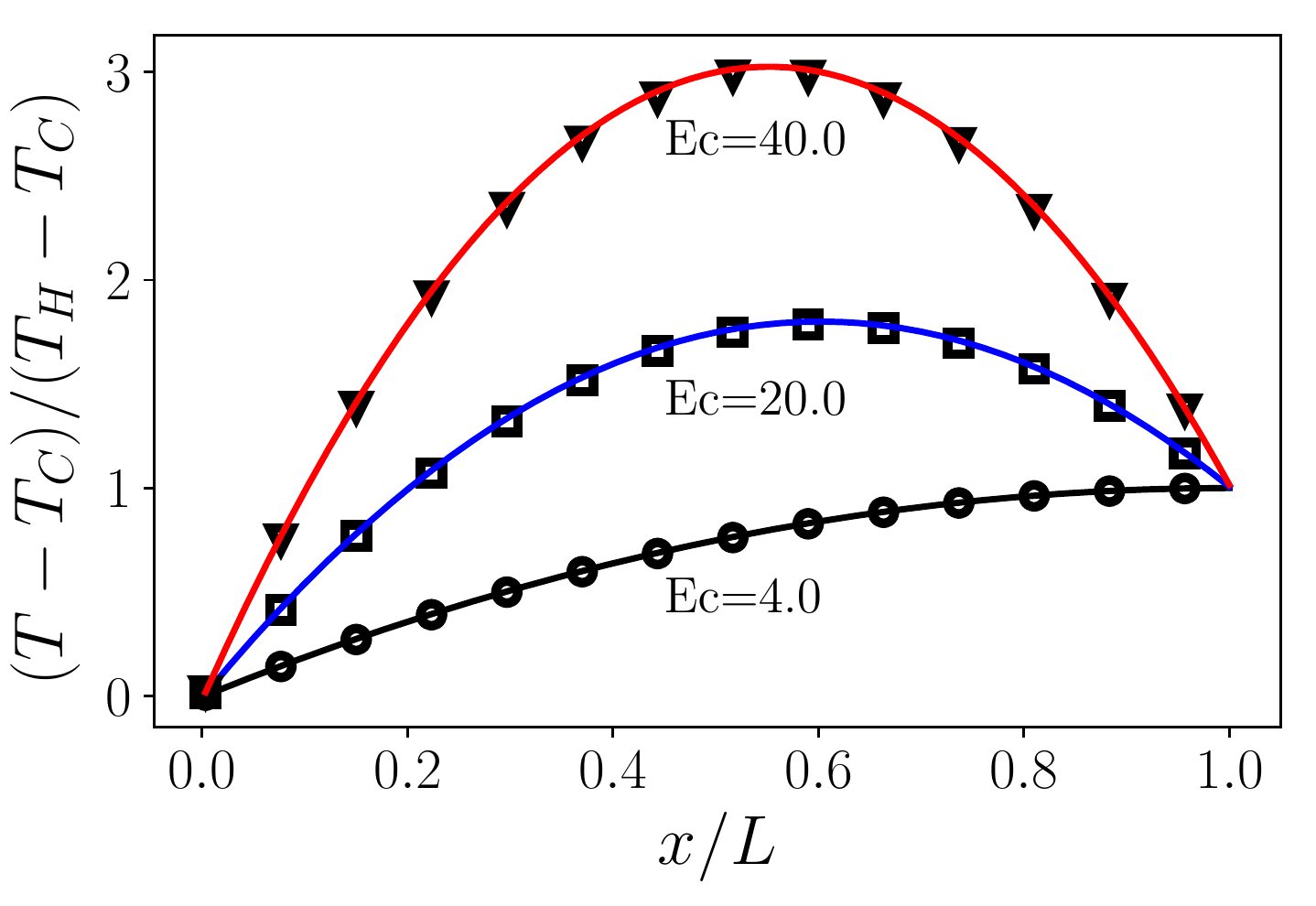}
    \includegraphics[width=0.45\textwidth]{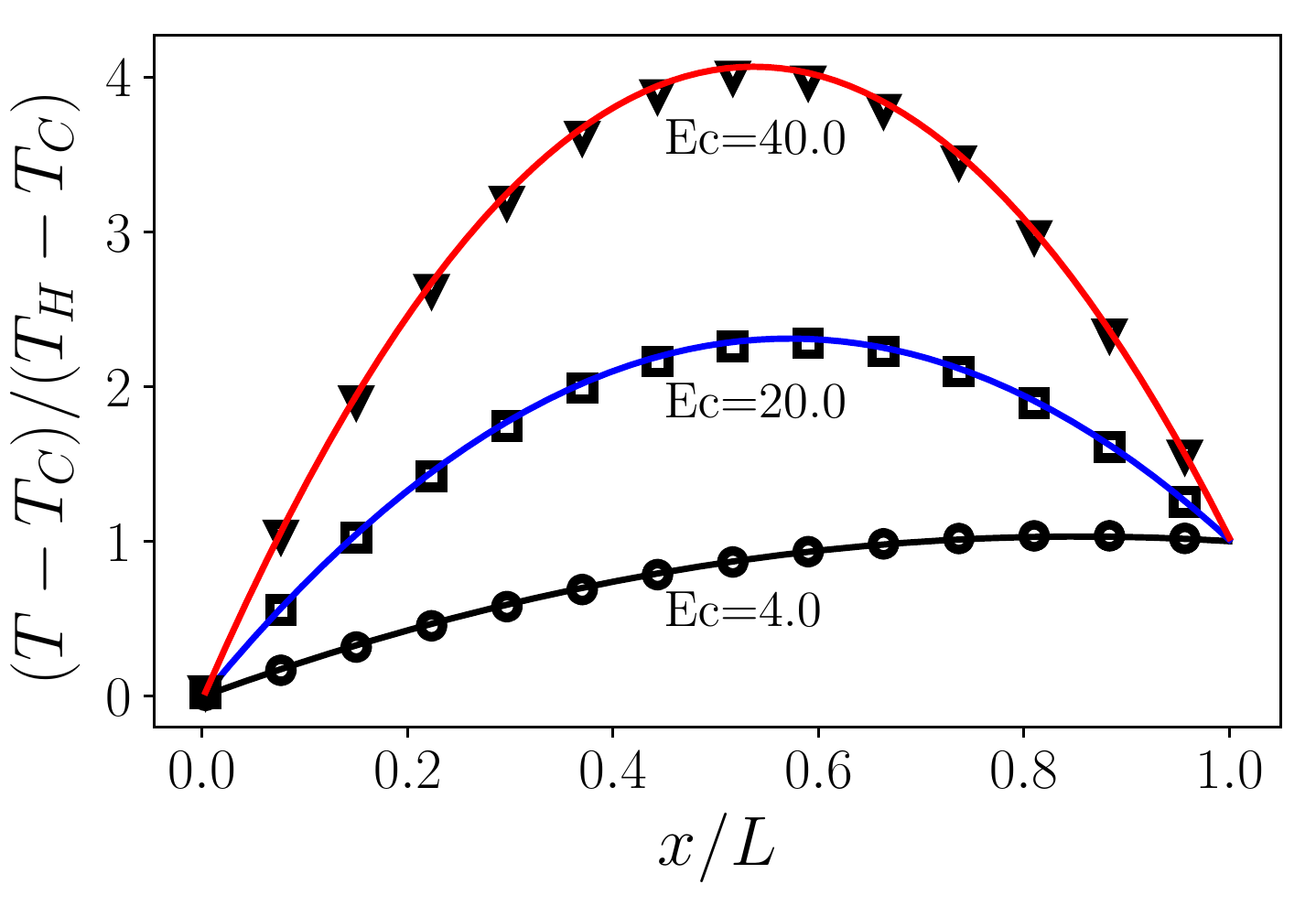}
    \includegraphics[width=0.45\textwidth]{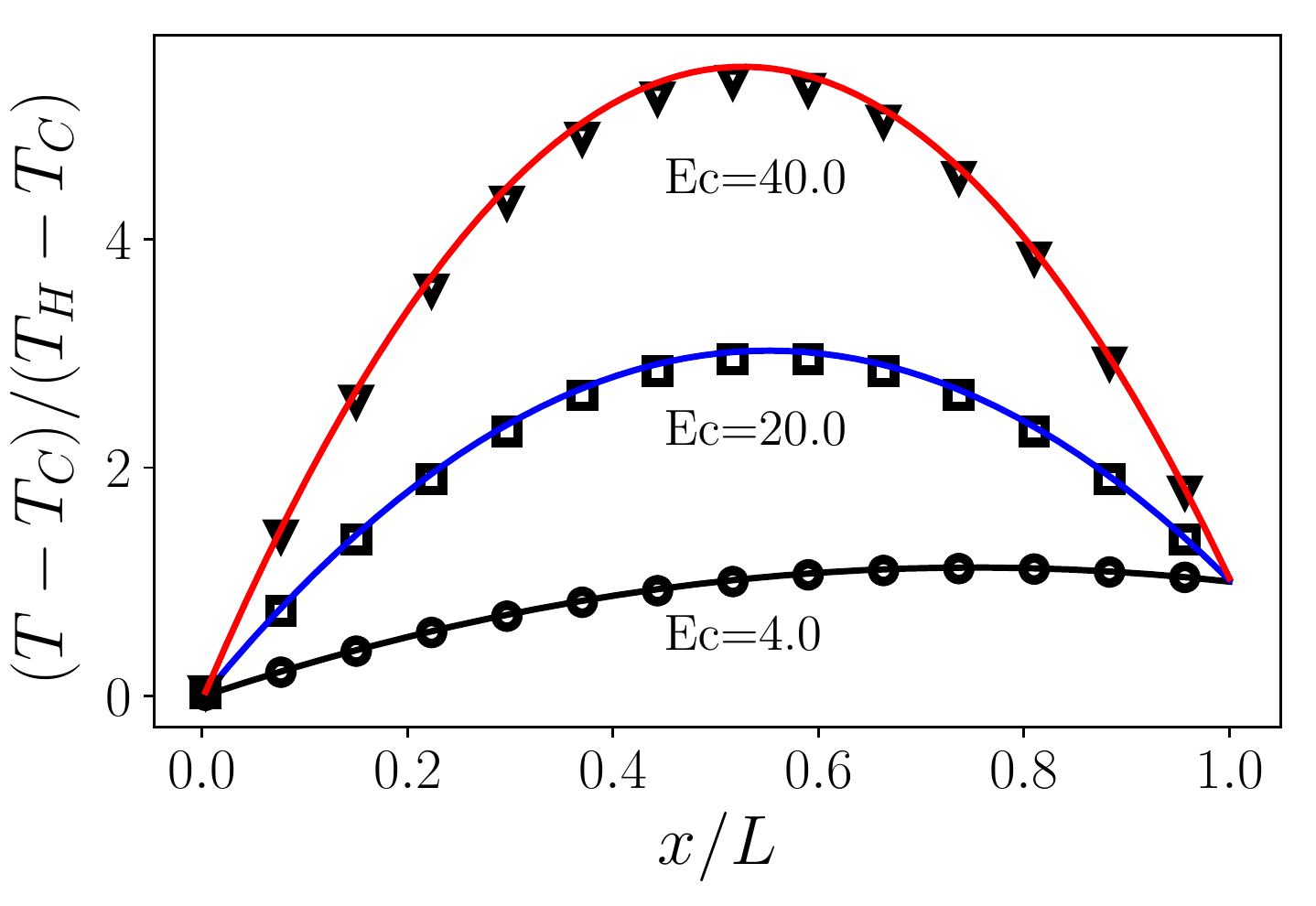}
    \caption{Thermal Couette flow problem. Results are shown for three different Prandtl numbers: $\mathrm{Pr}=0.5$ (top), $\mathrm{Pr}=0.7$ (middle), $\mathrm{Pr}=1$ (bottom). For each $\mathrm{Pr}$ number, the simulation is performed for Eckert numbers $\mathrm{Ec}=4, 20, 40$ (symbols) and compared with analytical solution (solid lines).}
    \label{fig::ThermalCouette}
\end{figure}

The thermal Couette flow is a benchmark test case to probe the viscous heat dissipation and the Prandtl number. The upper wall with the higher temperature $T_H$ is in motion with a constant speed $u_0$, while the lower wall is at rest and at a temperature $T_C$. The analytical solution for the temperature is,
\begin{equation}
    \frac{T-T_C}{T_H-T_C}=\frac{x}{L}+\frac{\rm Pr \cdot Ec}{2}\frac{x}{L}\left(1-\frac{x}L\right),
\end{equation}
where ${\rm Ec}=u_0^2/(C_p\Delta T)$ is the Eckert number and $\Delta T=T_H-T_C$. No slip and constant temperature BCs are applied at the top and bottom walls, while periodic BCs are enforced in the horizontal direction. The parameters for the simulations are ${\rm Ma}=u_0/\sqrt{\gamma T_C}=0.5, L=150,\ {\rm Re}=\rho u_0 L/ \mu = 100,\ T_C=1$. Fig.\ \ref{fig::ThermalCouette} shows the temperature profiles for three different Prandtl numbers (${\rm Pr}= 0.5, 0.7, 1.0$) and different Eckert numbers (${\rm Ec}= 4, 20, 40$), which are in very good agreement with the analytical solution. We note that the simulations have been performed with the $D2Q16$ lattice and thus a third-order Grad's projection frame transformation. The accuracy of the results suggest that the error term in the energy equation \eqref{3order_Energy_equation_OE2} is negligibly small.

\subsubsection{Viscous shock tube}

In this problem we probe the performance of our model with the viscous shock tube test, proposed by Daru and Tenaud \cite{DARUviscous}. A 2D shock tube $[0, 1]\times [0, 1]$ is initialized with the following conditions,
\begin{equation}
    (\rho,u_x,u_y,p)=\begin{cases}
    (120, 0, 0, 120/\gamma),& 0 \leq x<0.5, \\
    (1.2, 0, 0, 1.2/\gamma),& 0.5 \leq x \leq 1, \\
    \end{cases}
\end{equation}
where $\gamma=1.4$, the Prandtl number is set to $\rm Pr = 0.73$ and the viscosity is set such that Reynolds number is $\rm Re = 200$. No slip and adiabatic BCs are applied at the walls of the shock tube. Due to the symmetric configuration of the problem, the actual simulated geometry consists of the $[0, 1]\times [0, 0.5]$ domain, with symmetric conditions applied on the top boundary. The initial flow conditions create a right propagating shock wave of $\rm Ma=2.37$, a contact discontinuity and an expanding rarefaction wave towards both directions. It is noted that the motion of the shock wave induces a non-negligible boundary layer along the horizontal wall of the tube. The boundary layer interacts with the incident and reflected shock, forming a complicated flow pattern.

\begin{figure*}
    \centering
   \includegraphics[width=0.8\textwidth]{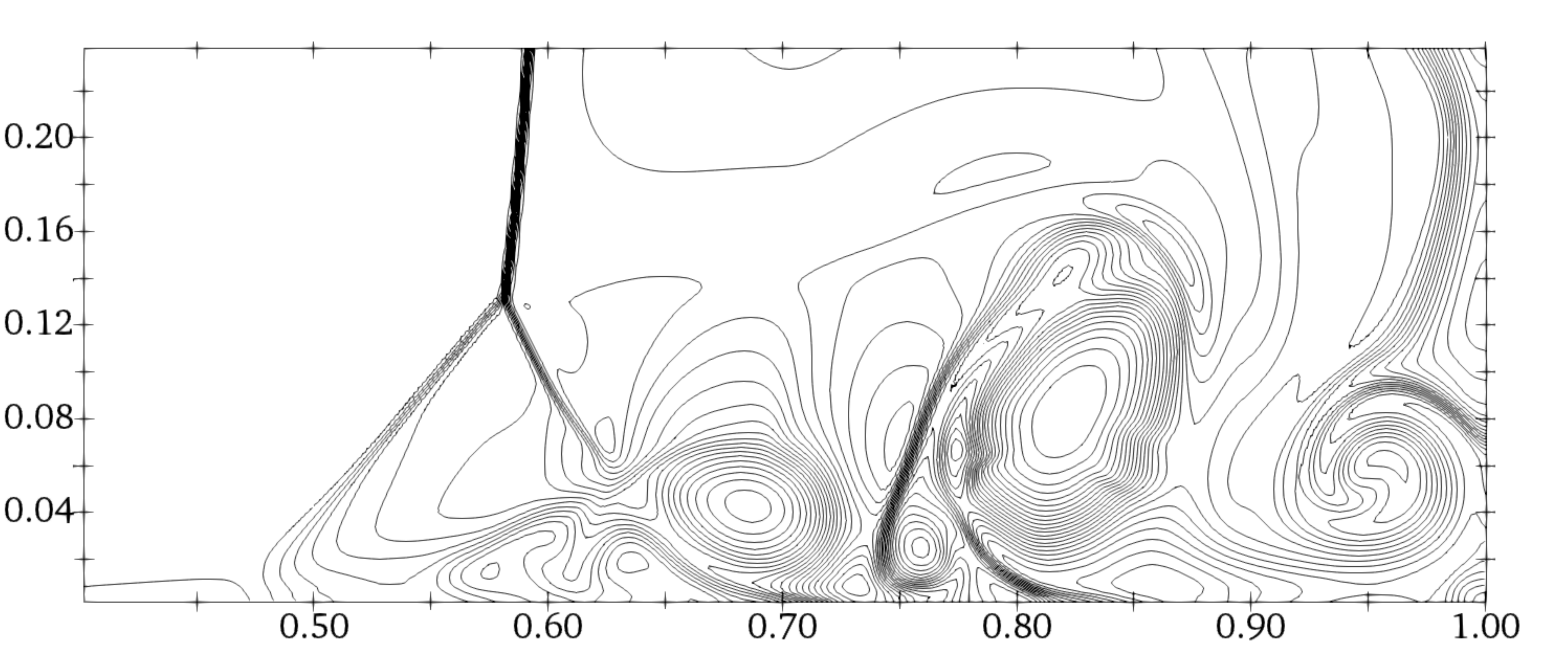}
   \includegraphics[width=0.8\textwidth]{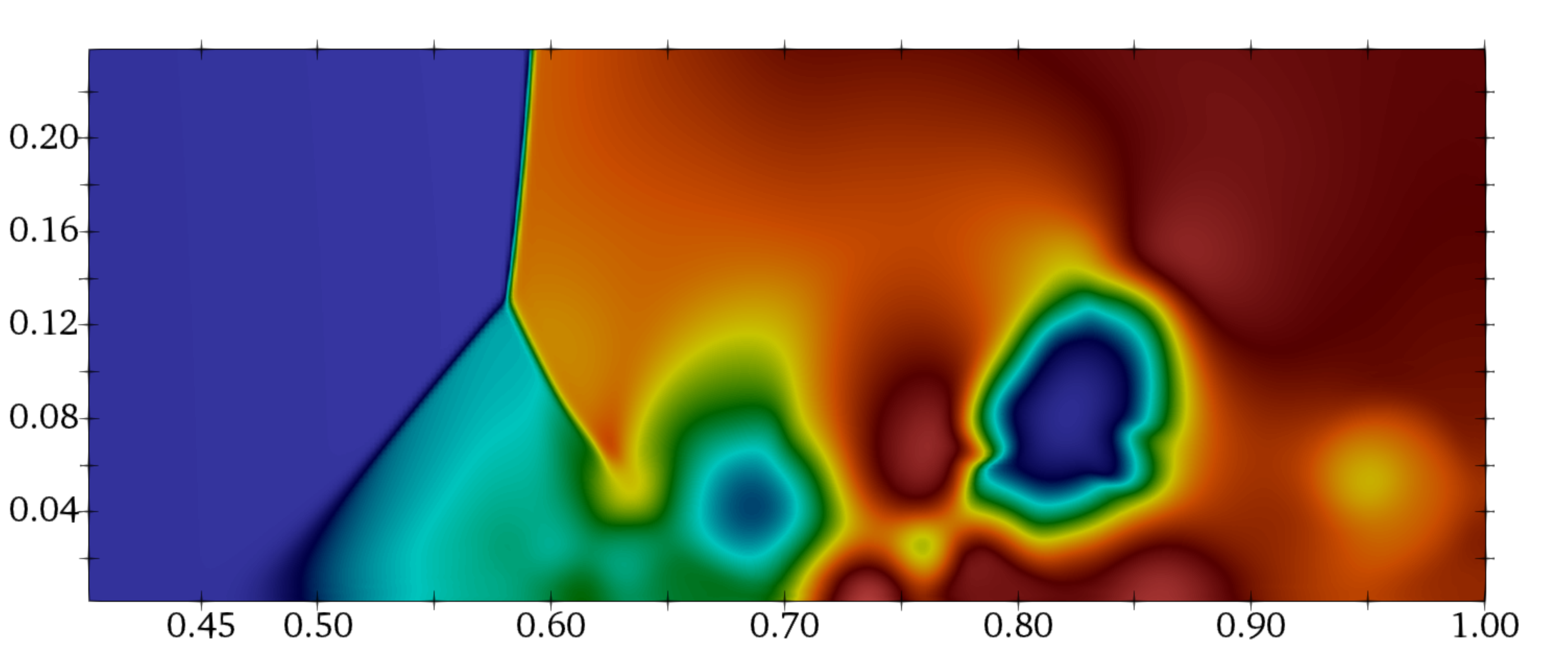}
   \includegraphics[width=0.8\textwidth]{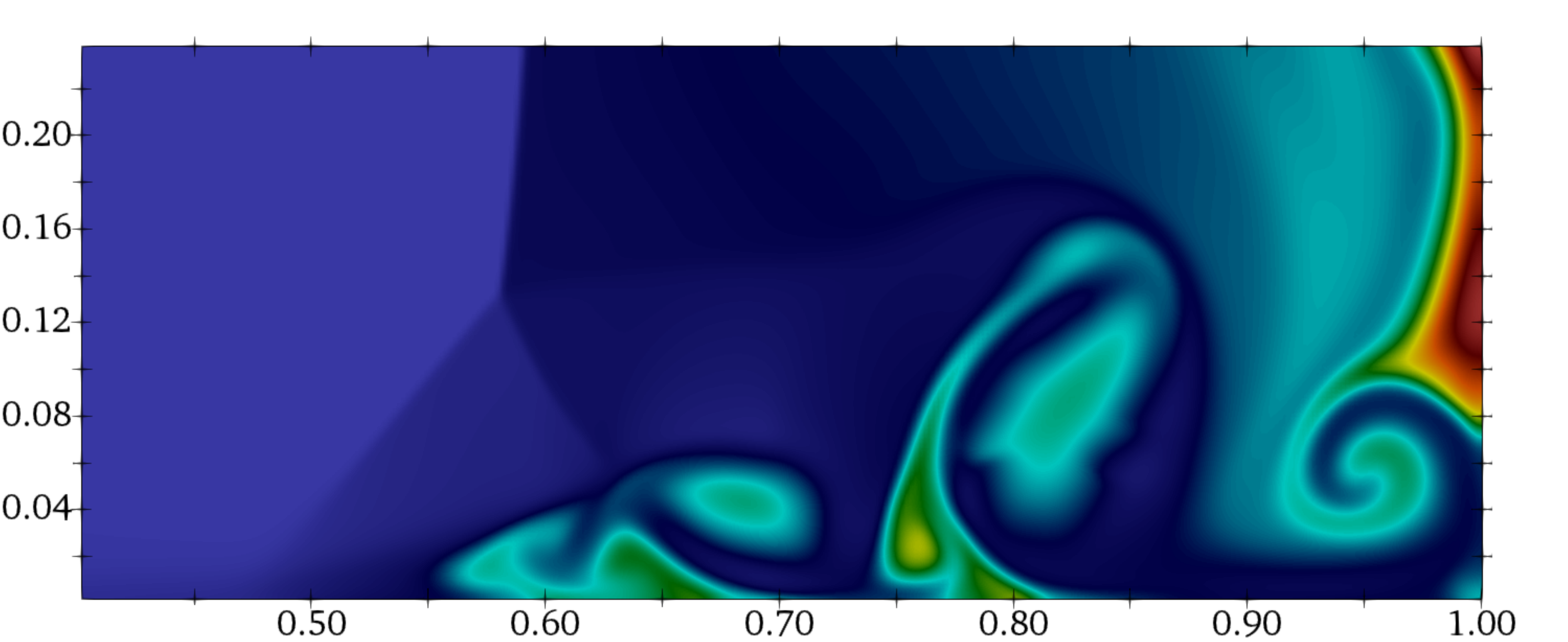} 
    \caption{Viscous shock tube problem. Top: 30 equidistant density contours. Middle: pressure field. Bottom: Temperature field.}
    \label{ViscousTube_Fields}
\end{figure*}

\begin{figure}
    \centering
    \includegraphics[width=0.45\textwidth]{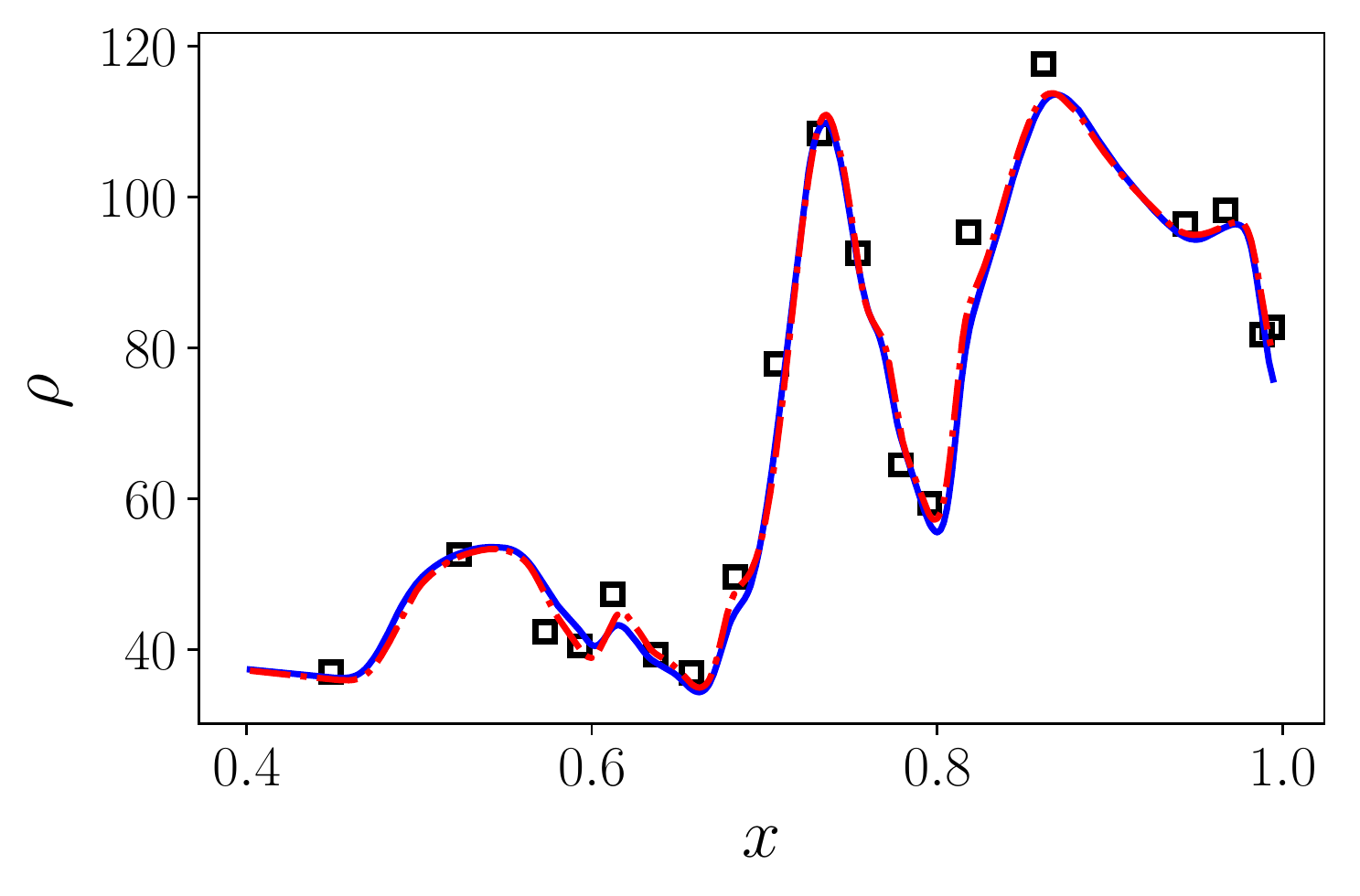} 
    \caption{Viscous shock tube problem. Density distribution along the solid wall. Solid line: $D2Q16$ model. Dotted line: $D2Q25$. Symbols: reference data from \cite{Zhou_viscous}}
    \label{ViscousTube_Comparison}
\end{figure}

The results of the density contours, the pressure and temperature fields for resolution of $[500, 250]$ at $t=1$, are shown in Fig.\ \ref{ViscousTube_Fields}. Additionally, we repeat the simulation with the D2Q25 and a fourth-order frame invariant moment system and compare the results. For the comparison, we report metrics suggested from \cite{Zhou_viscous}. In particular, Table \ref{Table::ViscousTube} summarizes the coordinates associated with the triple point and the primary vortex. The comparison with the reference data show a very good match of both $D2Q16$ and $D2Q25$ simulations. The same conclusion is drawn from Fig.\ \ref{ViscousTube_Comparison}, which plots the density distribution along the solid wall.

\begin{table}[] \centering
\caption{Accuracy criteria for viscous shock tube, according to \cite{Zhou_viscous}. x-TP (y-TP) correspond to the x-coordinate  (y-coordinate) of the triple point. x-PV corresponds to the horizontal axis intersection of the line passing through the primary vortex. y-PV corresponds to the height of the primary vortex. } 
\begin{tabular}{cccccl}
\hline \hline
     &  x-TP       & y-TP       & x-PV  & y-PV     \\ \hline
Reference &  0.58     & 0.137   & 0.78   & 0.166     \\
D2Q16    & 0.58   & 0.133      & 0.774    & 0.168       \\ 
D2Q25    & 0.58     & 0.134      & 0.775     & 0.168      \\
\hline \hline
\label{Table::ViscousTube}
\end{tabular}
\end{table}

\subsubsection{Shock structure problem}

The problems so far have demonstrated very good accuracy of the third-order frame invariant moment system and the associated $D2Q16$ lattice. The following benchmark involves steep hydrodynamic gradients and clearly demonstrates the limitations of the $D2Q16$. At the same time, the expansion of the frame invariant moment system from third-order to fourth-order ($D2Q25$) restores the accuracy.

The shock structure problem is a classical problem in kinetic theory of gases, in which non-equilibrium effects dominate the flow \cite{CercignBook}. We consider a quasi one-dimensional plane shock wave, with an initial step of density, velocity and temperature at the center of the computational domain. 
The upstream and downstream flow values are connected through the Rankine--Hugoniot conditions \cite{Anderson}. The upstream mean free path for hard sphere molecules is defined as,
\begin{equation}
\lambda_1=\frac{16}{5\sqrt{2\pi\gamma}} \left( \frac{\mu_1\alpha_1}{p_1} \right),
\end{equation}
where $p_1,\alpha_1,\mu_1$ are the pressure, the speed of sound and the viscosity of the gas upstream of the shock, respectively. The viscosity varies with the temperature as,
\begin{equation}
    \mu=\mu_1\left(\frac{T}{T_1}\right)^s,
\end{equation}
where for the case of hard spheres $s=0.5$.
The steady-state non-dimensional density, temperature, normal stress and heat flux are defined as follows,
\begin{equation}
\rho_n=\frac{\rho-\rho_1}{\rho_2-\rho_1}, T_n=\frac{T-T_1}{T_2-T_1}, \hat{\sigma}_{xx}=\frac{\sigma_{xx}}{p_1}, \hat{q}_x=\frac{q_x}{p_1\sqrt{2T_1}},
\end{equation}
where the subscripts $1$ and $2$ indicate the upstream and downstream values, respectively. The Prandtl number is set to ${\rm Pr}=2/3$ and the adiabatic exponent of monoatomic ideal gas to $\gamma=5/3$. The results reported for this case are the steady-state solutions and compared with the results of Ohwada \cite{ohwada}. The origin of the coordinate system is the point with $\rho_n=0.5$ and {$x_n=x/0.5\sqrt{\pi}\lambda_1$ is used as the reduced coordinate}.

\begin{figure*}
    \centering
   \includegraphics[width=0.45\textwidth]{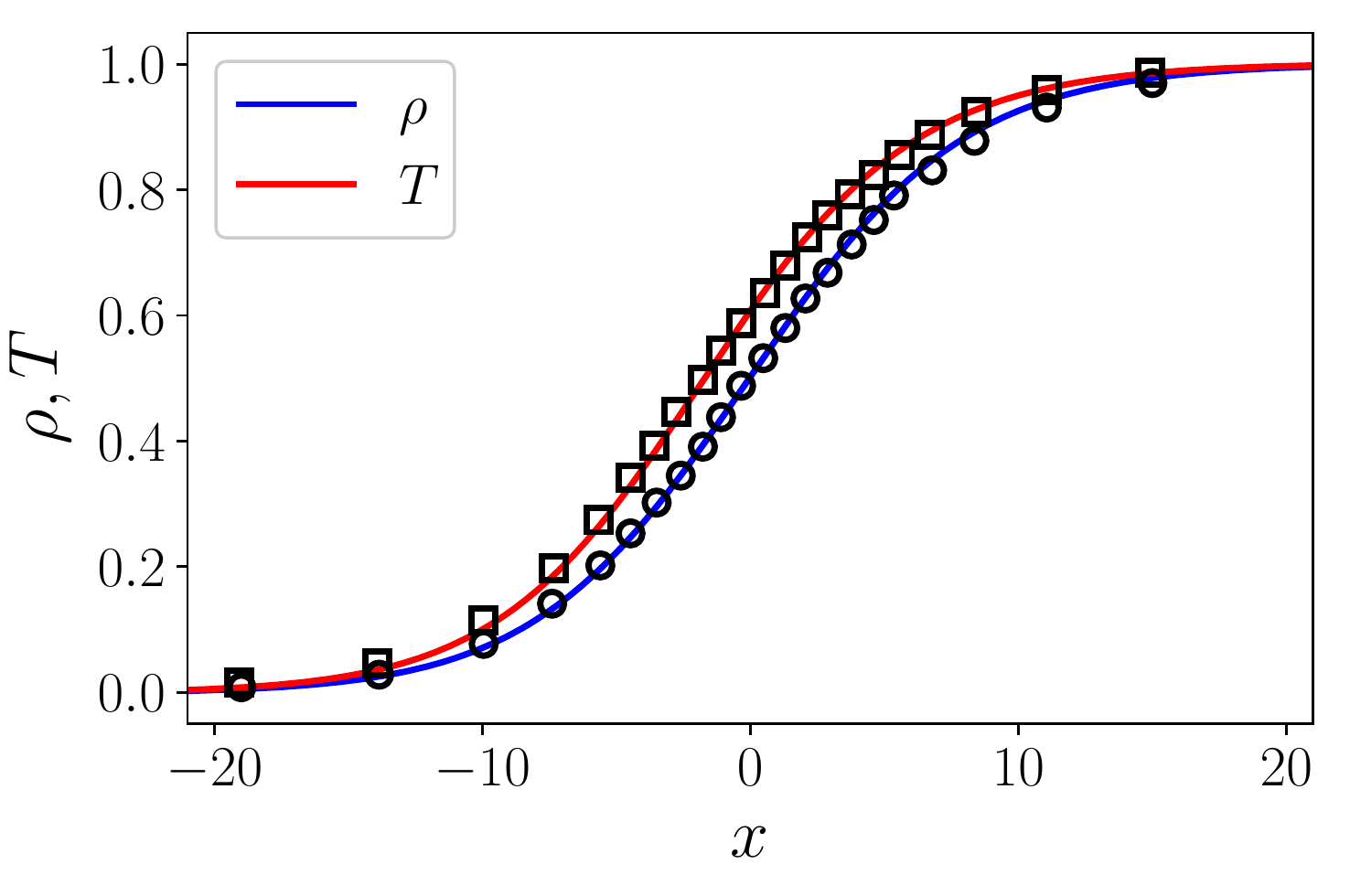}
   \includegraphics[width=0.45\textwidth]{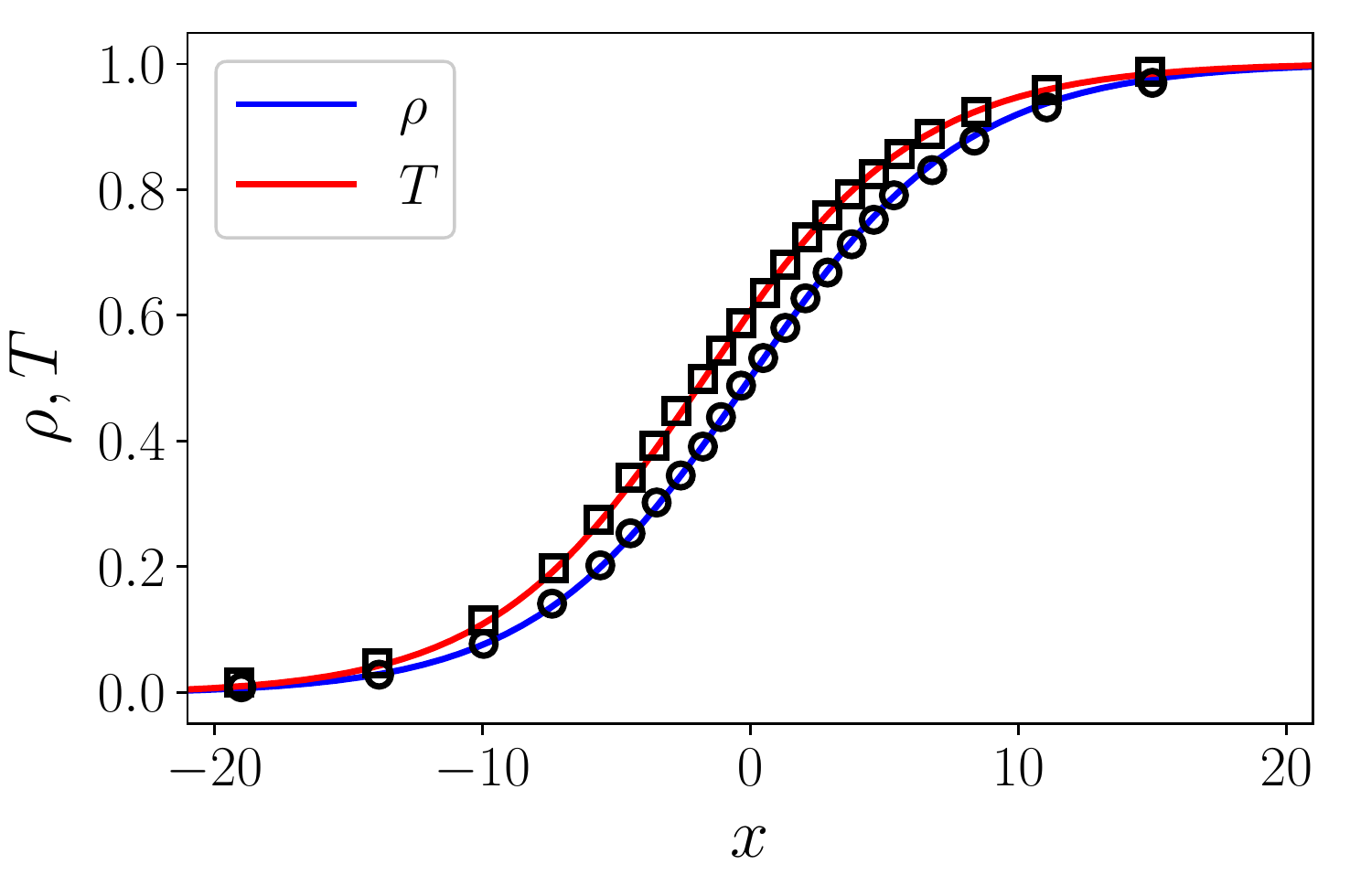}
    \includegraphics[width=0.45\textwidth]{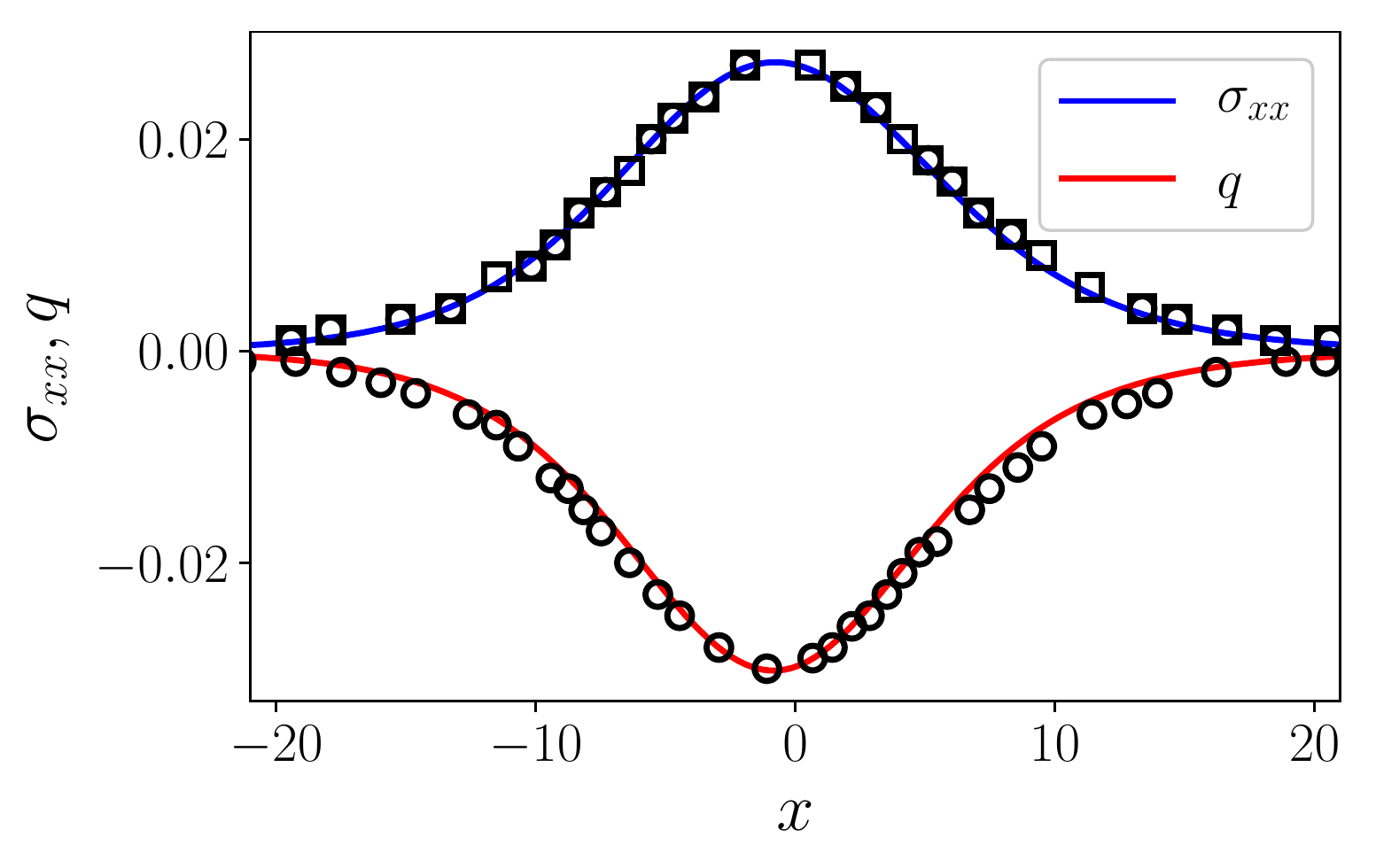}
    \includegraphics[width=0.45\textwidth]{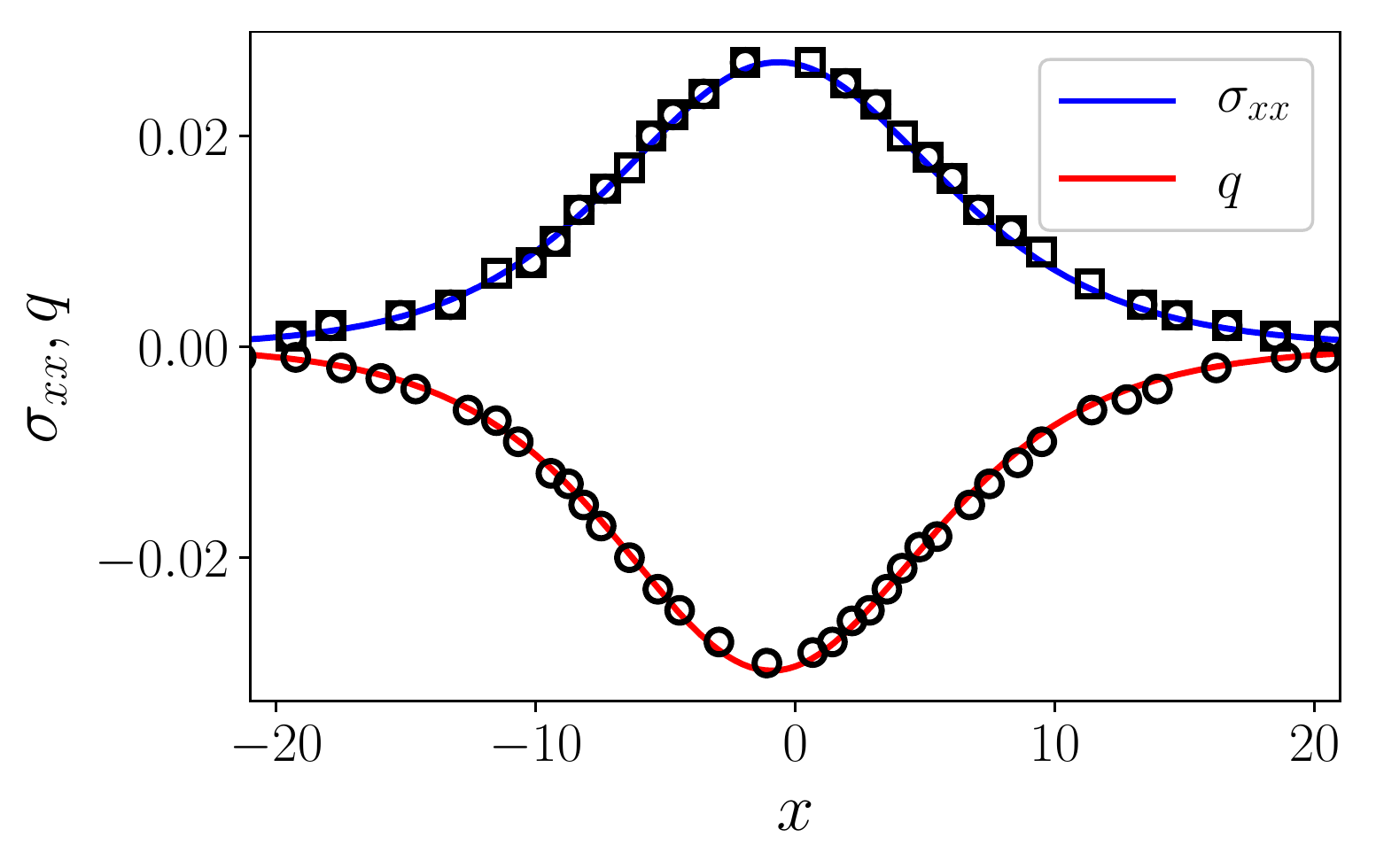}
    \caption{The shock structure problem with Ma 1.2. Density, temperature profiles (top), normal stress and heat flux profiles (bottom). Third-order transformation at the left column, fourth-order at the right column.}
    \label{fig::ShockStructure1.2}
\end{figure*}

We consider first the shock structure profiles for a Mach number $\rm Ma=1.2$. Two simulations are performed with different reference frame transformation orders. In particular, we compare the performance of the $D2Q16$ and $D2Q25$ lattices, using third- and fourth-order Grad's projection respectively. The results for the density, temperature, normal stress and heat flux profiles are shown in Fig.\ \ref{fig::ShockStructure1.2}. It is evident that both models perform very accurately, compared with the reference data.

\begin{figure*}
    \centering
   \includegraphics[width=0.45\textwidth]{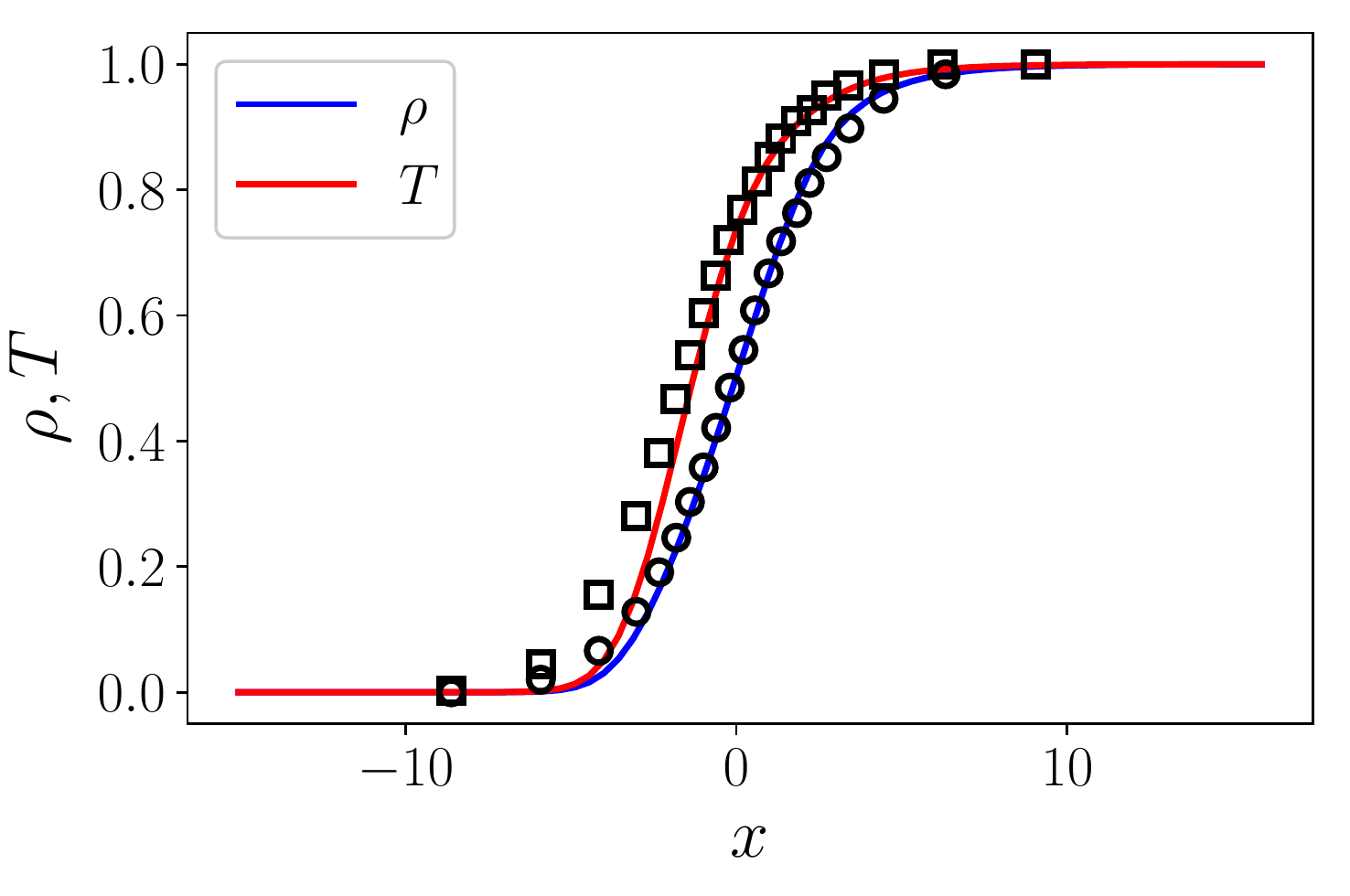}
   \includegraphics[width=0.45\textwidth]{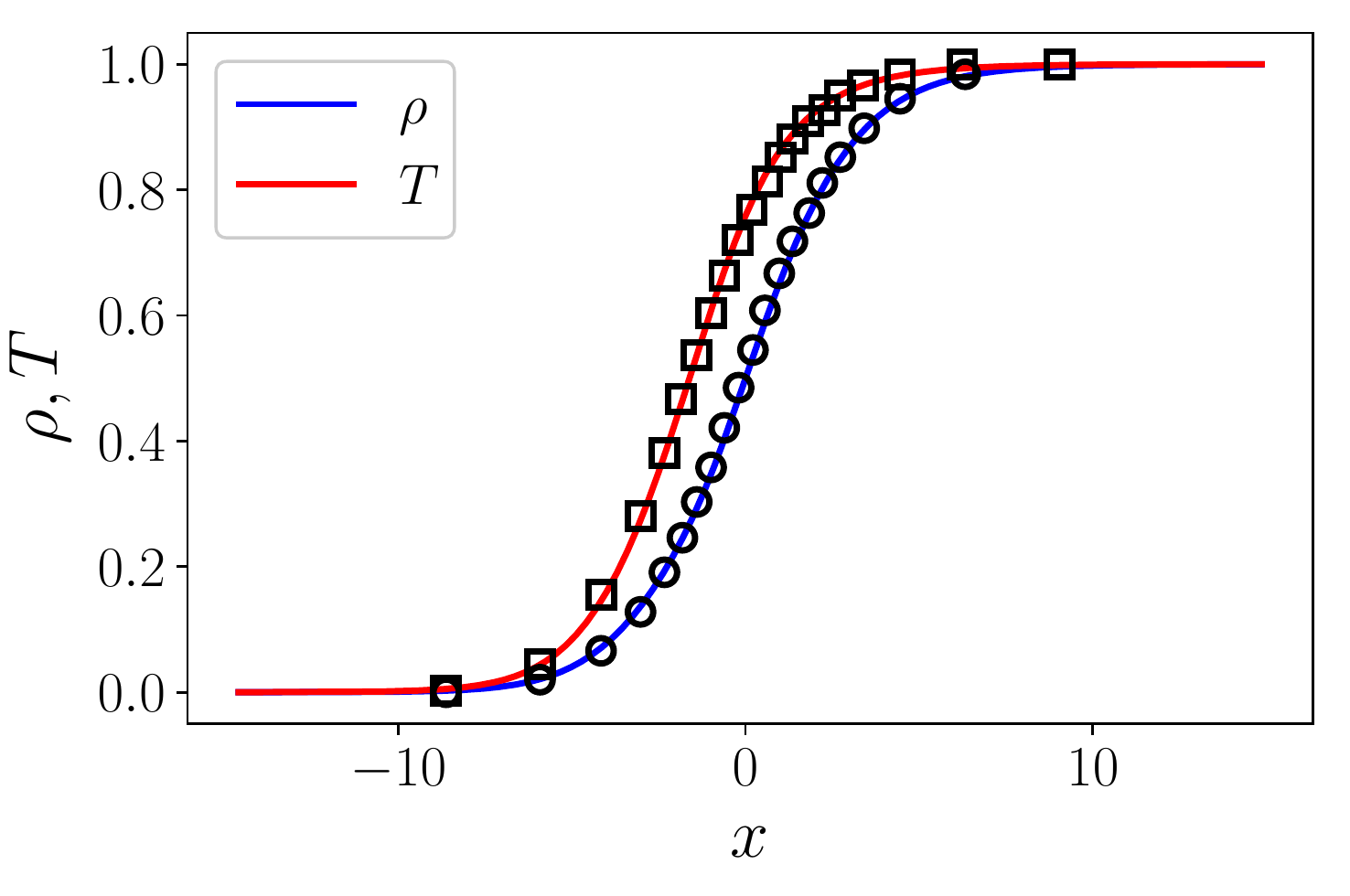}
    \includegraphics[width=0.45\textwidth]{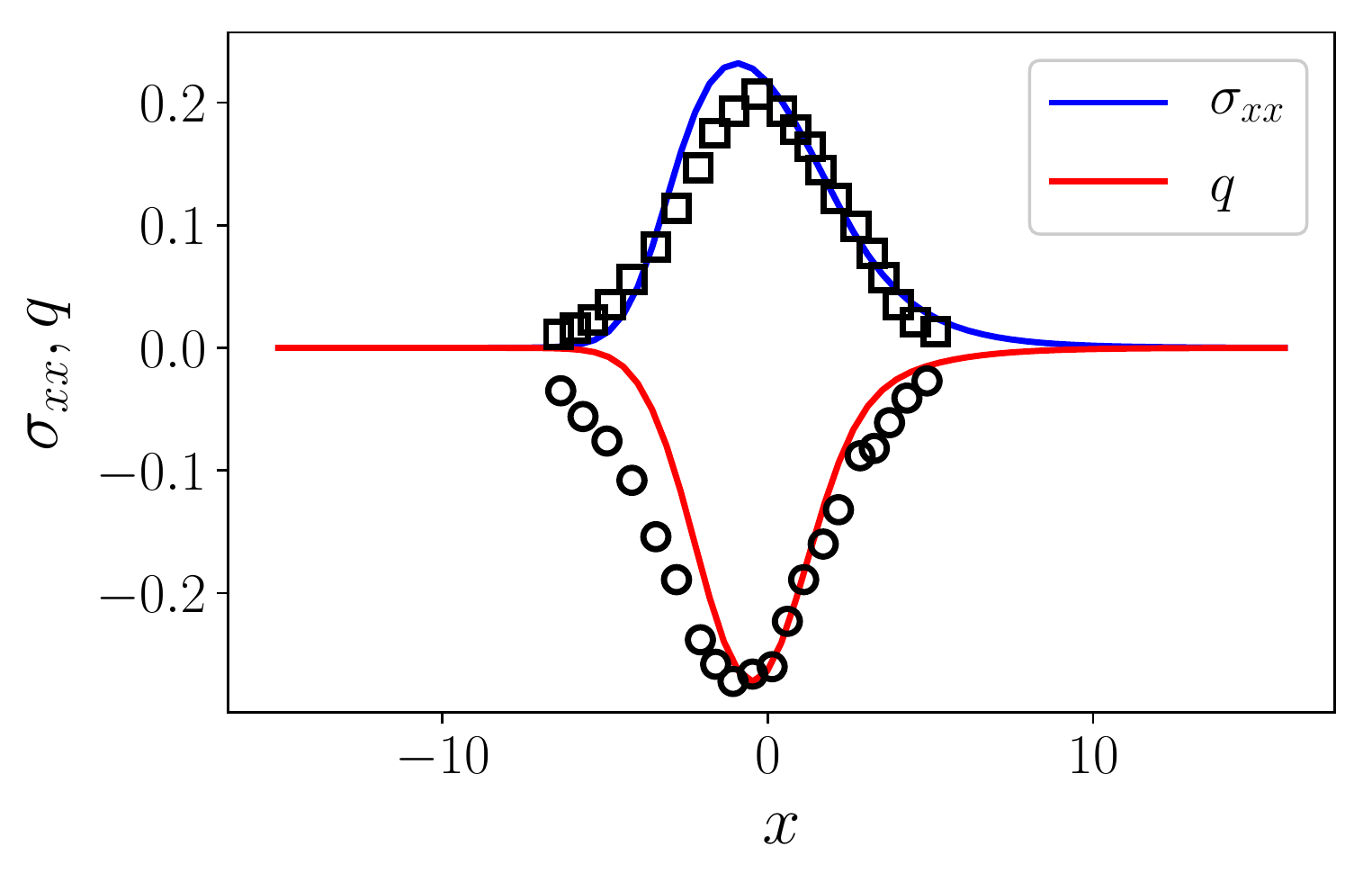}
    \includegraphics[width=0.45\textwidth]{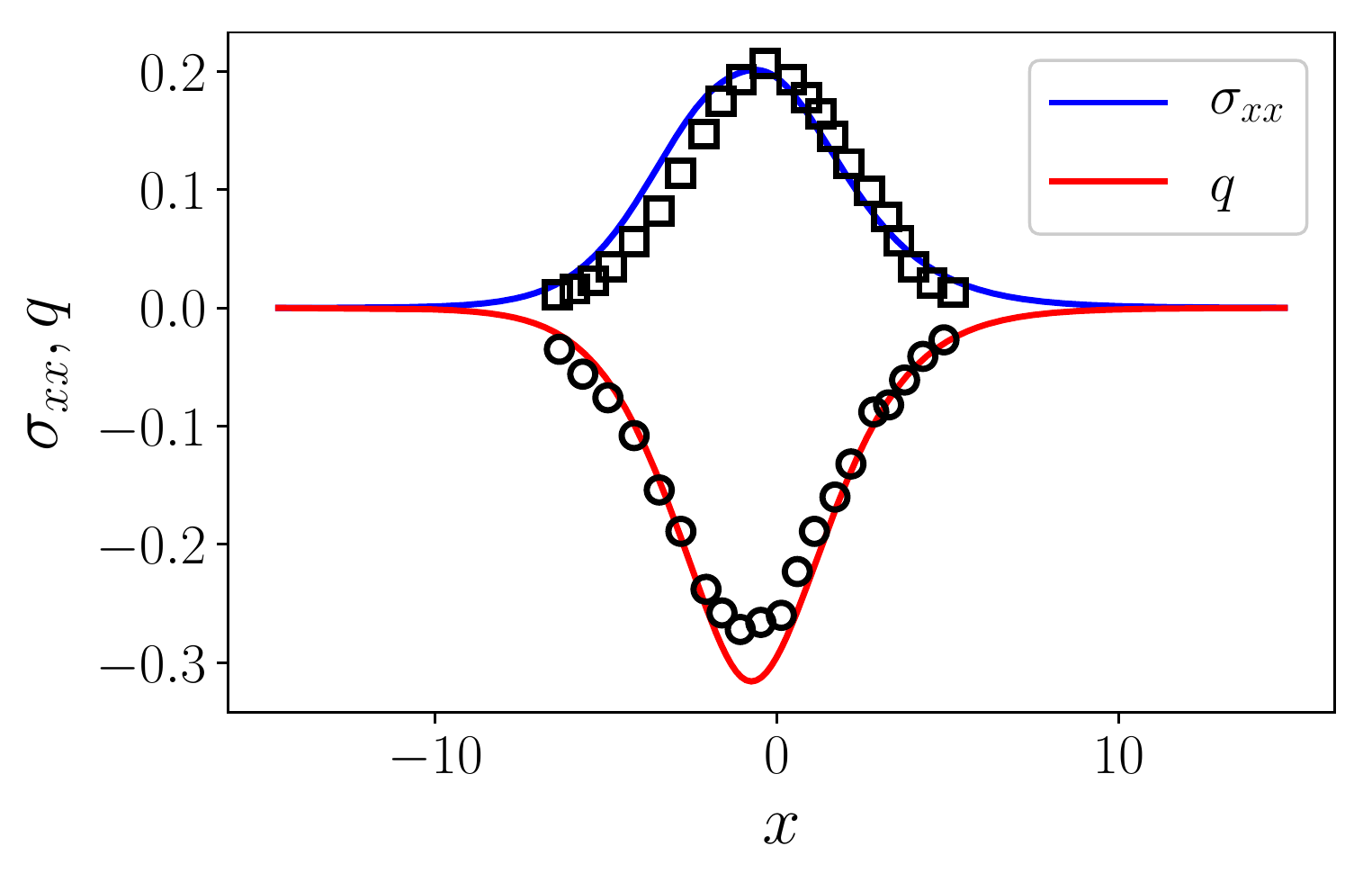}
    \caption{The shock structure problem with Ma 1.6. Density, temperature profiles (top), normal stress and heat flux profiles (bottom). 3rd order transformation at the left column, 4th order at the right column.}
 \label{fig::ShockStructure1.6}    
\end{figure*}

We continue with the shock structure at a higher Mach of $\rm Ma=1.6$ and repeat the numerical experiments with the different frame transformation orders. The results are summarized in Fig.\ \ref{fig::ShockStructure1.6}. Here, the third-order model clearly shows deviations in all the profiles, with the errors being prominent in the temperature and heat flux profiles, at the upstream part of the shock. In this case, the deviation terms due to the frame variant $\bm{R}^{\rm eq}$ moment are sustained, due to the steep gradients of velocity and temperature within the shock profile. Including the $\bm{R}^{\rm eq}$ moment list into the frame invariant list, i.e. the fourth-order model, recovers the accuracy of the model and achieves very good agreement with the reference results.

\subsection{Multiscale framework}

The final topic of interest is the multiscale extension of the scheme, with the deployment of different lattices across the domain. The switching criterion is a threshold on the local flow velocity and temperature gradients. The high-order lattice is activated at the portion of the domain with high gradients, while the low-order lattice everywhere else.

\subsubsection{Lax tube}

\begin{figure}
    \centering
   \includegraphics[width=0.45\textwidth]{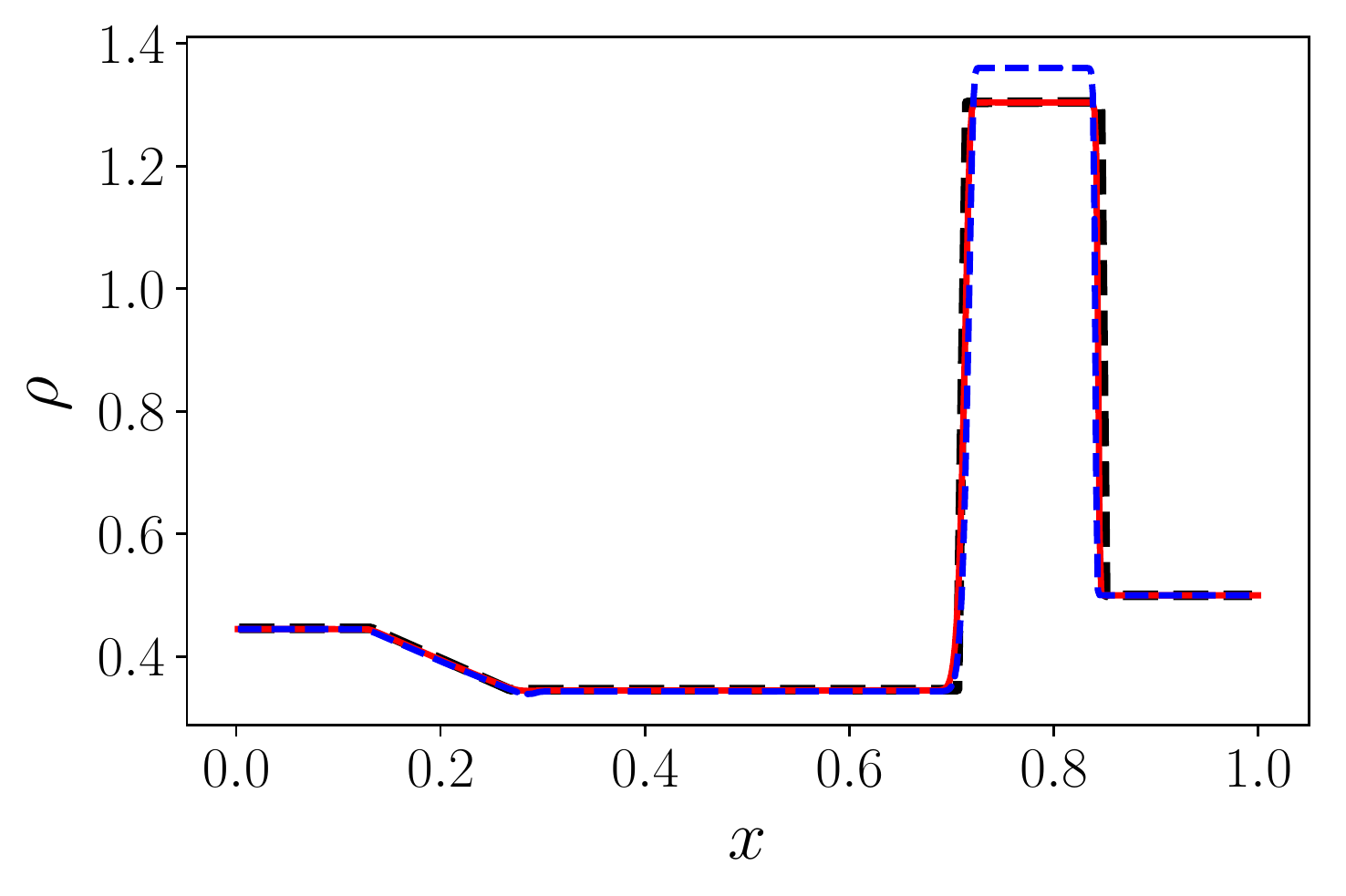}
    \caption{Density profile for the Lax tube problem, at $t=0.14$. The red line corresponds to the $D2Q16$ lattice, the blue dashed line to the $D2Q9$ and the black dashed line to the analytical solution.}
    \label{Multiscale Lax tube_Comparison}
\end{figure}

\begin{figure}
    \centering
    \includegraphics[width=0.45\textwidth]{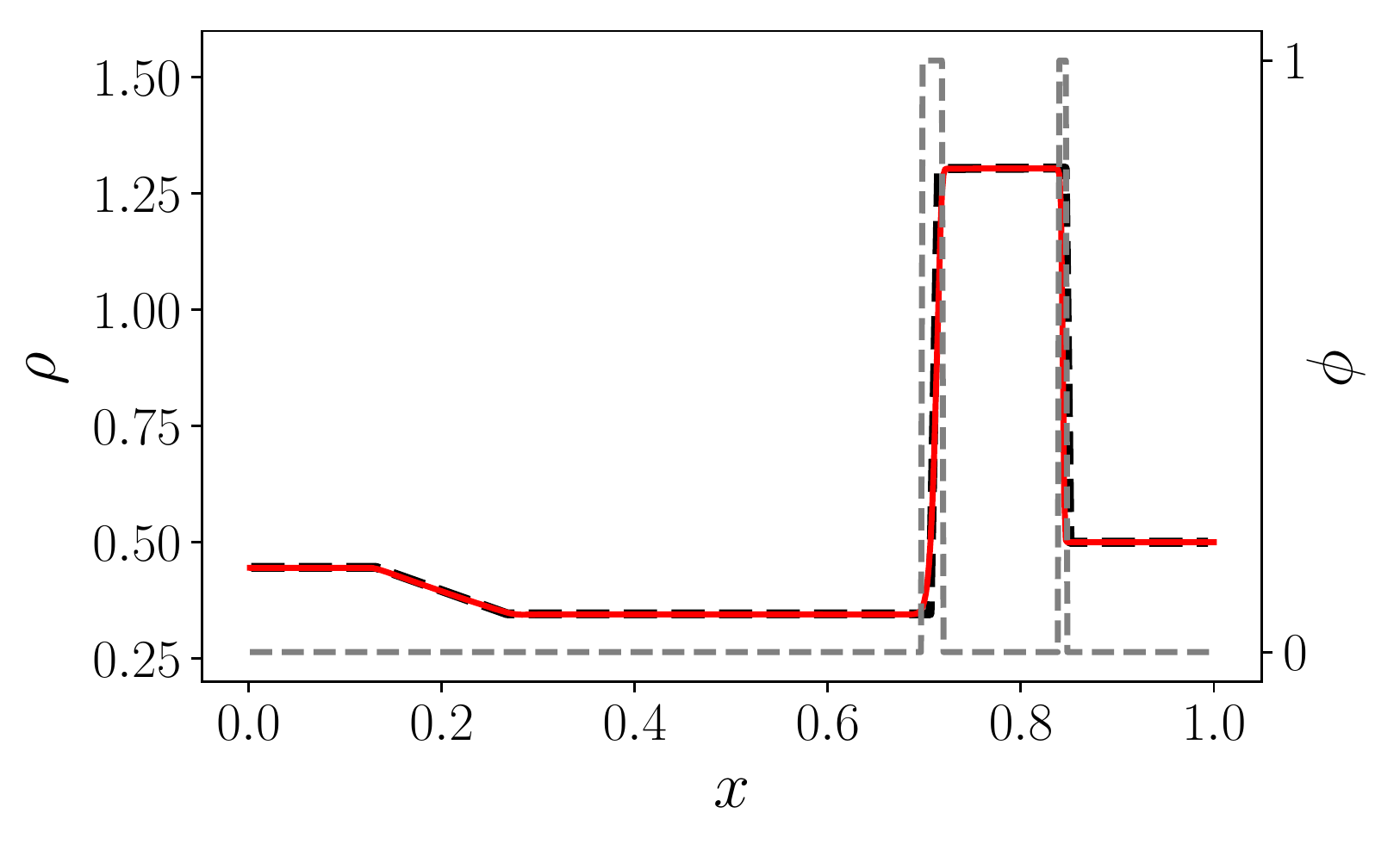}
    \caption{Density profile for the Lax tube problem, at $t=0.14$. The red line corresponds to the $D2Q9/D2Q16$ model and the black dashed line to the analytical solution. The grey dashed line indicates the occupancy regions of the $D2Q16$ lattice ($\phi=1$) and of the $D2Q9$ lattice ($\phi=0$).}
    \label{Multiscale Lax tube}
\end{figure}

We demonstrate a $D2Q9/D2Q16$ model, with the simulation of the Lax problem \cite{LaxTube}. The initial conditions are the following,
\begin{equation}
    (\rho,u_x,p)=\begin{cases}
    (0.445, 0.698, 3.528),& 0 \leq x<0.5, \\
    (0.5, 0, 0.571),& 0.5 \leq x \leq 1. \\
    \end{cases}
\end{equation}
The simulation is performed with $L=600$, until $t=0.14$. Fig.\ \ref{Multiscale Lax tube_Comparison} shows the solution obtained by the $D2Q9$ and $D2Q16$ lattices independently. While the $D2Q16$ model is in excellent agreement with the analytical solution, the $D2Q9$ model develops deviations, which manifest as overestimated density between the shock wave and the contact discontinuity. The discrepancy in the Euler level is expected for the case of $D2Q9$ and therefore a second-order moment invariant system. Fig.\ \ref{Multiscale Lax tube} shows the results of the multi-scale $D2Q9/D2Q16$ model and the regions of deployment of the two lattices. In particular, the $D2Q16$ is active in two thin regions, centered at the shock wave and the contact discontinuity.
The results of the multiscale model match again very well with the analytical solution.

\subsubsection{Shu-Osher problem}

\begin{figure}
    \centering
  \includegraphics[width=0.45\textwidth]{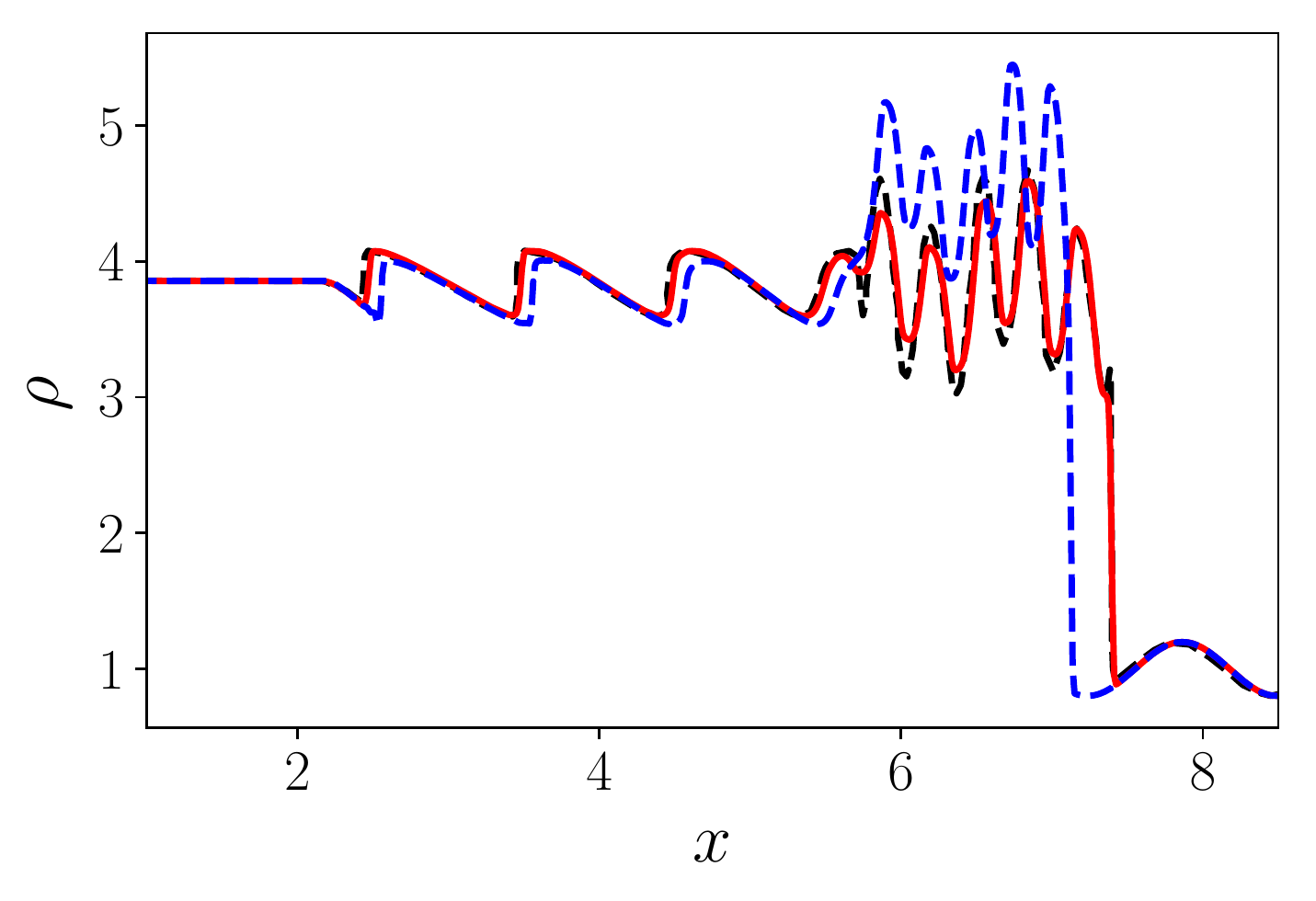}  
    \caption{Density profile for the Shu-Osher problem, at $t=1.8$. The red line corresponds to the $D2Q16$ lattice, the blue dashed line to the $D2Q9$ and the black dashed line to the reference solution \cite{ShuOsherReference}.}
     \label{Multiscale Shu-Osher tube-comparison}
\end{figure}

\begin{figure*}
    \centering
   \includegraphics[width=0.45\textwidth]{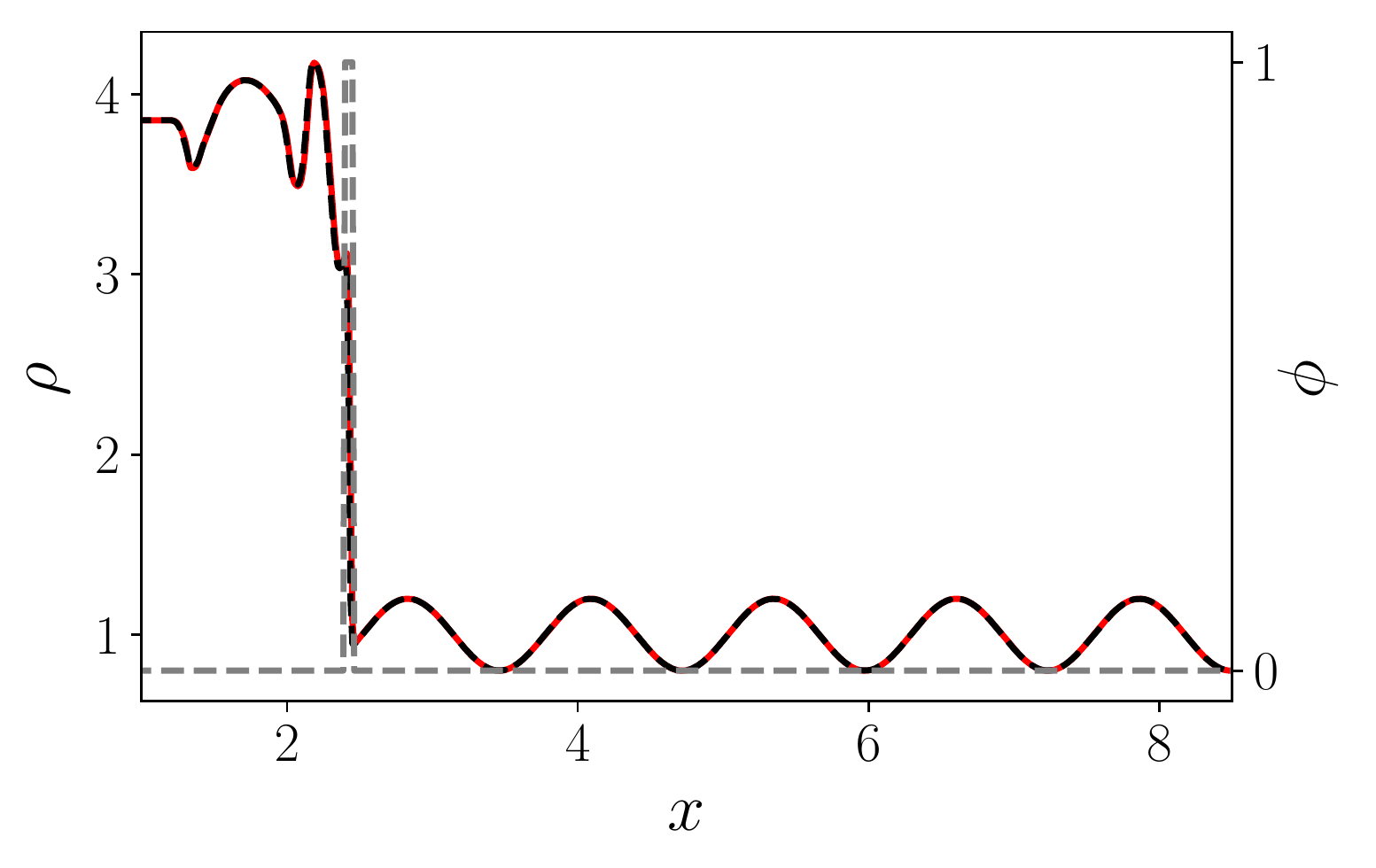}
 \includegraphics[width=0.45\textwidth]{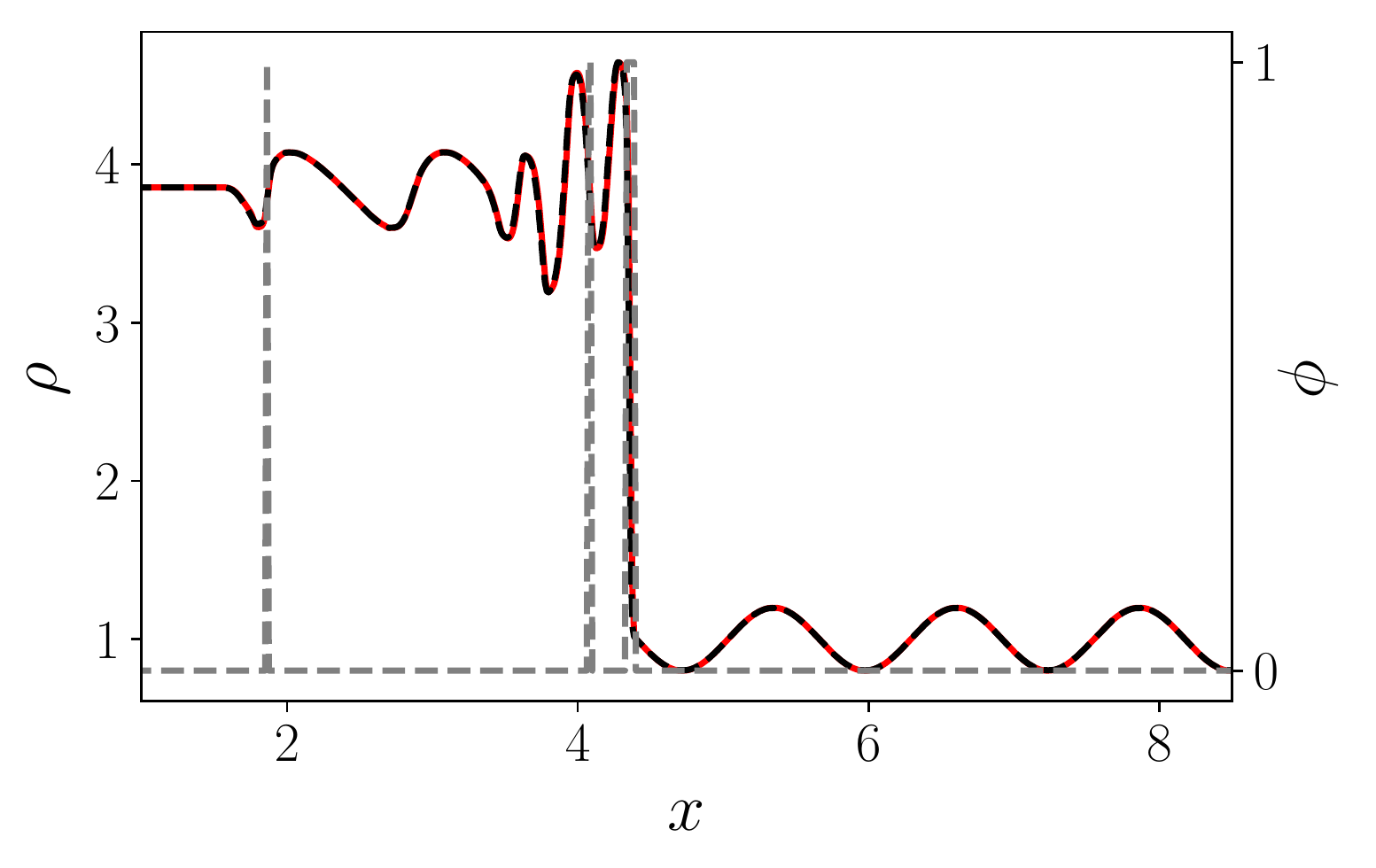}
  \includegraphics[width=0.45\textwidth]{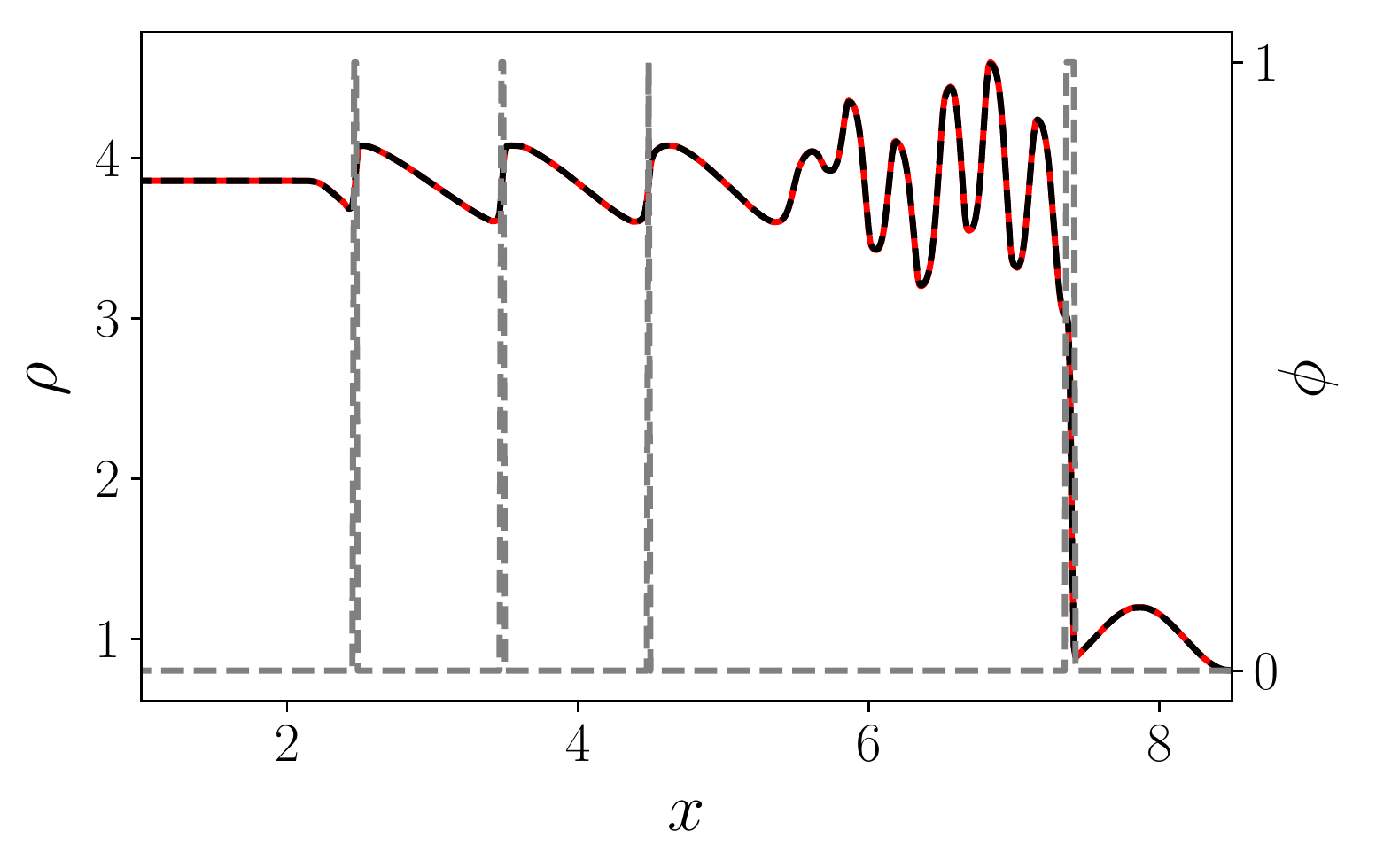}
    \caption{Density profile for the Shu-Osher problem, at different times. The red line corresponds to the $D2Q9/D2Q16$ model and the black dashed line to the $D2Q16$ solution. The grey dashed line indicates the occupancy regions of the $D2Q16$ lattice ($\phi=1$) and of the $D2Q9$ lattice ($\phi=0$).}
    \label{Multiscale Shu-Osher}
\end{figure*}

The $D2Q9/D2Q16$ model is further tested with the Shu-Osher problem \cite{ShuOsherProblem}. In this setup, a Mach 3 shock wave interacts with a perturbed density field. The interaction leads to discontinuities and the formation of small structures. The initial conditions are,
\begin{equation}
    (\rho,u_x,p)=\begin{cases}
    (3.857, 2.629, 10.333),& 0 \leq x<1, \\
    (1+0.2 \sin (5(x-5)), 0, 1),& 1 \leq x \leq 10. \\
    \end{cases}
\end{equation}
The results for the density profile are presented at $t=1.8$ and $L=800$. Fig.\ \ref{Multiscale Shu-Osher tube-comparison} shows the solutions of the $D2Q9$ and $D2Q16$ models and the comparison with a reference solution, obtained with characteristic-based 5th order WENO, RK4 temporal integration and resolution of 5000 points \cite{ShuOsherReference}. Apart from a small underestimation of the post-shock waves amplitudes, it is evident that the $D2Q16$ captures very well the shock location and the high frequency waves. In contrast, the $D2Q9$ model clearly deviates from the reference solution. Fig.\ \ref{Multiscale Shu-Osher} captures the evolution of the $D2Q9/D2Q16$ solution and compares it with the pure $D2Q16$ solution. The multiscale model is almost indistinguishable from the $D2Q16$ model and thus with the reference solution also. It is also interesting to observe that $D2Q16$ is activated only in narrow regions of the domain, as shown by the spikes in Fig.\ \ref{Multiscale Shu-Osher}.

\subsubsection{High Mach Astrophysical jet}

\begin{figure*}
    \centering
  \includegraphics[width=0.4\textwidth]{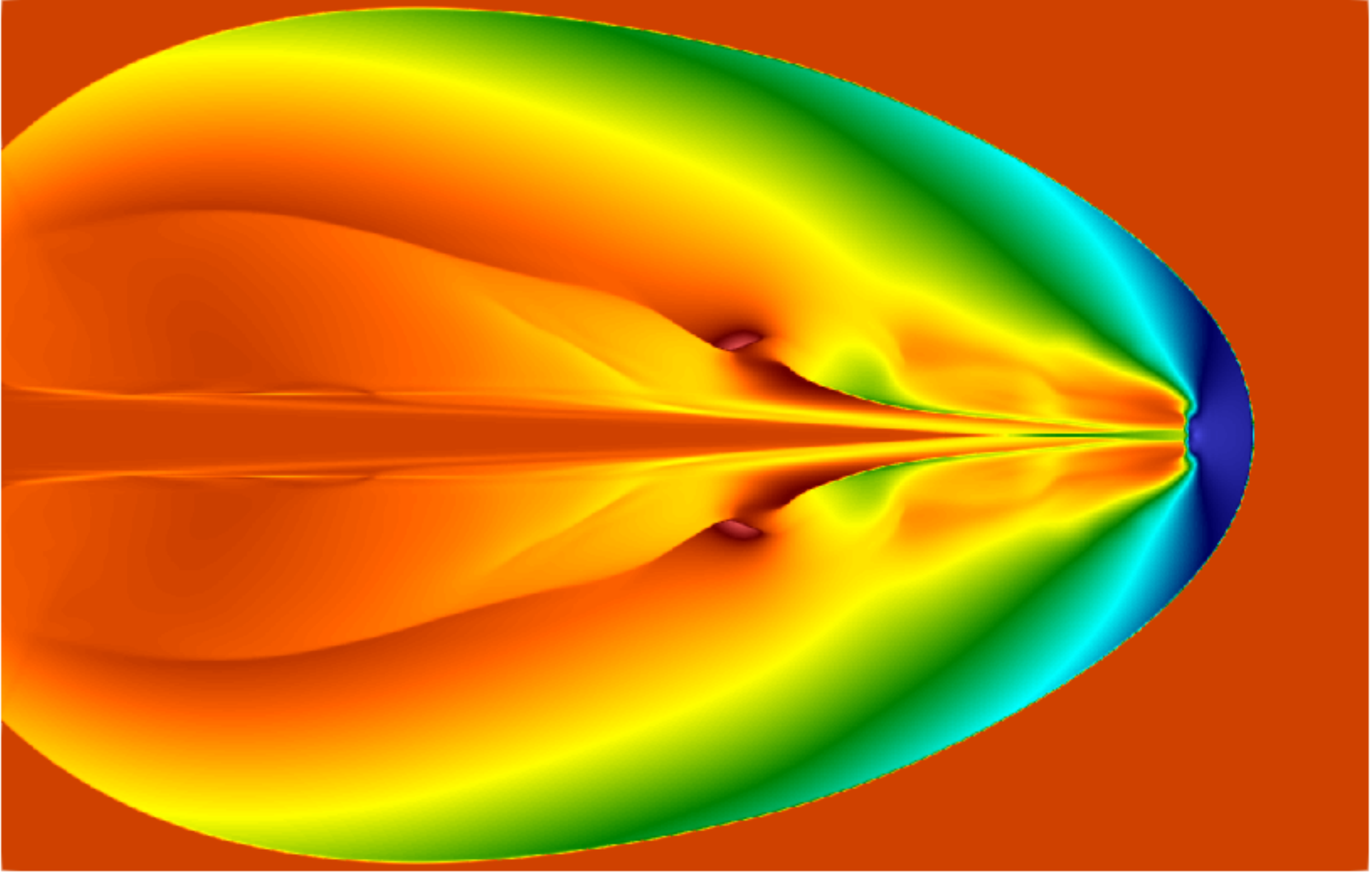}
   \includegraphics[width=0.4\textwidth]{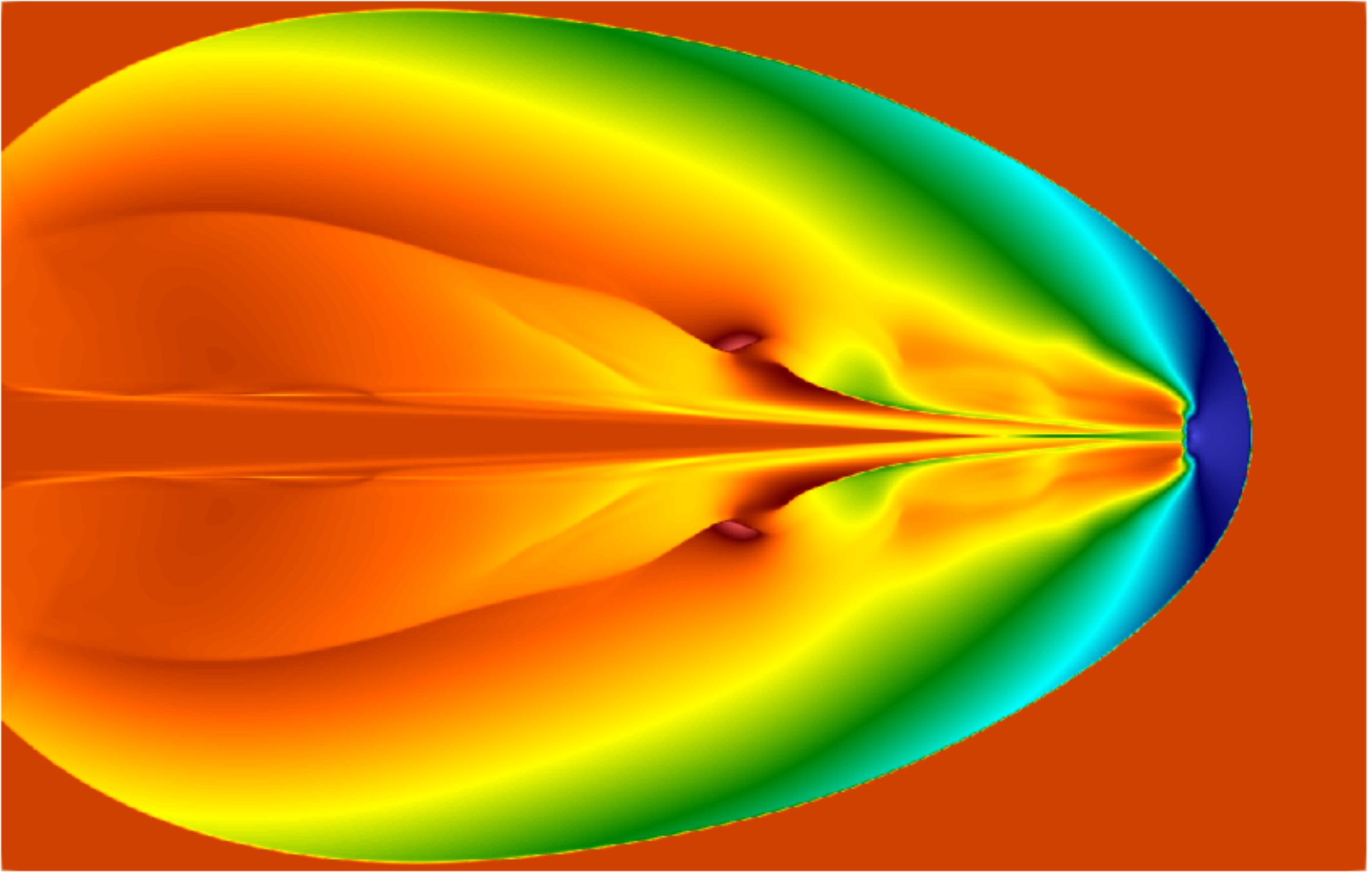}\\
    \includegraphics[width=0.4\textwidth]{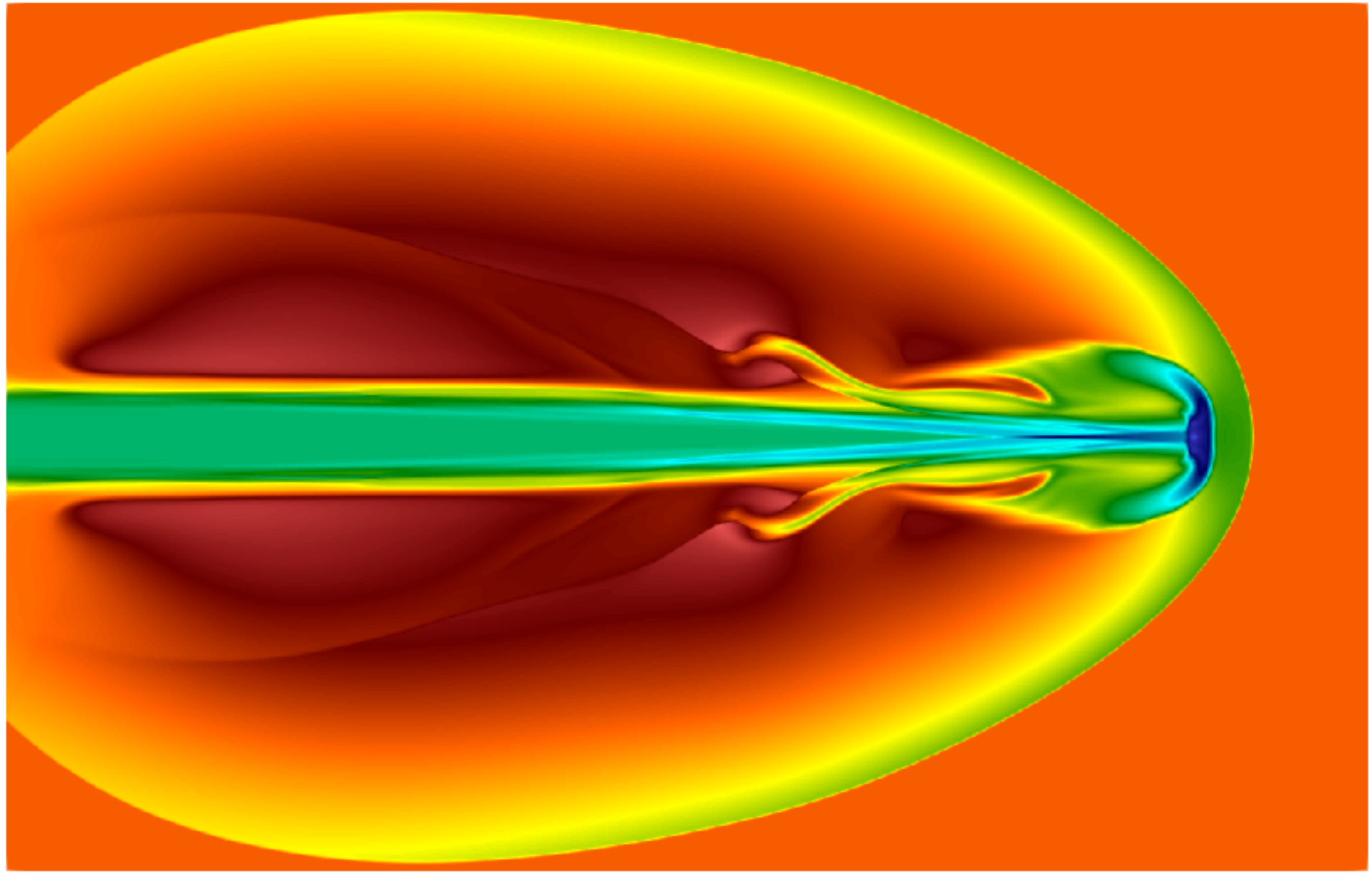}
     \includegraphics[width=0.4\textwidth]{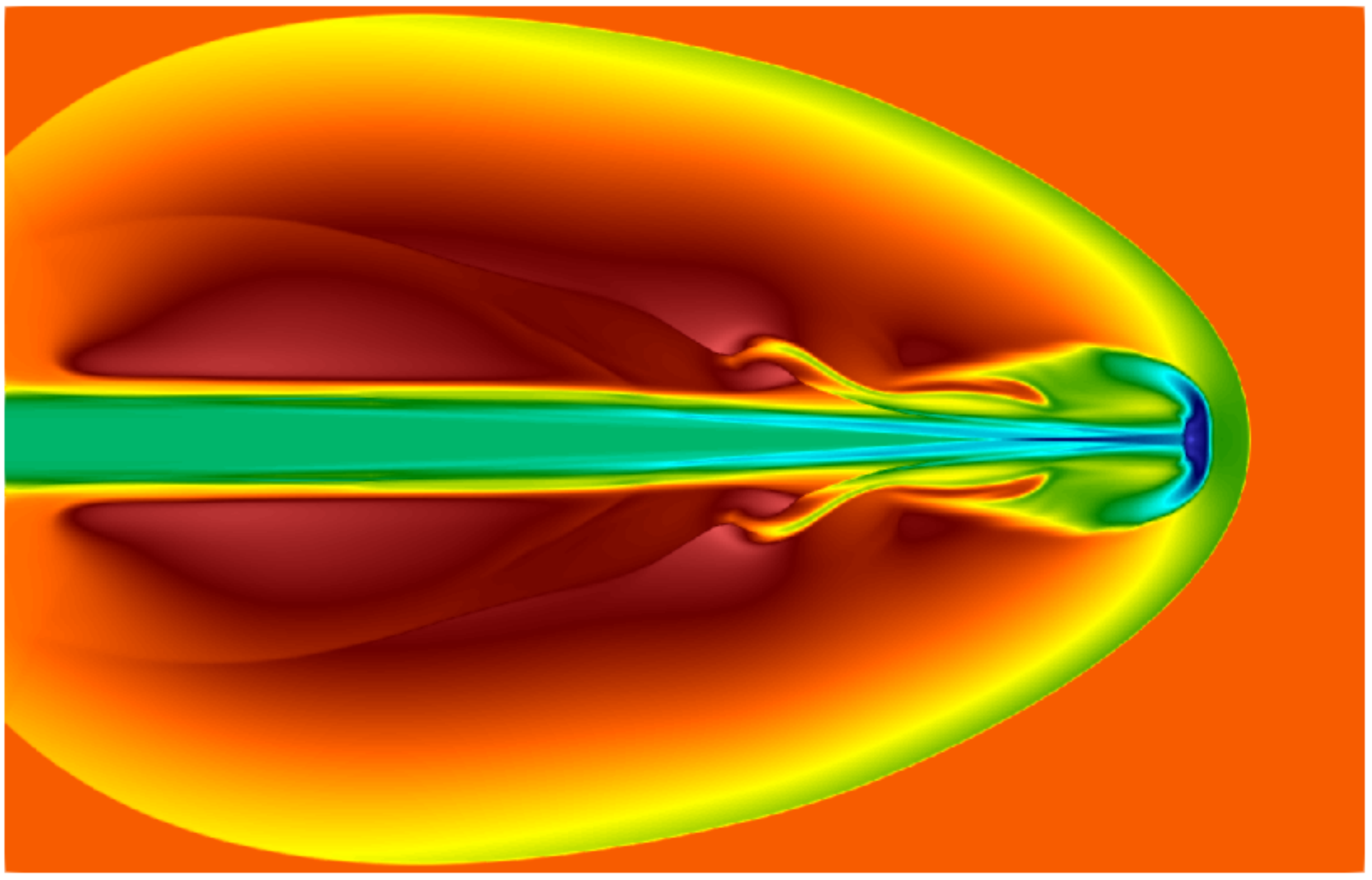}\\
    \includegraphics[width=0.4\textwidth]{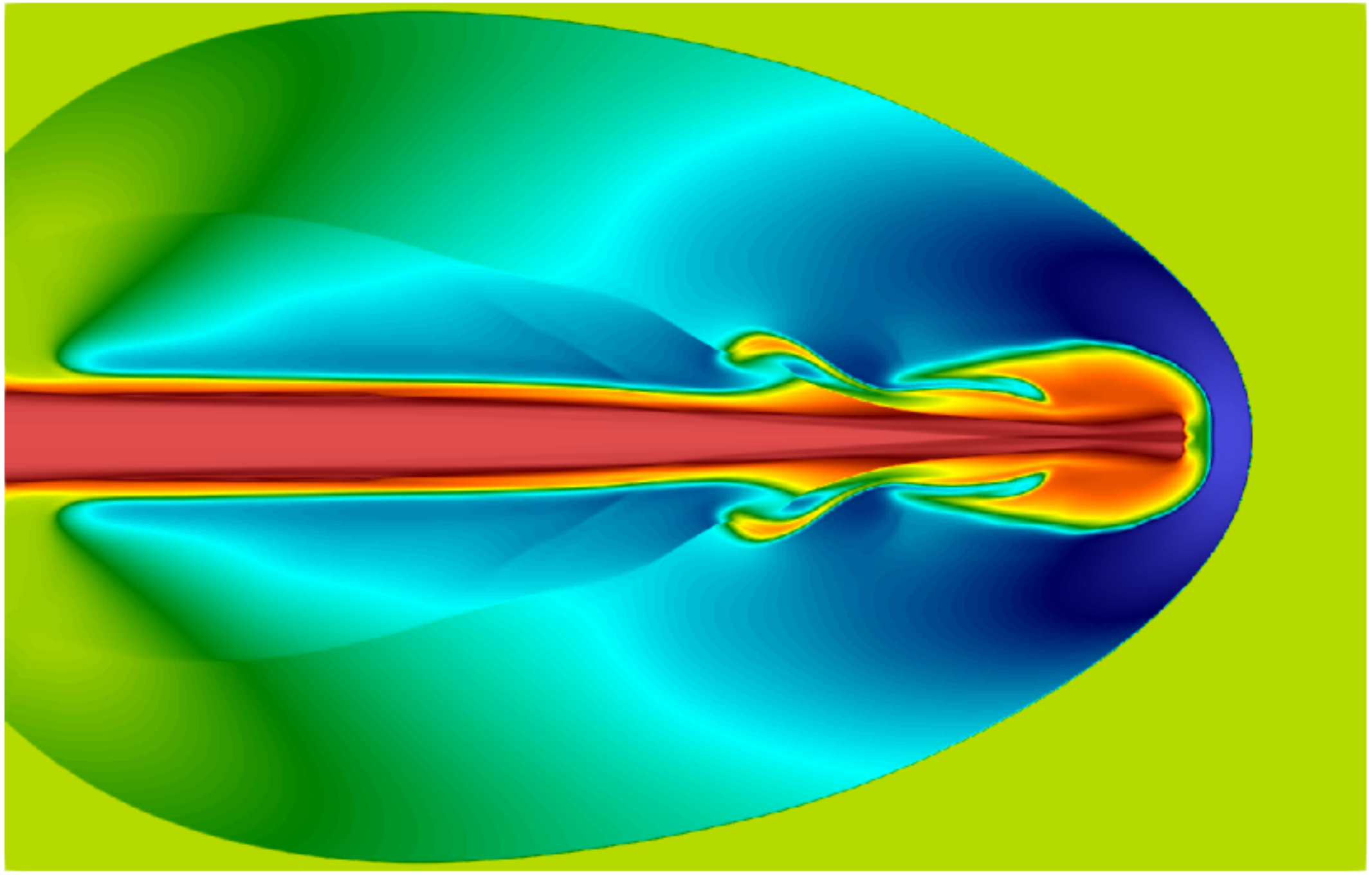} 
    \includegraphics[width=0.4\textwidth]{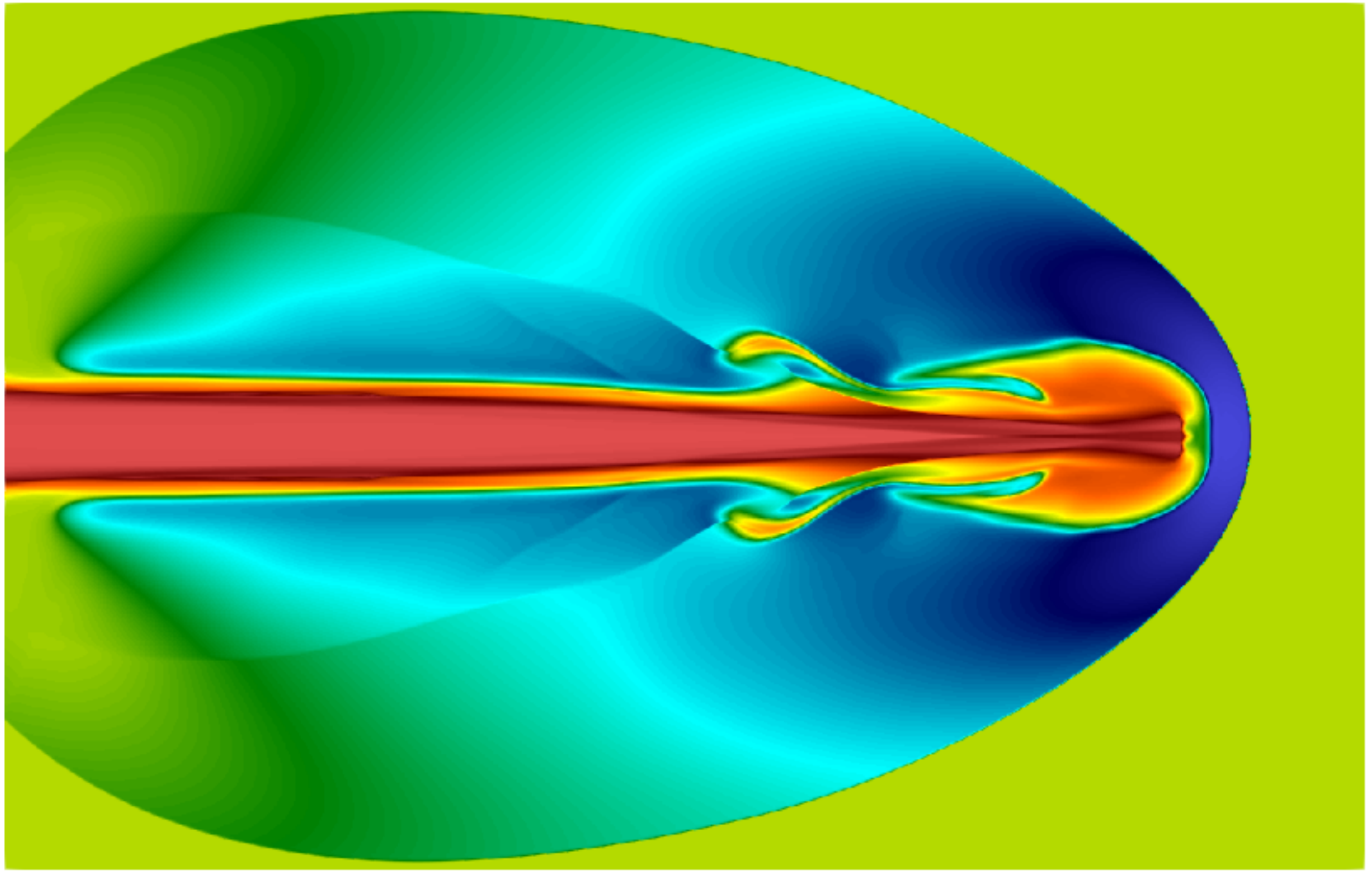}
    \caption{Mach 30 astrophysical jet problem. Top: pressure, middle: density, bottom: temperature. Left column: $D2Q16$. Right column: $D2Q9/D2Q16$.}
    \label{Jet_FieldsComparison}
\end{figure*}

\begin{figure}
    \centering
   \includegraphics[width=0.45\textwidth]{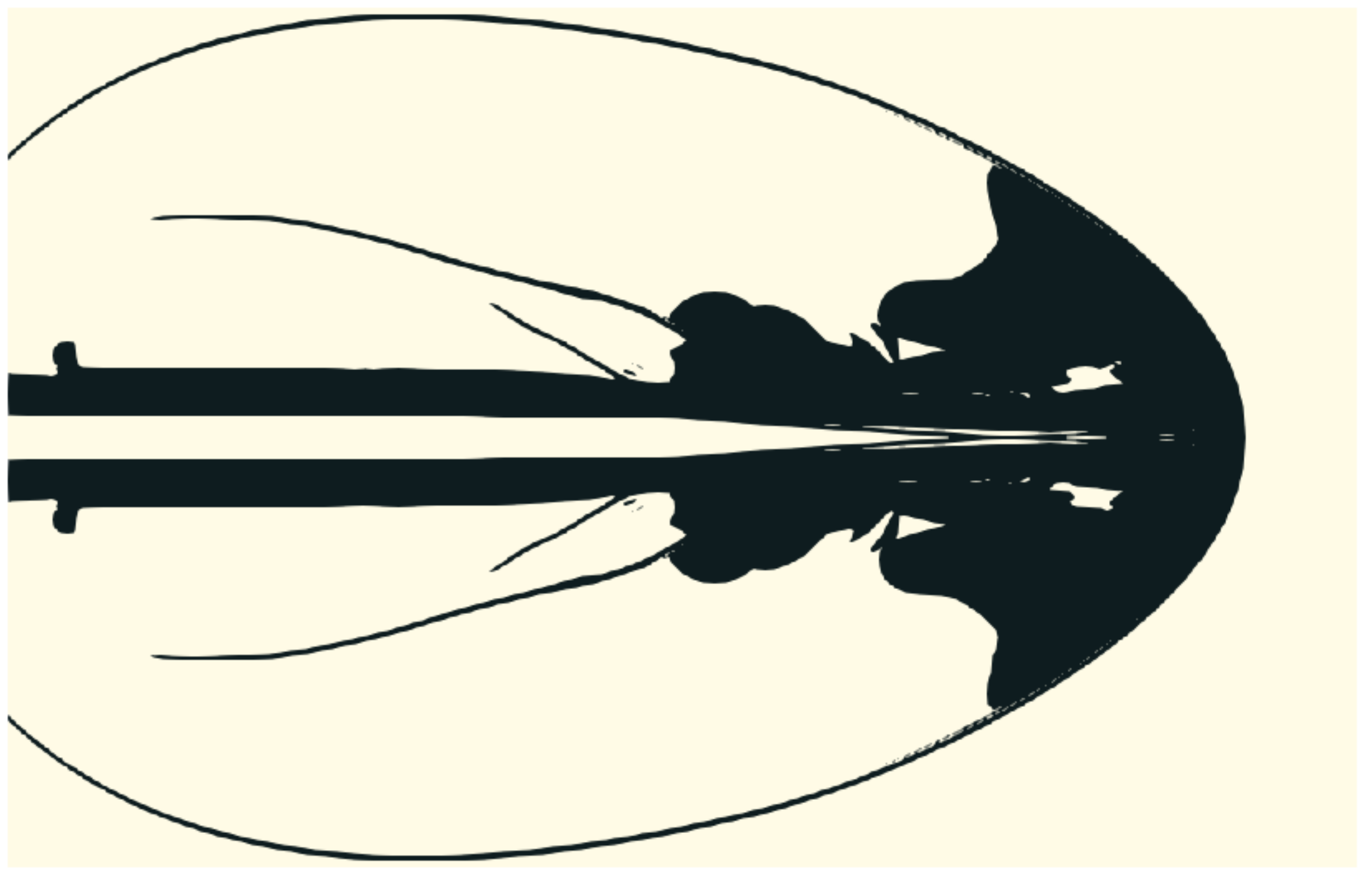}
    \caption{Mach 30 astrophysical jet problem. Lattice distribution for the $D2Q9/D2Q16$ model. Black regions indicate the $D2Q16$ lattice. }
    \label{Jet_Lattice}
\end{figure}

\begin{figure}
    \centering
   \includegraphics[width=0.45\textwidth]{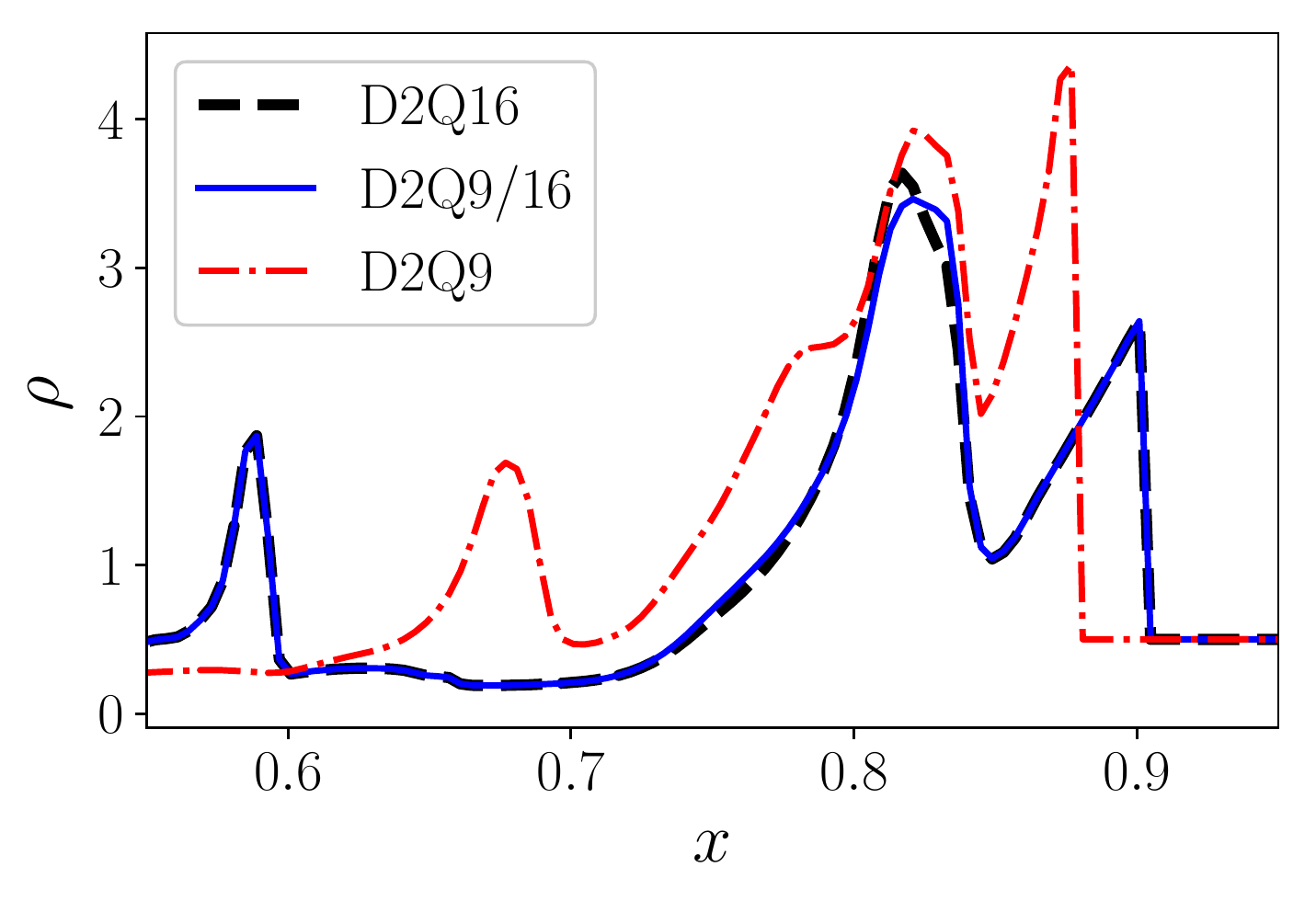} \\ 
    \includegraphics[width=0.45\textwidth]{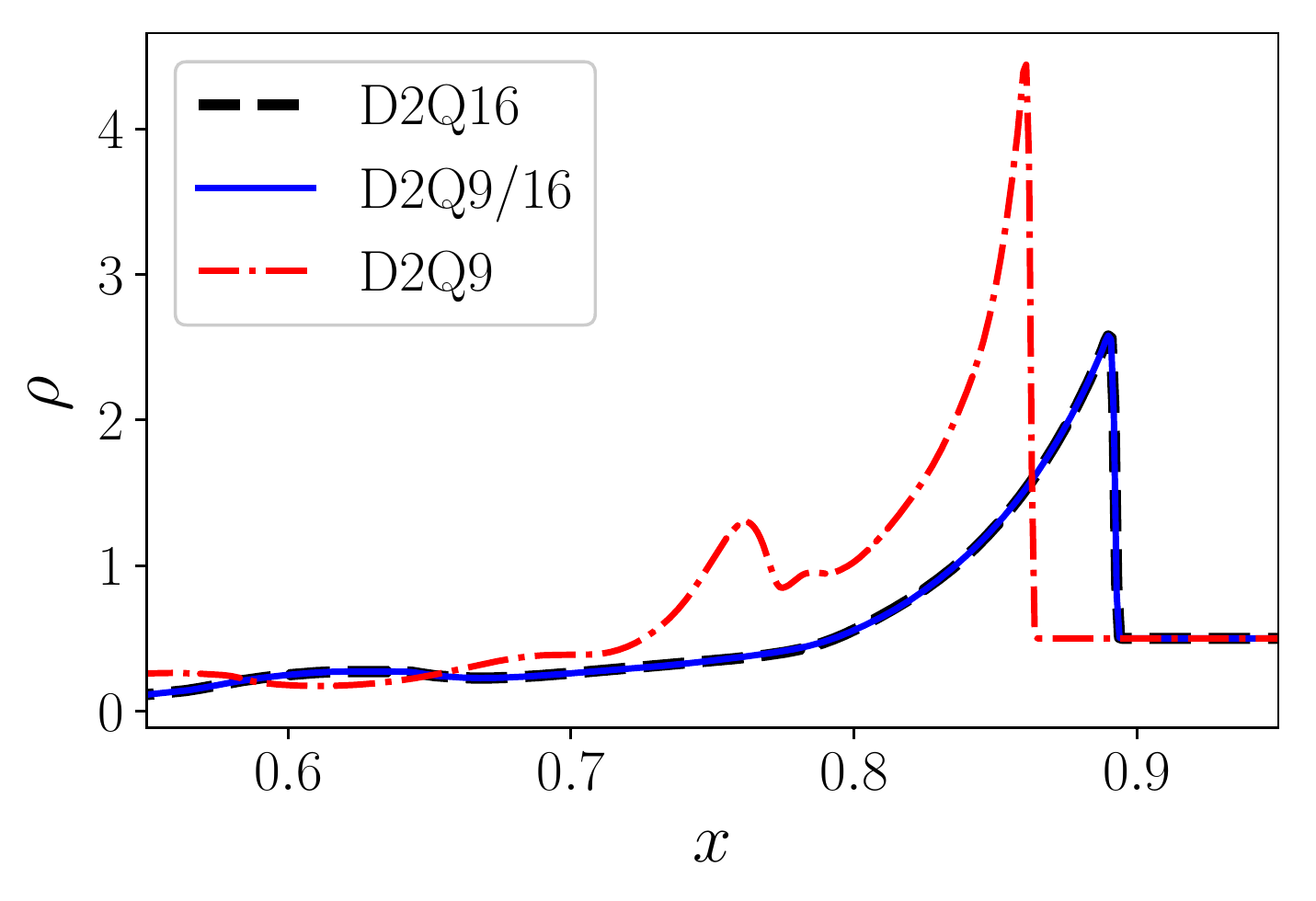} \\
    \includegraphics[width=0.45\textwidth]{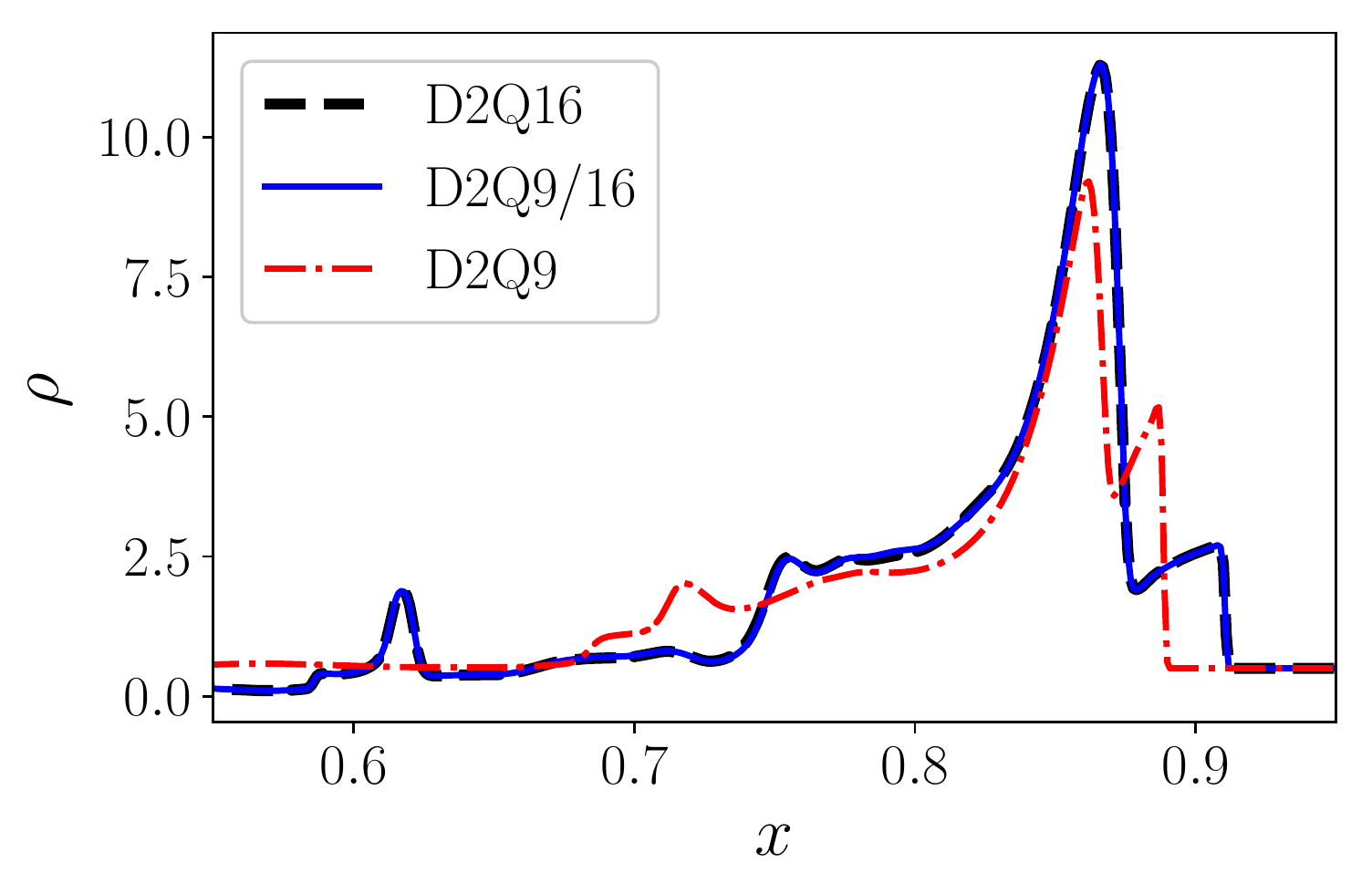} \\     
    \caption{Mach 30 astrophysical jet problem. Density profiles of the $D2Q16$ (black dashed line) , $D2Q9$ (red dashed dotted line) and $D2Q9/D2Q16$ (blue solid line) models, across horizontal cuts of the domain. The horizontal cuts intercept the $y$ axis at $y/L_y=0.66$ (top), $y/L_y=0.7$ (middle) and $y/L_y=0.73$ (bottom).}
    \label{Jet_Comparison}
\end{figure}

As a final test case, we consider an astrophysical jet of Mach 30, without radiative cooling \cite{ZhangShu2010}. This case is an example of actual gas flows revealed from images of the Hubble Space Telescope and therefore is of high scientific interest. Following the configuration in \cite{ZhangShu2010}, we initialize the computational domain $[0,2]\times [-0.5,0.5]$ with the following conditions,
 \begin{equation}
 \begin{split}
    & (\rho,u_x,u_y,p) \\& =\begin{cases}
     (5, 11.2, 0, 0.4127),& \text{if } x=0, \ -0.05\leq y \leq 0.05, \\
     (0.5, 0, 0, 0.4127),& \text{ otherwise}. \\
     \end{cases}
      \end{split}
 \end{equation}
Outflow BCs are used around the domain, except the left boundary, where the prescribed fixed conditions are imposed. The simulation was performed with resolution $[1200,600]$. We compare the results between the $D2Q9$, $D2Q16$ and the multi-scale $D2Q9/D2Q16$ models. Fig.\ \ref{Jet_FieldsComparison} shows a comparison of the pressure, density and temperature fields between the $D2Q16$ and $D2Q9/D2Q16$ solution. The propagation of the bow shock into the surrounding medium, as well as the developed Rayleigh Taylor instabilities within the jet cocoon, are captured in very good agreement between the two simulations. Fig.\ \ref{Jet_Lattice} depicts the distribution of the $D2Q16$ lattice in the computational domain. For a quantitative comparison, Fig.\ \ref{Jet_Comparison} plots the density field across three horizontal cuts of the domain. The multi-scale $D2Q9/D2Q16$ model and the pure $D2Q16$ are in excellent agreement. On the contrary, the pure $D2Q9$ model evolves with clear deviations, as shown in  Fig.\ \ref{Jet_Comparison}.

\subsection{Discussion}
\label{subsec::ResultsDiscussion}

We summarize the main strategies that we adopted to increase the efficiency of the PonD method, with minimal sacrifice of accuracy. The pivotal point is the identification of the frame invariant moment system for the $f$- and $g$- populations, according to the target hydrodynamic system. According to the analysis of Sec.\ \ref{subsec::HydroLimit}, the frame invariant moment system for the $g$- populations should include up to second order moments. Hence, irrespective of the $f$- lattice, we used in all simulations in this work the $D2Q9$ lattice for the $g$-populations, decreasing the computational cost for both the $g-$ populations update and the $g$- reference frame transformations. Numerical experiments with different $f$- and $g$- lattices did not reveal any appreciable effect on  the stability and the accuracy of the scheme.

The multiscale formulation enables the deployment of a low-order lattice for the $f$- populations, in regions with smooth flow velocity and temperature variations. In accordance with observations in \cite{Multiscale2021}, the stability and accuracy of the solutions are well-maintained. The efficiency gains from this approach are naturally case dependent. We note that the different lattices communicate solely through the Grad's reference frame transformation, which renders the transition from a single lattice to a multiscale model easy to program and highly efficient.

\begin{figure*}
    \centering
   \includegraphics[width=0.32\textwidth]{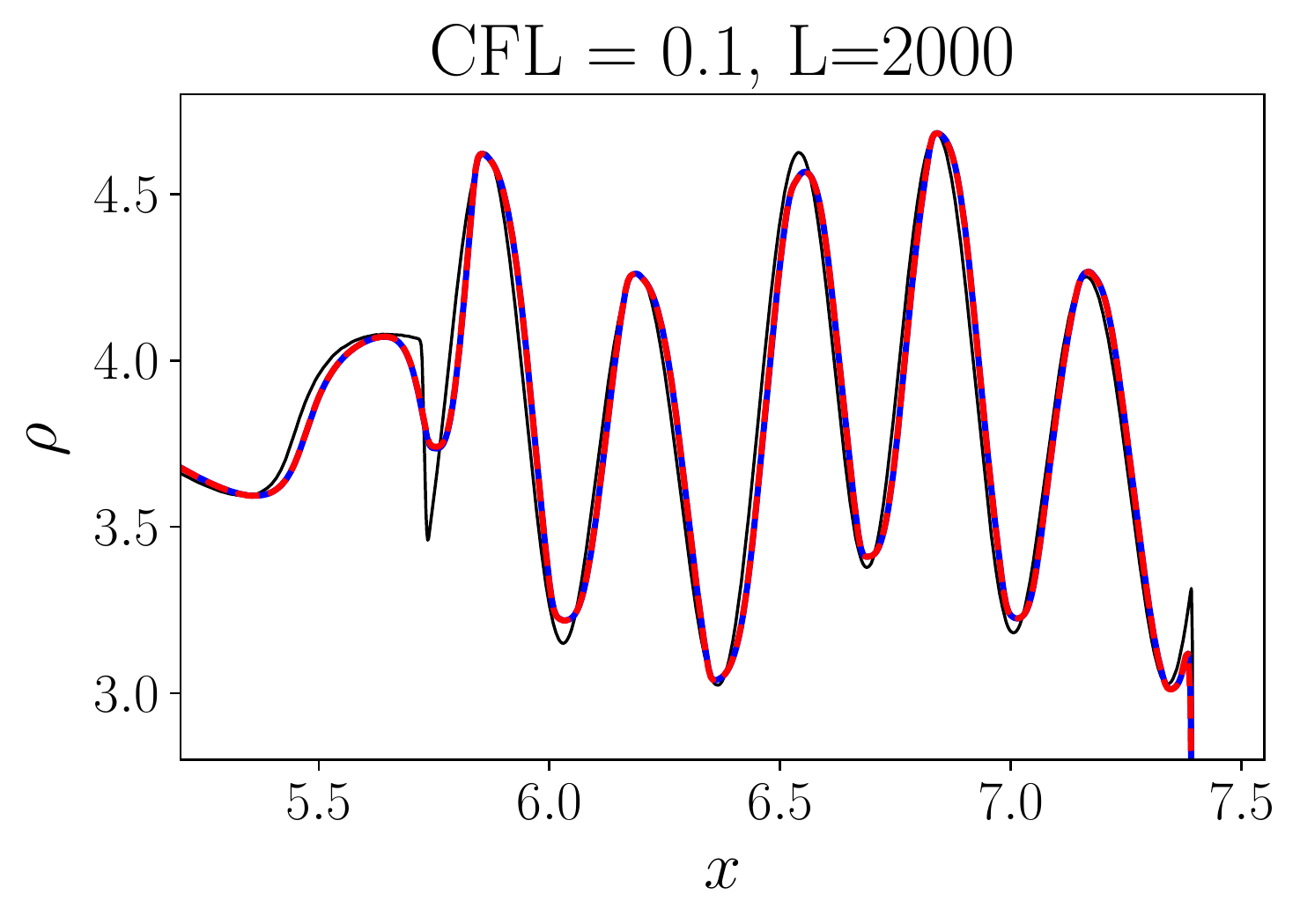}
 \includegraphics[width=0.32\textwidth]{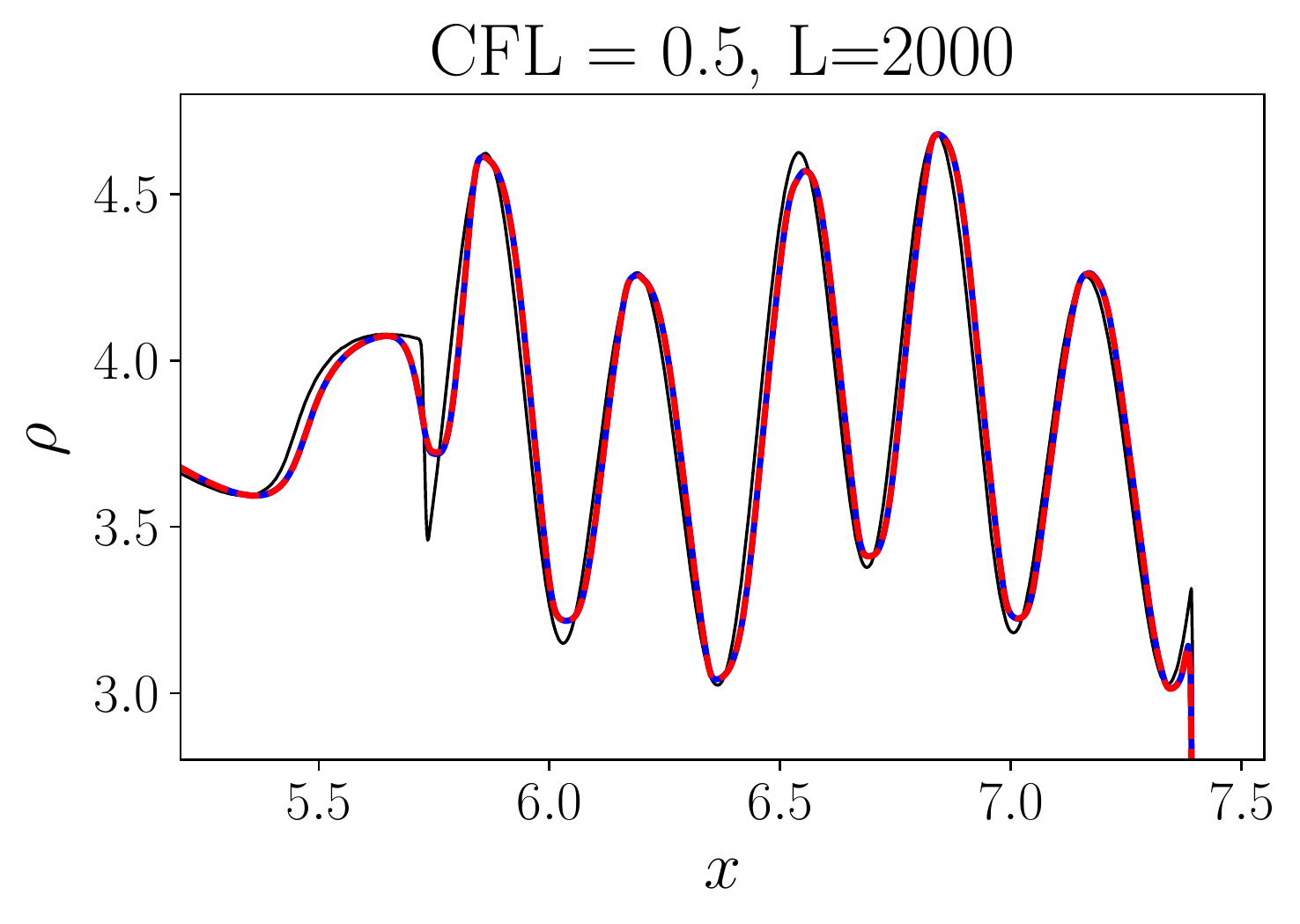}
  \includegraphics[width=0.32\textwidth]{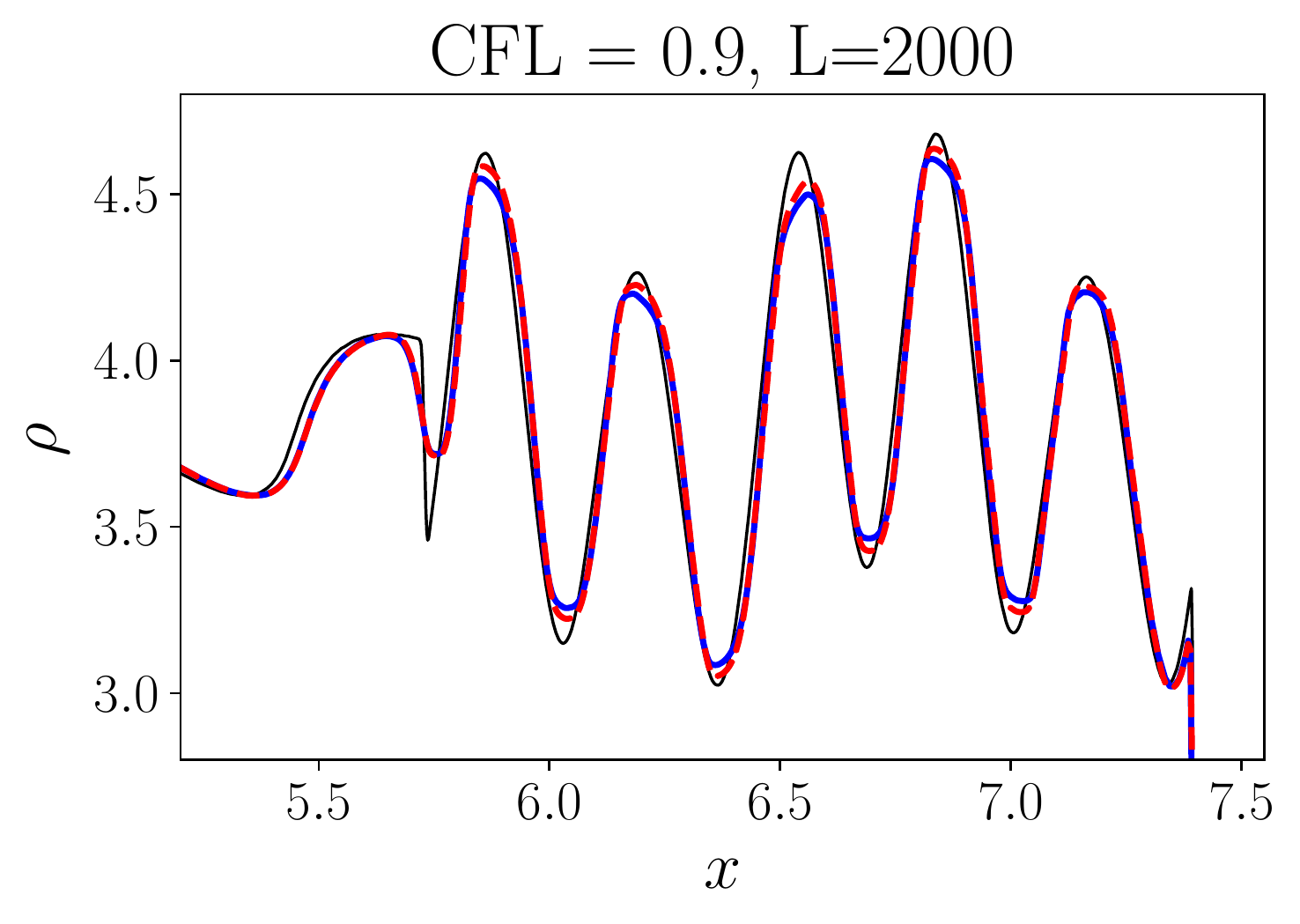}
\includegraphics[width=0.32\textwidth]{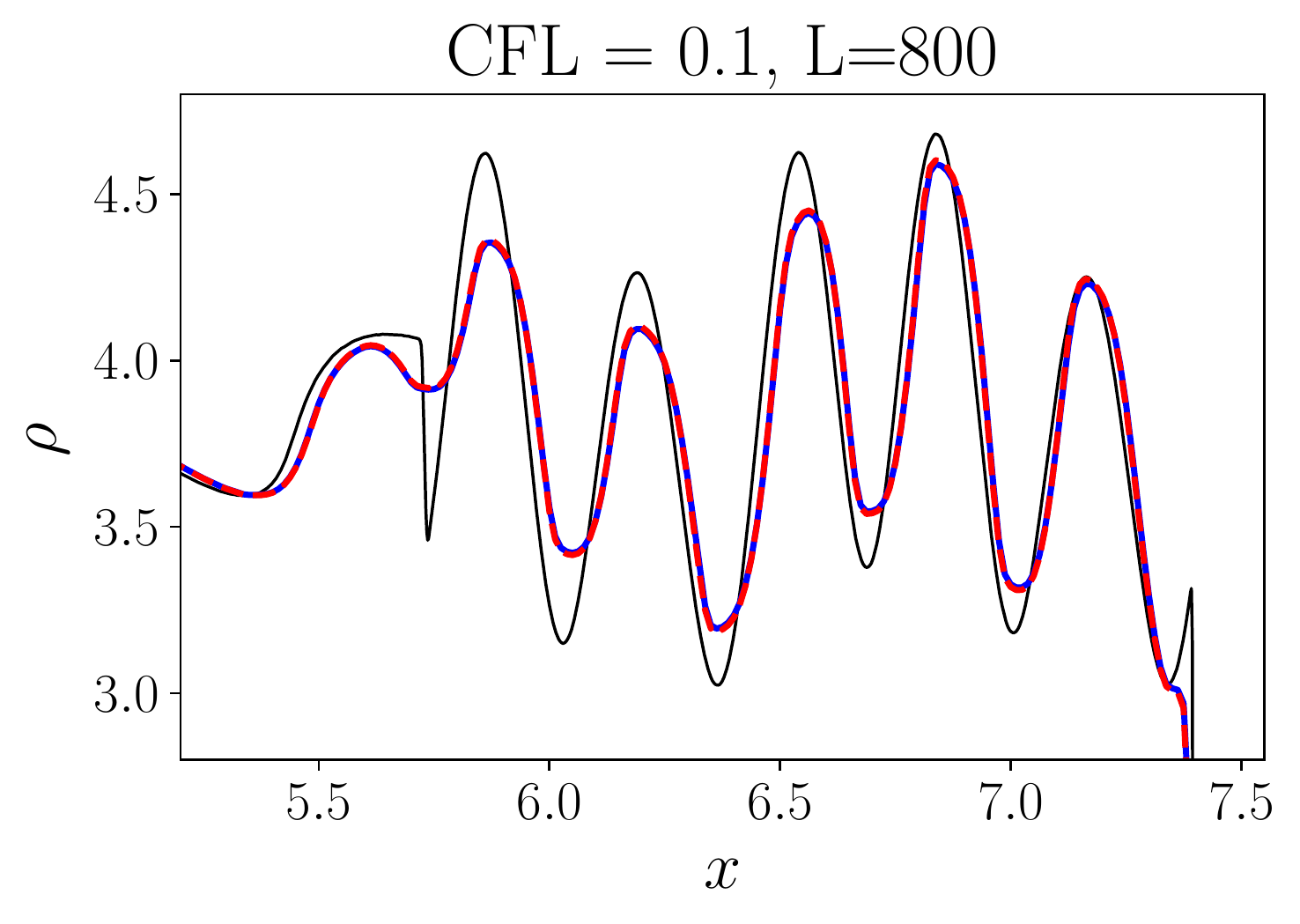}
 \includegraphics[width=0.32\textwidth]{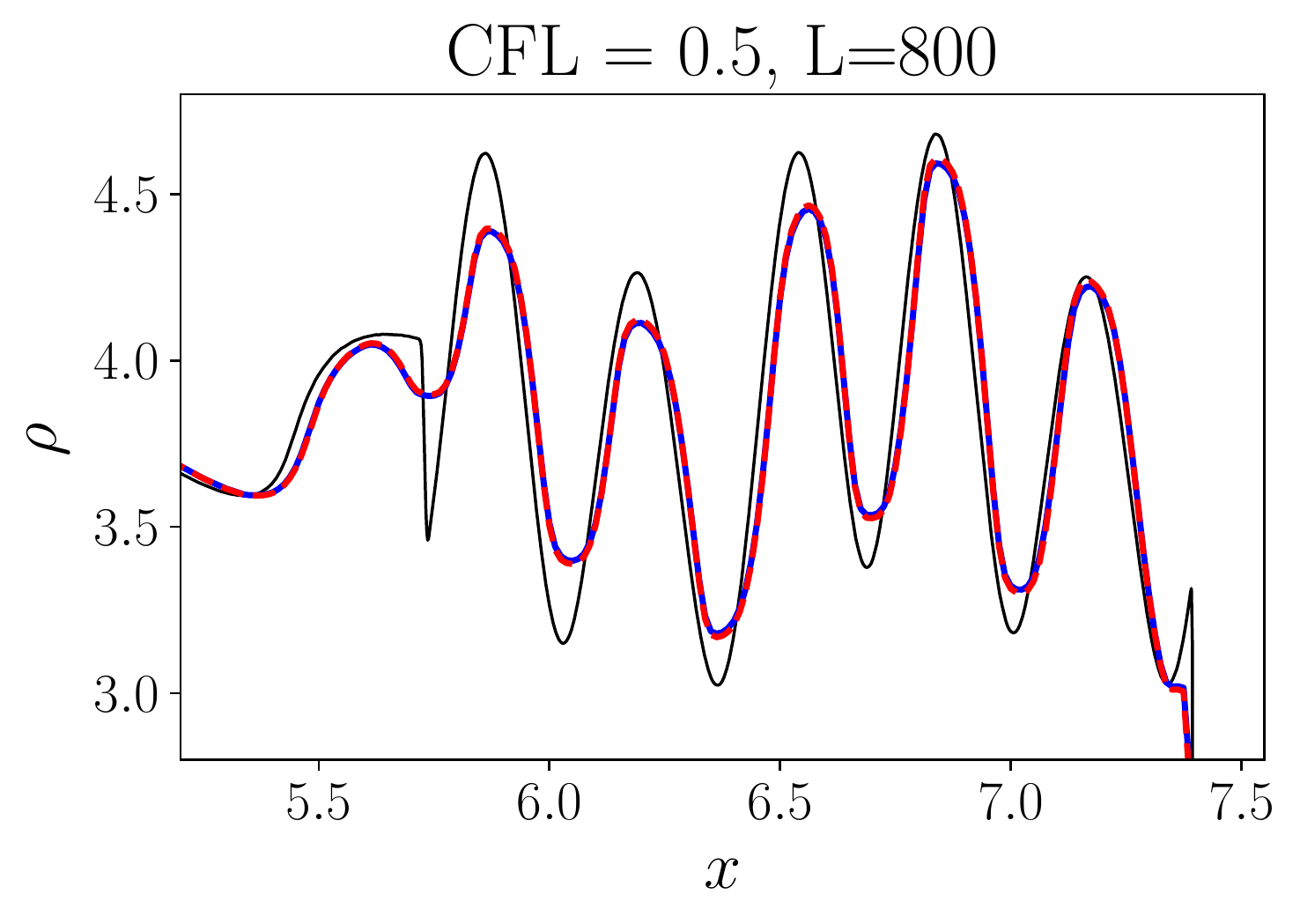}
\includegraphics[width=0.32\textwidth]{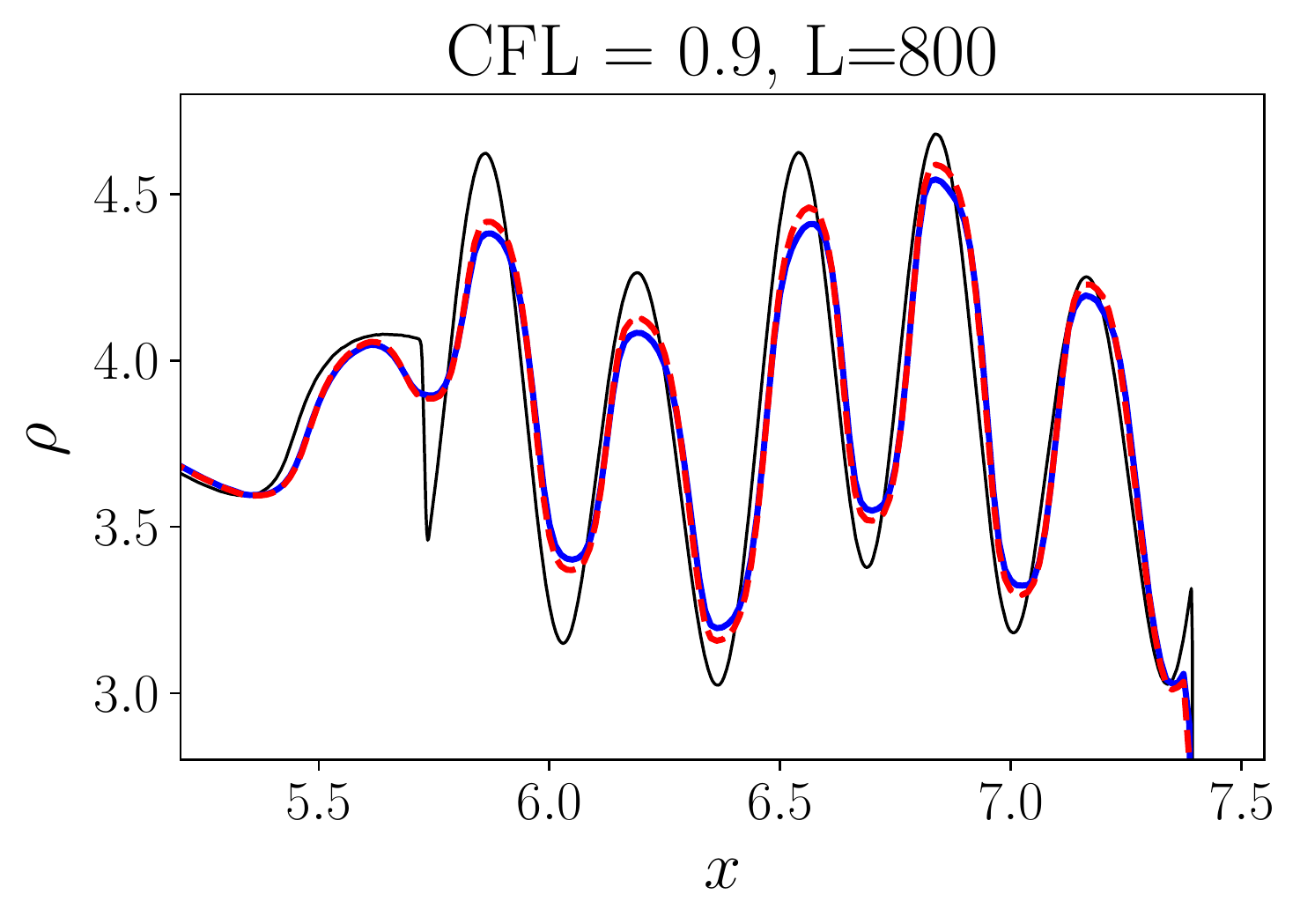}
  \includegraphics[width=0.32\textwidth]{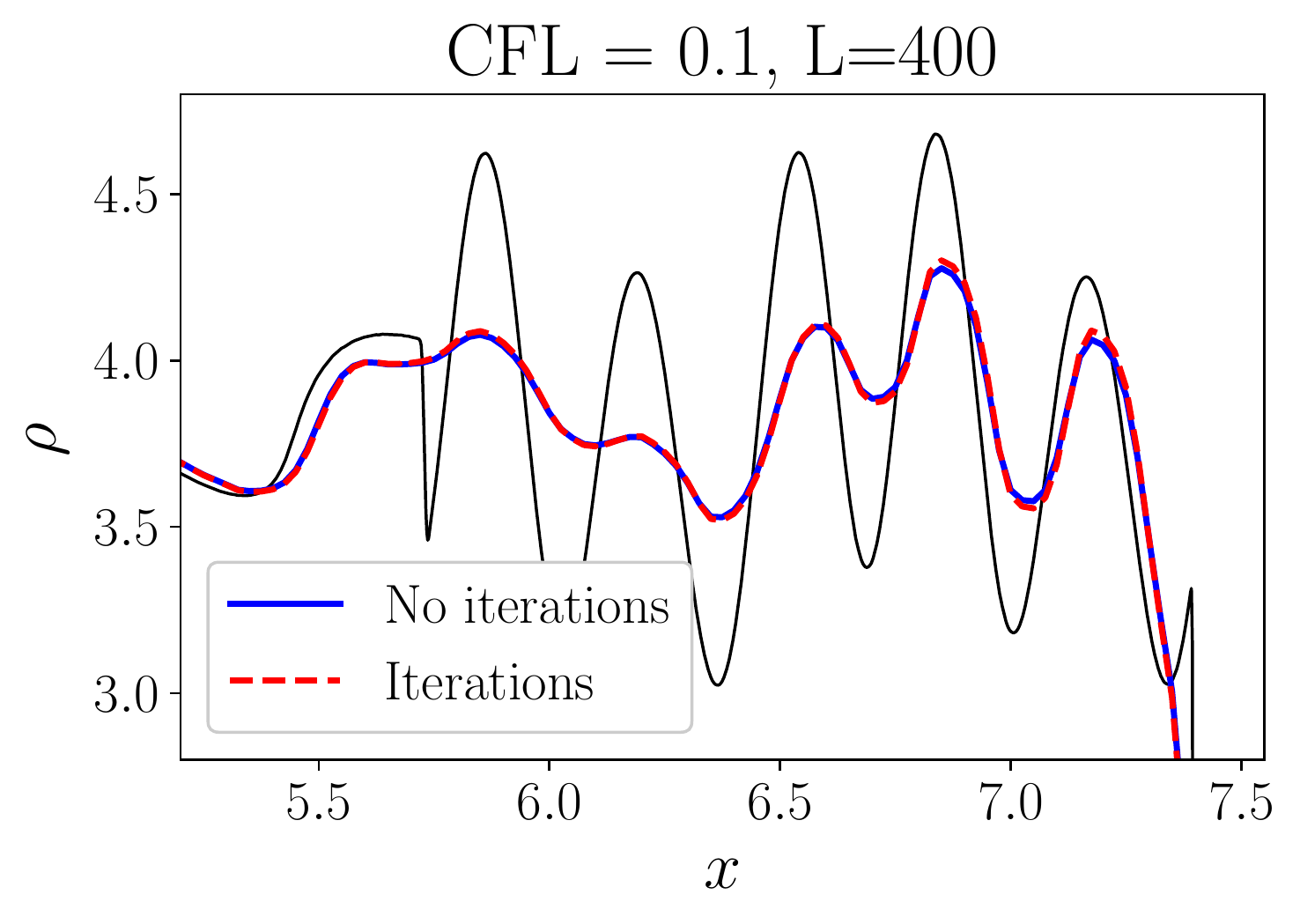}
 \includegraphics[width=0.32\textwidth]{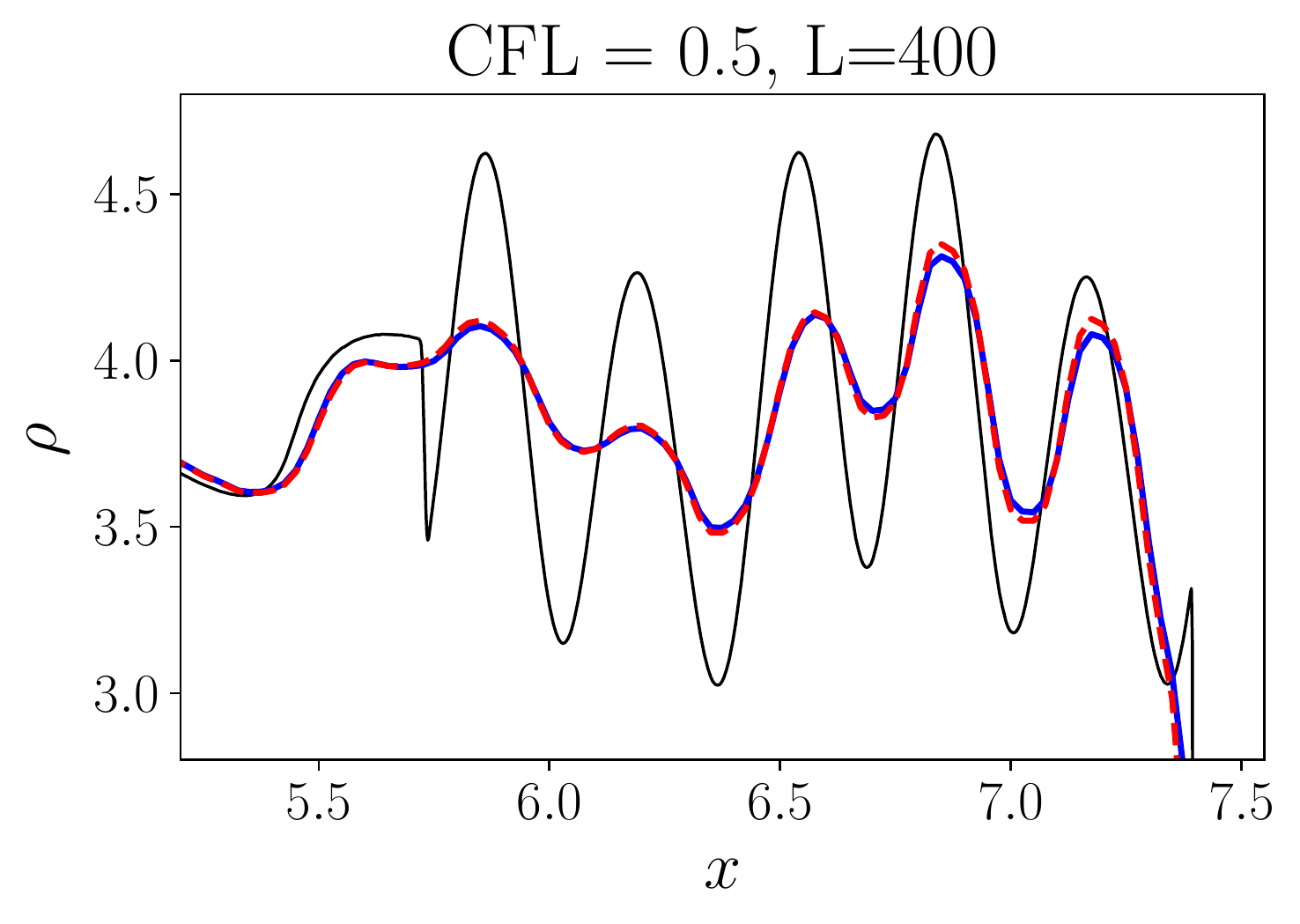}
  \includegraphics[width=0.32\textwidth]{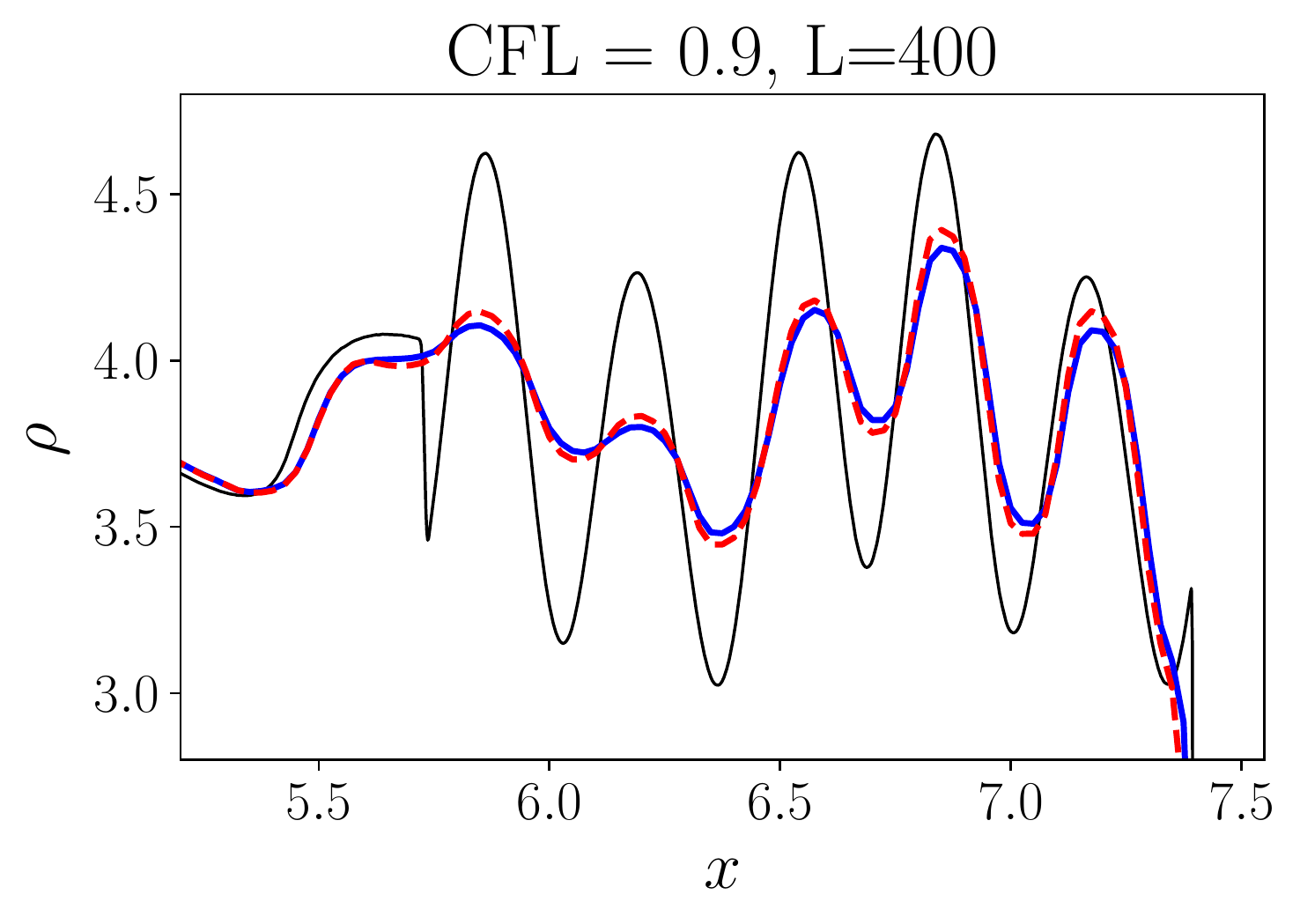}
    \caption{Density profile for the Shu-Osher problem, comparing the scheme with (red dashed line) and without (blue solid line) iterations. Black solid line corresponds to the reference results \cite{ShuOsherReference}. Results are shown for different CFL numbers and resolution.}
    \label{Iterations_Shu-Osher}
\end{figure*}

The last element which differs from the PonD formulations in \cite{PonD_DUGKS} is the absence of iterations within the flux calculation, as discussed in \ref{subsubsec::Summary}. We demonstrate a comparison between the iterative and the current formulation through the Shu-Osher problem \cite{ShuOsherReference}. Fig.\ \ref{Iterations_Shu-Osher} shows the results from the two schemes, for different $\rm{CFL}$ numbers and resolutions. One observes that for high $\rm{CFL}$ and coarse domains, the iterative scheme is marginally more accurate than its explicit counterpart. For moderate $\rm{CFL}$ and resolved domains the two solutions are almost indistinguishable. Additional numerical experiments confirm the above observations. We can conclude that for resolved simulations (spatially and temporally), the non-iterative flux calculation can be safely employed.

\section{Conclusions}
\label{sec::Conclusions}

In this work, we presented the PonD formulation with an emphasis on the requirements of the reference frame transformation. According to the target hydrodynamic equations, conventional LBM models on a static reference frame require a set of equilibrium moment constraints. In constrast, PonD utilizes an adaptive comoving reference frame with the pertinent equilibrium moment constraints being automatically satisfied by exact equilibrium populations. However, the target hydrodynamic equations introduce requirements on the frame invariant moment system of the reference frame transformation. The framework presented on this work is a finite volume discretization of the governing kinetic equations in an adaptive reference frame. In comparison with conventional finite volume LBMs (such as conventional DUGKS), the cost to be paid for the adaptive formulation amounts to the reference frame transformations. The benefit of this approach is enhanced accuracy, stability and an increased operating window in terms of Mach number and temperature. Additionally, a multiscale extension can easily be incorporated and results in further efficiency gains. Further high Mach $3D$ simulations with the presence of curved boundaries shall be the focus of future work.

\begin{acknowledgments}
This work was supported by European Research Council (ERC) Advanced Grant  834763-PonD. 
Computational resources at the Swiss National  Super  Computing  Center  CSCS  were  provided  under the grant  s1066.
\end{acknowledgments}

\appendix

\section{Hermite polynomials}
\label{HermiteAppendix}

The Hermite polynomials, up to fourth order and with discrete velocities scaled such that $T_L=1$, are the following,

\begin{align}
    H_i^{(0)} &= 1, \\
    H_{i \alpha}^{(1)} &= c_{i \alpha}, \\
    H_{i \alpha \beta}^{(2)} &= c_{i \alpha}c_{i \beta}-\delta_{\alpha \beta}, \\
    H_{i \alpha \beta \gamma}^{(3)} &= c_{i \alpha}c_{i \beta}c_{i \gamma}-[c_{i \alpha} \delta_{\beta \gamma}]_{\rm cyc}, \\    
    H_{i \alpha \beta \gamma \delta}^{(4)} &= c_{i \alpha}c_{i \beta}c_{i \gamma}c_{i \delta}-[c_{i \alpha}c_{i \beta}\delta_{\gamma \delta}]_{\rm cyc}+[\delta_{\alpha \beta}\delta_{\gamma \delta}]_{\rm cyc},     
\end{align}
where $[]_{\rm cyc}$ stands for cyclic permutations without repetition over indices. The contracted fourth order polynomial is the following,

\begin{equation}
\begin{split}
    H_{i,\alpha \beta}^{(4)}=c_{i \alpha}c_{i \beta}(c_i^2-(D+4))- \delta_{\alpha \beta}(c_i^2-(D+2)).
    \end{split}
\end{equation}

Calculation of expansion coefficients, for given moments $\bm{m}=\{ M^{(0)}, M_{\alpha}^{(1)}, M_{\alpha \beta}^{(2)}, M_{\alpha \beta \gamma}^{(3)},  M_{\alpha \beta}^{(4)}  \}$ and target reference frame $\lambda=\{T, \bm{u} \}$. The population in the target reference frame is expanded in Grad series,

\begin{equation}
    f_i^{\lambda}=W_i \sum_{n=0}^{4}\frac{1}{n!}\bm{\alpha}^{(n)}(\bm{m};\lambda)H^{(n)}(\bm{c}_i),
\end{equation}

The constraints which enforce the reference frame invariance of the selected moments are:

\begin{align}
    M^{(0)} &=\sum_{i=0}^{Q-1}  f_i^{\lambda}, \\
    M^{(1)}_{\alpha} &=\sum_{i=0}^{Q-1}  f_i^{\lambda}v_{i \alpha}^{\lambda}, \\
    M^{(2)}_{\alpha \beta} &=\sum_{i=0}^{Q-1}  f_i^{\lambda}v_{i \alpha}^{\lambda}v_{i \beta}^{\lambda}, \\
    M^{(3)}_{\alpha \beta \gamma} &=\sum_{i=0}^{Q-1}  f_i^{\lambda}v_{i \alpha}^{\lambda}v_{i \beta}^{\lambda}v_{i \gamma}^{\lambda}, \\   M^{(4)}_{\alpha \beta } &=\sum_{i=0}^{Q-1}  f_i^{\lambda}v_{i \alpha}^{\lambda}v_{i \beta}^{\lambda}v_{i \gamma}^{\lambda}v_{i \gamma}^{\lambda}, 
\end{align}
where the discrete velocities are $v_{i \alpha}^{\lambda}=(\sqrt{T} c_{i \alpha}+u_{\alpha})$. The solution for the expansion coefficients is below,

\begin{widetext}
\begin{align}
   \alpha^{(0)} &= M^{(0)}, \\ 
   \alpha^{(1)}_{\alpha} &= \frac{1}{T^{1/2}}\left(M^{(1)}_{\alpha} -u_\alpha M^{(0)} \right) , \\ 
   \alpha^{(2)}_{\alpha \beta} &= \frac{1}{T}\left(M^{(2)}_{\alpha \beta} -M^{(0)}(T\delta_{\alpha \beta} -u_\alpha u_\beta M^{(0)}) -[u_{\alpha}(M_{\beta}^{(1)}-u_\beta M^{(0)} )]_{\rm cyc} ) 
 \right) , \\    
 \alpha^{(3)}_{\alpha \beta \gamma} &= \frac{1}{T^{3/2}}\left(M^{(3)}_{\alpha \beta \gamma} -u_\alpha u_\beta u_{\gamma} M^{(0)} -[ (M_{\alpha}^{(1)}-M^{(0)}u_\alpha)(T \delta_{\beta \gamma} -u_\beta u_{\gamma}) ]_{\rm cyc}-[ (M_{\alpha \beta}^{(2)}-M^{(0)}u_\alpha u_\beta)u_\gamma ]_{\rm cyc} ) 
 \right) , 
\end{align}
\end{widetext}

\section{PonD equation}

We start with the kinetic equation formulated at a constant, uniform monitoring reference frame $\bar \lambda$,

\begin{equation}
\label{StartingEquationPonD}
    \partial_tf_i^{\overline{\lambda}}+\bm{v}_{i }^{\overline{\lambda}}  \cdot \nabla f_i^{\overline{\lambda}}=\Omega_{i}^{\overline{\lambda}},
\end{equation}
The moments $\bm{m}^{\bar \lambda}$ at the monitoring frame $\bar \lambda$ are connected with the corresponding populations via a linear matrix $\mathcal{M}^{\bar \lambda}$,

\begin{align}
    m^{\bar \lambda}_i=\mathcal{M}^{\bar \lambda}_{i,j}f^{\overline{\lambda}}_j, \\
    \label{poptomom}
    f^{\bar \lambda}_i=[\mathcal{M}^{\bar \lambda}]^{-1}_{i,j}m^{\overline{\lambda}}_j,    
\end{align}

We insert Eq.\ \eqref{poptomom} into Eq.\ \eqref{StartingEquationPonD} and obtain the following equation,

\begin{equation}
\label{StartingEquationPonD_2}
\begin{split}
    \partial_t([\mathcal{M}^{\bar \lambda}]^{-1}_{i,j}m^{\overline{\lambda}}_j)+\bm{v}_{i }^{\overline{\lambda}}  \cdot \nabla ([\mathcal{M}^{\bar \lambda}]^{-1}_{i,j}m^{\overline{\lambda}}_j)= \\ [\mathcal{M}^{\bar \lambda}]^{-1}_{i,j}m^{\overline{\lambda}}_{\Omega,j},
    \end{split}
\end{equation}
where $m^{\overline{\lambda}}_{\Omega,j}$ denotes the moments from the collision operator. Next, we invoke the reference frame invariance of the moments,
\begin{equation}
\label{FrameInvarianceMoments}
    m^{\bar \lambda}_i=m^{\lambda(\bm{x},t)}_i,
\end{equation}
where $\lambda(\bm{x},t)$ denotes the local reference frame. We insert the moments evaluated from the local reference frame \eqref{FrameInvarianceMoments} into Eq.\ \eqref{StartingEquationPonD_2},

\begin{equation}
\label{StartingEquationPonD_3}
\begin{split}
    \partial_t([\mathcal{M}^{\bar \lambda}]^{-1}_{i,j}m^{\lambda(\bm{x},t)}_j)+\bm{v}_{i }^{\overline{\lambda}}  \cdot \nabla ([\mathcal{M}^{\bar \lambda}]^{-1}_{i,j}m^{\lambda(\bm{x},t)}_j)= \\ [\mathcal{M}^{\bar \lambda}]^{-1}_{i,j}m^{\lambda(\bm{x},t)}_{\Omega,j}.
\end{split}    
\end{equation}
Subsequently, we interchange the moments $m^{\lambda(\bm{x},t)}_j$ with their populations at the local reference frame, $ m^{\lambda(\bm{x},t)}_j=\mathcal{M}^{\lambda(\bm{x},t)}_{j,k}f^{\lambda(\bm{x},t)}_k,$ and retrieve the following equation,

\begin{equation}
\label{StartingEquationPonD_4}
\begin{split}
    \partial_t([\mathcal{M}^{\bar \lambda}]^{-1}_{i,j}\mathcal{M}^{\lambda(\bm{x},t)}_{j,k}f^{\lambda(\bm{x},t)}_k)+ \bm{v}_{i }^{\overline{\lambda}}  \cdot \nabla ([\mathcal{M}^{\bar \lambda}]^{-1}_{i,j}\mathcal{M}^{\lambda(\bm{x},t)}_{j,k}f^{\lambda(\bm{x},t)}_k) \\ =[\mathcal{M}^{\bar \lambda}]^{-1}_{i,j}\mathcal{M}^{\lambda(\bm{x},t)}_{j,k}\Omega^{\lambda(\bm{x},t)}_{k}.
\end{split}    
\end{equation}

The multiplication of the matrices, $[\mathcal{M}^{\bar \lambda}]^{-1}_{i,j}\mathcal{M}^{\lambda(\bm{x},t)}_{j,k}$, is the definition of the reference frame transformation, 
\begin{equation}
    [\mathcal{G}_{\lambda(\bm{x},t)}^{\bar \lambda}]_{i,k}=[\mathcal{M}^{\bar \lambda}]^{-1}_{i,j}\mathcal{M}^{\lambda(\bm{x},t)}_{j,k},
\end{equation}
from the local frame ${\lambda(\bm{x},t)}$ to the monitoring frame $\bar \lambda$. Thus, Eq.\ \eqref{StartingEquationPonD_4} is the final PonD equation,

\begin{equation}
  \begin{split}
\partial_t(\mathcal{G}_{i,\lambda(\bm{x},t)}^{\overline{\lambda}} f^{\lambda(\bm{x},t)})+\bm{v}_{i }^{\overline{\lambda}} \cdot \nabla (\mathcal{G}_{i,\lambda(\bm{x},t)}^{\overline{\lambda}} f^{\lambda(\bm{x},t)}) \\ = \mathcal{G}_{i,\lambda(\bm{x},t)}^{\overline{\lambda}}\Omega^{\lambda(\bm{x},t)},
\end{split}
\end{equation}
where for convenience, the summation over repeated indices is not explicit,
\begin{equation}
    \mathcal{G}_{i,\lambda(\bm{x},t)}^{\overline{\lambda}} f^{\lambda(\bm{x},t)}=[\mathcal{G}_{\lambda(\bm{x},t)}^{\bar \lambda}]_{i,k} f^{\lambda(\bm{x},t)}_k
\end{equation}

\section{Conservation properties}

Without loss of generality, we consider a face at $\bm{x}_I$, with a unit normal vector pointing at the x-direction. According to the presented scheme, the fluxes have been calculated with a reference frame $\overline{\lambda}$ and are calculated as,

\begin{align}
F_{f,i}^{\overline{\lambda}}(\bm{x}_I) &= v_{ix}^{\overline{\lambda}} f_{i}^{\overline{\lambda}}(\bm{x}_I),  \\
F_{g,i}^{\overline{\lambda}}(\bm{x}_I) &= v_{ix}^{\overline{\lambda}} g_{i}^{\overline{\lambda}}(\bm{x}_I).  
\end{align}
The $f-$ populations at the left and right neighbouring cells $\bm{x}_L$, $\bm{x}_R$ are updated due to the fluxes $F_{f,i}^{\overline{\lambda}}(\bm{x}_I)$ as,
\begin{align}
    f_i^{\lambda_L}(\bm{x}_L,t_{n+1}) &= f_i^{\lambda_L}(\bm{x}_L,t_{n})-\delta t \mathcal{G}_{i,\overline{\lambda}}^{\lambda_L} F_{f}^{\overline{\lambda}}(\bm{x}_I), \\
    f_i^{\lambda_R}(\bm{x}_R,t_{n+1}) &= f_i^{\lambda_R}(\bm{x}_R,t_{n})+\delta t \mathcal{G}_{i,\overline{\lambda}}^{\lambda_R} F_{f}^{\overline{\lambda}}(\bm{x}_I).
\end{align}
and accordingly the $g-$ populations,
\begin{align}
    g_i^{\lambda_L}(\bm{x}_L,t_{n+1}) &= g_i^{\lambda_L}(\bm{x}_L,t_{n})-\delta t \mathcal{G}_{i,\overline{\lambda}}^{\lambda_L} F_{g}^{\overline{\lambda}}(\bm{x}_I), \\
    g_i^{\lambda_R}(\bm{x}_R,t_{n+1}) &= g_i^{\lambda_R}(\bm{x}_L,t_{n})+\delta t \mathcal{G}_{i,\overline{\lambda}}^{\lambda_R} F_{g}^{\overline{\lambda}}(\bm{x}_I),    
\end{align}
By summing over the population, we obtain the updates of the mass, momentum and energy at the left $\delta \rho_L, \delta (\rho \bm{u})_L, \delta (\rho E)_L $ and right $\delta \rho_R, \delta (\rho \bm{u})_R, \delta (\rho E)_R $ neighbouring cells due to the fluxes of the interface,
\begin{widetext}
\begin{align}
    \delta \rho_L &=\sum_{i=0}^{Q-1} \left \{ f_i^{\lambda_L}(\bm{x}_L,t_{n+1})-f_i^{\lambda_L}(\bm{x}_L,t_{n}) \right \}=-\delta t \sum_{i=0}^{Q-1}  \mathcal{G}_{i,\overline{\lambda}}^{\lambda_L} F_{f}^{\overline{\lambda}}(\bm{x}_I), \\
    \delta (\rho \bm{u})_L &=\sum_{i=0}^{Q-1} \left \{ \bm{v}_i^{\lambda_L} f_i^{\lambda_L}(\bm{x}_L,t_{n+1})-\bm{v}_i^{\lambda_L} f_i^{\lambda_L}(\bm{x}_L,t_{n}) \right \}=-\delta t \sum_{i=0}^{Q-1}  \bm{v}_i^{\lambda_L} \mathcal{G}_{i,\overline{\lambda}}^{\lambda_L} F_{f}^{\overline{\lambda}}(\bm{x}_I), \\   
    \delta (\rho E)_L &=\sum_{i=0}^{Q-1} \left \{ \frac{({v}_i^{\lambda_L})^2}{2} f_i^{\lambda_L}(\bm{x}_L,t_{n+1})+g_i^{\lambda_L}(\bm{x}_L,t_{n+1})-\frac{({v}_i^{\lambda_L})^2}{2} f_i^{\lambda_L}(\bm{x}_L,t_{n})-g_i^{\lambda_L}(\bm{x}_L,t_{n}) \right \}    \\ \nonumber
    &= -\delta t \sum_{i=0}^{Q-1} \left \{  \frac{({v}_i^{\lambda_L})^2}{2} \mathcal{G}_{i,\overline{\lambda}}^{\lambda_L} F_{f}^{\overline{\lambda}}(\bm{x}_I)+\mathcal{G}_{i,\overline{\lambda}}^{\lambda_L} F_{g}^{\overline{\lambda}}(\bm{x}_I) \right \}, \\ \nonumber \\
    \delta \rho_R &=\sum_{i=0}^{Q-1} \left \{ f_i^{\lambda_R}(\bm{x}_R,t_{n+1})-f_i^{\lambda_R}(\bm{x}_R,t_{n}) \right \}=\delta t \sum_{i=0}^{Q-1}  \mathcal{G}_{i,\overline{\lambda}}^{\lambda_R} F_{f}^{\overline{\lambda}}(\bm{x}_I), \\
    \delta (\rho \bm{u})_R &=\sum_{i=0}^{Q-1} \left \{ \bm{v}_i^{\lambda_R} f_i^{\lambda_R}(\bm{x}_R,t_{n+1})-\bm{v}_i^{\lambda_R} f_i^{\lambda_R}(\bm{x}_R,t_{n}) \right \}=\delta t \sum_{i=0}^{Q-1}  \bm{v}_i^{\lambda_R} \mathcal{G}_{i,\overline{\lambda}}^{\lambda_R} F_{f}^{\overline{\lambda}}(\bm{x}_I), \\   
    \delta (\rho E)_R &=\sum_{i=0}^{Q-1} \left \{ \frac{({v}_i^{\lambda_R})^2}{2} f_i^{\lambda_R}(\bm{x}_R,t_{n+1})+g_i^{\lambda_R}(\bm{x}_R,t_{n+1})-\frac{({v}_i^{\lambda_R})^2}{2} f_i^{\lambda_R}(\bm{x}_R,t_{n})-g_i^{\lambda_R}(\bm{x}_R,t_{n}) \right \}     \\ \nonumber
    &= \delta t \sum_{i=0}^{Q-1} \left \{  \frac{({v}_i^{\lambda_R})^2}{2} \mathcal{G}_{i,\overline{\lambda}}^{\lambda_R} F_{f}^{\overline{\lambda}}(\bm{x}_I)+\mathcal{G}_{i,\overline{\lambda}}^{\lambda_R} F_{g}^{\overline{\lambda}}(\bm{x}_I) \right \}.    
\end{align}
\end{widetext}
The finite volume is strictly conservative with respect to mass, momentum and total energy when,
\begin{align}
 \delta \rho_R+\delta \rho_L &= 0, \\
 \delta  (\rho \bm{u})_R+\delta (\rho \bm{u})_L &= 0, \\
 \delta (\rho E)_R+\delta (\rho E)_L &= 0. 
\end{align}
Substituting from the above expressions we arrive at the following constraints,
\begin{align}
\sum_{i=0}^{Q-1} \mathcal{G}_{i,\overline{\lambda}}^{\lambda_L} F_{f}^{\overline{\lambda}} & = \sum_{i=0}^{Q-1} \mathcal{G}_{i,\overline{\lambda}}^{\lambda_R} F_{f}^{\overline{\lambda}} , \\
 \sum_{i=0}^{Q-1}  \bm{v}_i^{\lambda_L} \mathcal{G}_{i,\overline{\lambda}}^{\lambda_L} F_{f}^{\overline{\lambda}} & =  \sum_{i=0}^{Q-1}  \bm{v}_i^{\lambda_R} \mathcal{G}_{i,\overline{\lambda}}^{\lambda_R} F_{f}^{\overline{\lambda}},\\
 \begin{split}
 \sum_{i=0}^{Q-1}   \frac{(\bm{v}_i^{\lambda_L})^2}{2} \mathcal{G}_{i,\overline{\lambda}}^{\lambda_L} F_{f}^{\overline{\lambda}}+\mathcal{G}_{i,\overline{\lambda}}^{\lambda_L} F_{g}^{\overline{\lambda}}  & =  \sum_{i=0}^{Q-1}  \frac{(\bm{v}_i^{\lambda_R})^2}{2} \mathcal{G}_{i,\overline{\lambda}}^{\lambda_R} F_{f}^{\overline{\lambda}} \\ & +\mathcal{G}_{i,\overline{\lambda}}^{\lambda_R} F_{g}^{\overline{\lambda}}.
 \end{split}
\end{align}

The constraints are satisfied if the following moments of the $f-$ populations are invariant upon reference frame transformation ,
\begin{align}
\sum_{i=0}^{Q-1} v_{ix}^{\lambda_L}f_i^{\lambda_L} &= \sum_{i=0}^{Q-1} v_{ix}^{\lambda_R}f_i^{\lambda_R} ,\\
\sum_{i=0}^{Q-1} \bm{v}_{i}^{\lambda_L} v_{ix}^{\lambda_L}f_i^{\lambda_L} &= \sum_{i=0}^{Q-1} \bm{v}_{i}^{\lambda_R} v_{ix}^{\lambda_R}f_i^{\lambda_R} ,\\
\sum_{i=0}^{Q-1} (\bm{v}_{i}^{\lambda_L})^2 v_{ix}^{\lambda_L}f_i^{\lambda_L} &= \sum_{i=0}^{Q-1} (\bm{v}_{i}^{\lambda_R})^2  v_{ix}^{\lambda_R}f_i^{\lambda_R}.
\end{align}
and the following for the $g-$ populations,
\begin{equation}
    \sum_{i=0}^{Q-1} v_{ix}^{\lambda_L}g_i^{\lambda_L} = \sum_{i=0}^{Q-1} v_{ix}^{\lambda_R}g_i^{\lambda_R}.
\end{equation}

\section{ Forcing scheme }

We consider the continuous kinetic equation, with a forcing term,
\begin{equation}
\label{eq:f_equation}
{\partial_t f_i}+\bm{v}_i \cdot \nabla f_i=\frac{1}{\tau}(f_i^{\rm eq}-f_i)+F_i.
\end{equation}
The body force can be expressed as
\begin{equation}
    F_i=-\frac{\bm{a}\cdot(\bm{v}_i-\bm{u})}{T}f_i^{\rm eq}.
\end{equation}
The force can be incorporated by the Strang-Splitting approach,
\begin{align}
{\partial_t f_i}&=F_i,\ (\mathrm{for} \ \delta t/2)\\
{\partial_t f_i}+\bm{v}_i \cdot \nabla f_i&=\frac{1}{\tau}(f_i^{\rm eq}-f_i),\  (\mathrm{for} \ \delta t) \\
{\partial_t f_i}&=F_i \ (\mathrm{for} \ \delta t/2).
\end{align}
The intermediate step is the kinetic update without body force. In the two half-time forcing steps, the distribution function and the macroscopic velocity are updated as,
\begin{align}
    f_i^{\ast} &= f_i+\frac{1}{2}\delta t F_i,\\
    \bm{u}^{\ast} &= \bm{u}+\frac{1}{2}\delta t \bm{a}.
\end{align}

\newpage
\
\newpage

\bibliography{bibliogr}

\end{document}